\pdfoutput=1
\documentclass[11pt]{article}

\usepackage[T1]{fontenc}
\usepackage[margin=1in]{geometry}
\usepackage{color}
\usepackage{setspace}
\usepackage{amsmath,amsfonts,graphicx,amssymb,bm,amsthm}
\usepackage{appendix}
\usepackage{hyperref}
\usepackage{color}
\usepackage{xcolor}
\usepackage{multirow}
\usepackage{makecell}
\usepackage{booktabs}
\usepackage[ruled, vlined]{algorithm2e}
\usepackage{natbib}
\usepackage{mathtools}
\usepackage{cleveref}
\usepackage{titlesec}
\usepackage{thmtools}
\makeatletter
\ifdefined\newcounteralias
  \let\sscompat@newcounteralias\newcounteralias
  \renewcommand*\newcounteralias[2]{%
    \@ifundefined{c@#1}{\sscompat@newcounteralias{#1}{#2}}{}%
  }
\fi
\makeatother
\usepackage{caption,subfigure}
\newif\ifappendixlemma
\appendixlemmafalse
\usepackage{tikz}
\usetikzlibrary{arrows.meta,positioning}

\crefname{assumption}{Assumption}{Assumption}
\crefname{algorithm}{Algorithm}{Algorithm}
\newcommand{\vmax}{{\overline{v}}}
\newcommand{\coupling}{{(\mathtt{c})}}
\newcommand{\fluid}{{(\mathtt{e})}}
\newcommand{\historic}{{(\mathtt{h})}}
\newcommand{\resolving}{{(\mathtt{r})}}
\newcommand{\discretized}{{(\mathtt{d})}}
\newcommand{\online}{{(\mathtt{o})}}
\newcommand{\actual}{\online}
\newcommand{\ulower}{{\underline{u}}}
\newcommand{\uupper}{{\overline{u}}}

\newcommand{\betalower}{\underline{\beta}}

\newcommand{\bv}{{\mathbf{v}}}
\newcommand{\sumi}{{\sum_{i=1}^n}}
\newcommand{\sumt}{{\sum_{t=1}^T}}

\newcommand{\indi}[1]{{\mathbf{1}\left(#1\right)}}

\theoremstyle{plain}
\newtheorem{theorem}{Theorem}[section]

\newtheorem{lemma}[theorem]{Lemma}
\newtheorem{proposition}[theorem]{Proposition}

\newtheorem{assumption}{Assumption}
\newtheorem{definition}{Definition}

\titleformat{\paragraph}[runin]
  {\normalfont\bfseries}
  {\theparagraph}
  {1em}
  {}                       
  
\title{ Online Generalized-Mean Welfare Maximization:\\ Achieving Near-Optimal Regret from Samples}

\author{Zongjun Yang\thanks{Columbia University. Email: \texttt{zy2684@columbia.edu}}  \qquad Rachitesh Kumar\thanks{Carnegie Mellon University. Email: \texttt{rachitesh@cmu.edu}} \qquad Christian Kroer\thanks{Columbia University. Email: \texttt{ck2945@columbia.edu}}  }
\date{}
  
\begin{document}

\maketitle
\doublespacing

\begin{abstract}

We study online fair allocation of $T$ sequentially arriving items among $n$ agents with heterogeneous preferences, with the objective of maximizing generalized-mean welfare, defined as the $p$-mean of agents' time-averaged utilities, with $p\in (-\infty, 1)$. We first consider the \textit{i.i.d.}\ arrival model and show that the pure greedy algorithm---which myopically chooses the welfare-maximizing integral allocation---achieves $\widetilde{O}(1/T)$ average regret. Importantly, in contrast to prior work, our algorithm does not require distributional knowledge and achieves the optimal regret rate using only the online samples.

We then go beyond \textit{i.i.d.} arrivals and investigate a nonstationary model with time-varying independent distributions. In the absence of additional data about the distributions, it is known that every online algorithm must suffer $\Omega(1)$ average regret. We show that only a \textit{single} historical sample from each distribution is sufficient to recover the optimal $\widetilde{O}(1/T)$ average regret rate, even in the face of arbitrary non-stationarity. Our algorithms are based on the re-solving paradigm: they assume that the remaining items will be the ones seen historically in those periods and solve the resulting welfare-maximization problem to determine the decision in every period. Finally, we also account for distribution shifts that may distort the fidelity of historical samples and show that the performance of our re-solving algorithms is robust to such shifts.

\end{abstract}
\newpage
\tableofcontents
\newpage
\section{Introduction}
\label{sec: introduction}
Allocating sequentially arriving items to agents---at scale and with low latency---is a fundamental challenge in many marketplaces and Internet platforms. These settings typically require the decision maker to commit to irrevocable decisions in real time without knowledge of future arrivals. In the most canonical applications, items are time-sensitive resources that become available over time and must be assigned promptly to satisfy heterogeneous demands.
Examples include allocating computational resources to cloud users~\citep{wei2010game,wang2014multi,dinh2020online}, donations to food banks~\citep{alkaabneh2021unified,sinclair2022sequential}, and medical equipment or vaccines to hospitals~\citep{grigoryan2021effective,pathak2021fair}. Another class of applications in online economics treats ``items'' as opportunities for user impression upon their visit, as in online advertising~\citep{conitzer2022multiplicative,conitzer2022pacing}, notification feeds~\citep{kroer2023fair}, and mobilization platforms for food rescue~\citep{shi2020improving}, volunteering~\citep{manshadi2023redesigning}, and blood donation apps~\citep{mcelfresh2020matching}. For these applications, the scarce resources are the platform users who arrive online, and they need to be allocated appropriately to the stakeholders or communities who benefit from users' engagement. 

Across these settings, a central challenge is to balance efficiency and fairness considerations across agents that may have heterogeneous preferences. For example, a food-bank network may allocate perishable donations to the pantries that can distribute them in high volume and in short time, while it also needs to ensure that no county or neighborhood is systematically underserved in a distribution cycle. Similarly, in notification optimization, platforms like Instagram aim to maximize the total user engagement generated by the impressions, while they also have to maintain a meaningful share of user attention to every campaign or creator group. To what extent efficiency should be traded off for fairness, however, differs across applications, and even across operating periods and service types of the same platform. A decision maker is therefore rarely fully served by an algorithm tailored to one fixed objective, or to one fixed fairness criterion. To that end, we propose and analyze an algorithmic principle which performs well along an entire spectrum of efficiency--fairness tradeoffs, so that the platform can choose an appropriate balance without changing the algorithm or losing its performance guarantees.
 
A well-studied way to formalize the efficiency--fairness tradeoff in allocation problems is through the generalized-mean welfare family $\left(\sum_i u_i^{p}\right)^{1/p}$~\citep{moulin2004fair}, which aggregates agents' utilities via the $p$-H\"older mean function. The parameter $p \in (-\infty, 1]$ captures the extent of inequality aversion. As the parameter varies, the objective continuously interpolates between two extremes: at the efficiency end (corresponding to $p=1$), the objective is the utilitarian welfare, focusing solely on total surplus with no regard for equitable utility distribution across agents, while in the limit $p\to -\infty$ it approaches the egalitarian welfare criterion, which prioritizes the worst-off agent only and could be highly inefficient. A particularly important intermediate point is $p=0$ (the geometric mean of utilities), which corresponds to the well-known objective of Nash welfare~\citep{nash1950bargaining}. The optimal allocation under the Nash welfare objective is invariant to any single agent's utility scaling. In the setting with linear utilities, it corresponds to a competitive equilibrium (CE) of the linear Fisher market, and is known to imply strong fairness guarantees such as envy-freeness and proportionality~\citep{varian1974equity,moulin2004fair}.

In this paper, we study an online allocation problem with $n$ heterogeneous agents and a horizon of $T$ arrivals. In each round, a unit-supply item arrives, each agent's value for it is revealed, and the decision maker makes an irrevocable allocation before seeing future arrivals. The goal is to maximize the generalized-mean welfare over the $n$ agents. We measure the performance of an online algorithm by its time-averaged \textit{regret}, defined as the additive gap between the hindsight optimal and the algorithmic generalized-mean welfare on time-averaged utilities.

In real-world applications, arrival processes can range from idealized stationary patterns to strongly nonstationary ones, which motivates the investigation of different arrival models. For example, \citet{balseiro2022uniformly} studies online generalized-mean welfare maximization under an \textit{i.i.d.}\ model with a \emph{known} arrival distribution, which assumes precise and stable prior knowledge of a high-dimensional supply--demand environment (e.g., traffic, user population, product mix). In practice, such knowledge can quickly become stale due to seasonality, competitor actions, or exogenous events. At the other extreme, the worst-case deterministic arrivals considered in \citet{barman2022universal,huang2025long} can be overly pessimistic: many platforms operate in data-rich regimes where historical logs offer partial predictability, and the dominant nonstationarity is often gradual drift with recurring patterns rather than fully adversarial behavior. To this end, we focus on \textit{stochastic }arrivals with \emph{unknown} distributions, where the decision maker can only access the item \textit{samples}. We study two settings: 1) completely prior-free \textit{i.i.d.}\ arrivals, and 2) nonstationary arrivals with auxiliary samples from historical data.

\subsection{Our Contributions}
\label{sec: contribution}

\paragraph{Greedy Achieves No-Regret under Unknown Distribution, \textit{i.i.d.}\ Input.}

Our first result addresses the basic stochastic benchmark of stationary \textit{i.i.d.}\ arrivals with an \emph{unknown} distribution. We show that a remarkably simple greedy policy (\Cref{alg: greedy})---which, upon each arrival, assigns the item to the agent that maximizes the instantaneous increase in generalized-mean welfare---achieves near-optimal $\widetilde O(1/T)$ time-averaged regret for all $p\in(-\infty,1)$ (\Cref{thm: greedy-for-iid}). This yields the first near-optimal regret bound in the prior-free \textit{i.i.d.}\ setting for online generalized-mean welfare maximization, including the Nash welfare ($p=0$) case.

Beyond being prior-free, the guarantee is (essentially) as strong as what is known under substantially more restrictive assumptions: \citet{balseiro2022uniformly} obtain $O(1/T)$ average regret in the \textit{known} distribution, \textit{i.i.d.}\ setting, requiring full and precise knowledge of the arrival distribution. Their algorithm involves solving a fluid relaxation that is computationally viable only under strong finite-type assumptions. In contrast, our greedy algorithm works for continuous value space, runs in real time with minimal computation, and is integral---there is no need to split items fractionally---while losing only a logarithmic factor in the regret bound.

\paragraph{Re-solving under Nonstationary Arrivals with a Single Sample.}

We next move beyond the \textit{i.i.d.}\ model and consider independent but time-varying distributions. When the distributions are arbitrarily nonstationary and nothing is known about them, \citet{barman2022universal,huang2025long} show that any online algorithm must suffer from constant time-averaged regret, with an approximation ratio no better than $\operatorname{poly}(1/n)$. They show improved approximation factors under additional normalization, but this still gives constant average regret.

In real-world scenarios, fortunately, additional information is usually available in the form of historical data. Most online platforms operate over repeating operational windows, e.g., a
day or a week. Within an operational window, the arrivals could have a highly nonstationary pattern, with their distributions inherently varying over time. Across operational windows,
however, the patterns tend to recur, and are usually recorded in the platform's log data. For example, user traffic or activity usually fluctuates highly within a day from morning to evening, but the patterns of fluctuation are often similar for subsequent days, and it is
often meaningful to learn from the data from past days to better handle a new day's operations. 

In this paper, we consider the setting where the decision maker only has access to data from \emph{one} historical operational window. Formally, they are assumed to have access to one realization of the nonstationary arrival process, or equivalently, a single sample point for each of the $T$ time-varying distributions. Compared with learning from many past operational windows, this single-sample paradigm minimizes data dependency, and may improve data fidelity by using the historical window that best resembles the current one. More importantly, we show that this minimal amount of information is sufficient to achieve the optimal regret rate. 

In this single-sample setting, we extend the same algorithmic principle behind greedy---allocating each item by solving the welfare maximization problem using the best information currently available---to develop a class of re-solving algorithms (\Cref{alg: re-solving (primal),alg: re-solving (dual)}). Concretely, at each time step the algorithm forms a welfare maximization program which, given the irrevocable past, greedily maximizes over future allocations using historical data as a forecast, and then allocates the current item as suggested by an optimizer of this proxy problem. We show that the re-solving algorithms achieve the near-optimal $\widetilde O(1/T)$ regret rate as in the \textit{i.i.d.}\ case despite nonstationarity, using only one sample per distribution (\Cref{thm: adaptive-with-density}).

\paragraph{Re-solving is Robust to Distribution Shifts.}

In practice, historical logs are rarely perfectly representative samples of the environment faced online. User populations change, supply and demand drift, and selection effects can introduce systematic bias, causing arrival distributions to slowly shift over time. To capture this reality, we allow the historical trace and the online process to be generated from different distributions, and quantify the distribution shift by the time-averaged Wasserstein discrepancy $\mathcal{W}$ (in $\ell_1$ metric). We show that under distribution shifts, the guarantees degrade smoothly: the average regret is $\widetilde O(1/\sqrt{T}+\mathcal{W})$ (\Cref{thm: adaptive}), achieved by the same re-solving algorithm without making any adaptations to the distribution drift. This provides a robust performance certificate in the face of potential distribution shifts.

\subsection{Technical Highlights}
To analyze the greedy algorithm, we use a \textit{dual charging} framework where we interpret the greedy algorithm as a primal-dual algorithm using shadow prices. We leverage its greedy nature to argue that the algorithm \textit{self-stabilizes}, with all primal iterates bounded after logarithmic rounds with high probability. This enables effective smoothness in the analysis, despite the fact that the objective itself is non-Lipschitz and non-smooth.

To analyze the re-solving algorithm, we use a \textit{clairvoyant coupling} framework. Intuitively, we ask: 
\textit{Given the decisions made by the algorithm so far, what is the best possible outcome if we perform all future allocations optimally?} This gives a sequence of coupling programs, and their consecutive differences
sum up to total suboptimality. We then apply refined uniform concentration analysis to bound the number of changes in the coupling programs in expectation. For this proof, we also establish monotonicity and stability properties for generalized-mean welfare maximizers (\Cref{lem: stability}) and corresponding concentration properties (\Cref{lemma: concentration-single}), which are new findings even for the $p=0$ case, and could be of broader interest.

To further incorporate distribution shifts, we generalize the coupling framework to allow for arbitrary coupling arrivals. By choosing the coupling sequence appropriately, this enables decomposition of the regret into 1) an unbiased stochastic estimation term, and 2) an explicit distribution-shift term linear in the time-averaged Wasserstein discrepancy. This approach potentially extends to broader classes of structured online convex optimization problems. 

\subsection{Relationship to Literature}
\label{section: related}

\paragraph{Online Generalized-Mean maximization.}

In the online allocation setting with sequentially arriving items, the problem of maximizing generalized-mean welfare $(p<1)$ of linear-utility agents has been studied under deterministic, worst-case arrivals, with the goal of achieving constant approximation ratios \textit{w.r.t.} the hindsight optimum. When no assumptions are made on the deterministic arrivals, the best possible approximation ratio is $\Theta(\operatorname{poly}(1/n))$~\citep{barman2022universal,huang2025long}. Under a normalization assumption that the monopolistic utility (total value over the entire horizon, $\sum_t v_{t,i}$) of each agent is $1$, \citet{barman2022universal} develops a threshold-based algorithmic framework that yields competitive approximation ratios for all $p<1$ and $p=-\infty$ (egalitarian welfare). Using the same normalization assumption, \citet{huang2025long} sharpens the guarantees with a more fine-grained characterization of how the optimal competitive ratio evolves with $p$; for $p < 1/\log n$ they use a greedy-style algorithm with mixed Nashian and auxiliary components and for $p\in (1/\log n,1)$ they argue optimality of the existing algorithm template for general concave-return objectives~\citep{devanur2012online}. Notably in these works, the best achievable approximation ratio is bounded away from $1$: no better than $\max\{1/\log n, p\}$ for $p\in [-1/\log n, 1)$ and $\mathrm{poly}(1/n)$ for $p\in [-\infty, -1/\log n)$. In contrast to these worst-case approximation results, we seek \textit{no-regret} type of guarantees ($1-o(1)$ approximation ratio) under arrival models with stronger stochasticity or information assumptions; we do not consider the $p = -\infty$ case (egalitarian welfare).

For the ideal \textit{i.i.d.}\ input model with the distribution \textit{known} to the decision maker, \citet{balseiro2022uniformly} achieves $O(1/T)$ time-averaged regret for generalized-mean welfare with $p\in (-\infty, 1)$ with a fluid-based algorithm. Their algorithm critically relies on the knowledge of the entire arrival distribution and optimizes over an expectation \textit{w.r.t.} the distribution per step, which can be computationally inefficient or even intractable for continuous distributions. In contrast, our greedy algorithm does not require any prior knowledge on the distributions to achieve $\widetilde O(1/T)$ time-averaged regret for \textit{i.i.d.} arrivals, which is only worse than their bound by a logarithmic factor. This significantly increases data and computational efficiency with little compromise on regret. \citet{cohen2024near} considers the uniform permutation arrival model. Their $\widetilde{O}(1/T)$ bound applies only for $p<-1$ and requires a stringent assumption that the \textit{minimum}-achievable welfare (which can be $0$ in most practical cases) is sufficiently large. 

For the specific case $p=0$ (Nash welfare), \citet{gao2021online,yang2024onlinefairallocationbestofmanyworlds} show an efficient, greedy-based algorithm that converges to the competitive equilibria of the underlying Fisher market with $\widetilde{O}(1/\sqrt{T})$ rate in \textit{i.i.d.}\ and stationary-like settings; they focus on fairness properties such as envy-freeness and proportionality rather than regret analysis. For the $p= -\infty$ case (egalitarian welfare), which we do not cover, \citet{kawase2022online} develops specialized algorithms that achieve near-optimal regret guarantees under \textit{i.i.d.}\ and worst-case inputs, respectively. 

\paragraph{Comparisons with Online Resource Allocation.} In this paper, we study a fair allocation problem where the item \textit{supply} arrives online, and we need to allocate them equitably to $n$ agents to maximize the generalized-mean welfare. This is fundamentally different from the classical \textit{online resource allocation} problem, where the supply is fixed under constraints, and the online arrivals are \textit{demands} to consume the resources. In online resource allocation, the objective is typically utilitarian and time-separable, with no notions of consumer agents and fairness among them. As a result, fairness-aware objectives such as generalized-mean on the consumer side have not been discussed in the online resource allocation literature, despite existing literature on \textit{supply} side fairness, e.g., \citet{balseiro2021regularized}. Therefore, online resource allocation algorithms are not applicable to our setting. The achievable regret also differs in the two problems: the optimal time-averaged regret is $\widetilde{O}(1/\sqrt{T})$ for the \textit{i.i.d.} setting with general unknown distribution~\citep{balseiro2023best}, while the target regret bound in this paper for generalized-mean maximization is $\widetilde{O}(1/T)$.

\paragraph{The Single Sample Paradigm.} A growing literature studies online decision-making under unknown, time-varying distributions, provided with a \emph{single sample per distribution} (equivalently, a single historical sequence). This paradigm is first studied in prophet inequalities and secretary problems with unknown distributions~\citep{azar2014prophet}, with a stream of literature on the problem variants and improvements~\citep{rubinstein2019optimal,kaplan2020competitive,caramanis2022single,gravin2022optimal,cristi2024prophet,ezra2026prophet}. In operations and market motivated settings, \cite{banerjee2020uniform,banerjee2020constant} considers a class of online decision-making problems, including bin packing, and achieves near-optimal regret with a single sample trace. Their results require the set of possible request types to be finite. For the online resource allocation problem with fixed supply and arriving demands, \citet{balseiro2023robust,ghuge2025single,feng2026non} gives robust-to-bias regret bounds by combining the expenditure plans learned from the history with dynamic rate controllers. Different from these works, our algorithm for the single sample setting is simply based on re-solving.

\paragraph{Online Convex Optimization.}
One might hope to investigate generalized-mean welfare maximization through general 
online convex optimization (OCO) frameworks such as \citet{agrawal2014fast}. 
However, their $\widetilde{O}(1/T)$ time-averaged regret results critically rely on 
the Lipschitzness and smoothness of the objective, which is not satisfied by the generalized-mean 
welfare functions. For non-smooth and non-strongly concave objectives such as ours, 
\citet{agrawal2014fast} only achieves $\widetilde{O}(1/\sqrt{T})$ time-averaged regret 
even after omitting the Lipschitzness issue. Our coupling method tracks the performance 
of the algorithm using a sequence of clairvoyant programs, and is similar to the Compensated 
Coupling methodology by \citet{vera2019bayesian}. Compared with their work, we extend the 
coupling sequences from the exact hindsight optimizer to general sequences, thus enabling 
our analysis under distribution shifts. For online stochastic optimization,
\citet{jiang2025online} uses Wasserstein distance to measure the nonstationarity of the arrival distributions across 
periods, whereas our Wasserstein metric measures the discrepancy between the sample and actual distributions.

\section{Model and Preliminaries}
\label{section: preliminaries}
\paragraph{Notations.} We use $\mathbb{R}_+^n$ to denote the set of non-negative $n$-dimensional real vectors, and $\mathbb{R}_{++}^n$ to denote the space of strictly positive $n$-dimensional real vectors. 
For any integer $k \in \mathbb{N}_+$, let $[k]$ denote the set $\{1, \cdots, k\}$. 
For any $q>0$ or $q = \infty$, let $\|\cdot \|_q$ denote the $\ell_q$-norm. 
For any real numbers $y<z$, we let $[y,z]^n$ denote the $n$-dimensional hyper-rectangle $[y, z]\times \cdots\times [y,z]$.  

\subsection{Online Fair Allocation with Generalized-Mean Welfare Objective}
In this paper, we consider the problem of dynamically allocating $T$ online items to $n$ agents. Each item $t\in [T]$ is described as a value vector $\bm{v}_t  = (v_{t,1}, \cdots, v_{t,n})\in [0, \vmax]^n$, where $v_{t,i}$ is the unit valuation of agent $i$ for item $t$ (i.e., utility gained when receiving one unit of the item); $\vmax >0$ is a constant. At the $t$'th time step, item $t$ arrives and reveals its value vector $\bm v_t$. The decision maker then commits to an irrevocable allocation $\bm{x}_t  = (x_{t,1}, \cdots, x_{t,n})\in \mathbb{R}_+^n$ without seeing the future arrivals, subject to unit supply constraint $\|\bm{x}_t\|_1\leq 1$. By the end of the $T$ time steps, the time-averaged utilities of agents are given by $\bm{u} = (u_1, \cdots, u_n) \in \mathbb{R}_+^n$ where $u_{i} = (1/T) \sum_{t=1}^T v_{t,i} x_{t,i}$. 

For notational simplicity, we denote $\bm v_{s:t} := (\bm v_{s}, \cdots, \bm v_{t}) \in [0, \vmax]^{n\times t-s+1}$. The input to the problem is denoted by $\mathbf{v} := \bm v_{1:T}$, which we refer to as an item sequence.

We say the allocation is integral if $\bm x_t \in \{0,1\}^n$ for all $t\in[T]$. We require $T\geq n$ throughout the paper so the decision maker can guarantee at least one item for each $i\in [n]$ even when restricted to integral allocations.\footnote{Our algorithms \Cref{alg: greedy} and \Cref{alg: re-solving (dual)} can do integral allocations only, but we still allow the competing hindsight benchmark to use fractional allocation.} 

\paragraph{Welfare Measures.}
We evaluate the outcomes by $f(\bm u)$, where $f:\mathbb{R}^n \rightarrow \mathbb{R}$ is a \textit{social welfare} function that aggregates the utility levels of agents and is meant to capture a certain balance of fairness and efficiency.
We motivate our generalized-mean welfare objective as the symmetric case of the CES (constant elasticity of substitution) welfare family; see the following \Cref{definition: welfare}. For simplicity, we prove our results for generalized-mean welfare, but all analysis extends to general CES welfare by re-weighting the utilities. 

\begin{figure}[t]
\centering
\scalebox{0.8}{%
\begin{tikzpicture}[
  font=\large,
  bluebar/.style={teal, very thick},
  tick/.style={black, line width=0.9pt},
  >=Latex
]
\def\xL{0}
\def\xmone{4}
\def\xzero{6.5}
\def\xone{9}
\def\xR{12}

\def\ybar{0}
\def\tickH{0.35}

\node[anchor=south west] at (\xL-0.2, 1.05) {modulus $p$:};
\node[anchor=south] at (\xL, 0.55) {$-\infty$};
\node[anchor=south] at (\xmone, 0.55) {$-1$};
\node[anchor=south] at (\xzero, 0.55) {$0$};
\node[anchor=south] at (\xone, 0.55) {$1$};
\node[anchor=south] at (\xR, 0.55) {$+\infty$};

\draw[bluebar, <-] (\xL, \ybar) -- (\xone, \ybar);
\draw[bluebar, dotted, ->] (\xone, \ybar) -- (\xR, \ybar);

\draw[tick] (\xmone, \ybar-\tickH) -- (\xmone, \ybar+\tickH);
\draw[tick] (\xzero, \ybar-\tickH) -- (\xzero, \ybar+\tickH);
\draw[tick] (\xone,  \ybar-\tickH) -- (\xone,  \ybar+\tickH);

\node[teal, anchor=east] at (\xL-0.1, \ybar+0.15) {fairer, less efficient};
\node[teal, anchor=west] at (\xR+0.2, \ybar+0.05) {less fair};
\node[align=left, anchor=east] at (\xL-0.6, -1.4)
{};

\node[align=center] at (\xL,   -1.15) {egalitarian \\(min)};
\node[align=center] at (\xmone,-1.05) {harmonic};
\node[align=center] at (\xzero,-1.05) {Nash \\(geometric)};
\node[align=center] at (\xone, -1.05) {utilitarian \\(arithmetic)};
\node[align=center] at (\xR,   -1.05) {max};

\end{tikzpicture}%
}
\caption{Different generalized-mean welfare functions, parametrized by different values of $p$.}
\label{fig: welfare}
\end{figure}
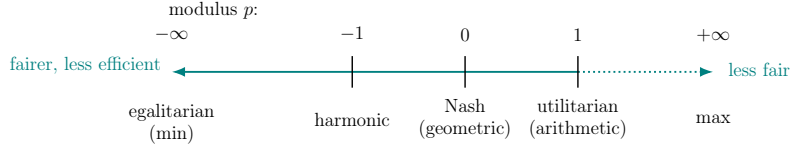
\begin{definition} \textnormal{(CES and Generalized-Mean Welfare)}
\label{definition: welfare}
    The family of CES (constant elasticity of substitution) welfare functions is the class of all $f_{\bm B, p}$ of the following form, 
    \begin{equation*}
        f_{\bm B, p}(\bm u) = \begin{cases}
        \left(\sum_{i=1}^n B_i u_i^p\right)^{1/p} & p\neq 0, \\
        \prod_{i=1}^n u_i^{B_i} & p = 0,
        \end{cases}
    \end{equation*}
    where $p \in \mathbb{R}, \bm B = (B_1, \cdots, B_n)\in \mathbb{R}_{++}^n.$ The family of generalized-mean welfare is the class of all CES welfare functions with $B_i = 1/n$ for all $i\in [n]$. 
\end{definition}

\Cref{definition: welfare} covers a broad spectrum of social preferences. In fact, it characterizes \textit{all} welfare functions that simultaneously satisfy a set of classical economic axioms; see the following proposition.

\begin{proposition}\textnormal{(\citet{hardy1952inequalities,moulin2004fair})}
\label{prop: welfare}
    If a welfare function $f: \mathbb{R}_+^n \to \mathbb{R}_+$ satisfies the following $4$ axioms, then it is in the CES family. 
    \begin{enumerate}
        \item Strict monotonicity: if $\bm u^\prime \geq \bm u$ and $\bm u^\prime \neq \bm u$, then $f(\bm u^\prime)>f(\bm u)$.
        \item Continuity: $f$ is a continuous function on $\mathbb{R}_+^n$.
        \item Independence of irrelevant agents: if $f(u_1^\prime, u_2^\prime, \cdots, u_n^\prime) \geq f(u_1^{\prime \prime}, u_2^\prime, \cdots, u_n^\prime )$, then for all $\bm u \in \mathbb{R}_n^+$ we have $f(u_1^\prime, u_2, \cdots, u_n) \geq f(u_1^{\prime\prime}, u_2, \cdots, u_n)$; the same property holds for each other entry $i=2, \cdots, n$. 
        \item Positive 1-Homogeneity: $f(\alpha \bm u) = \alpha f(\bm u)$ for any $\alpha >0$. 
    \end{enumerate}
    Moreover, $f$ satisfies the above $4$ axioms and the following principle simultaneously if and only if it is a CES welfare function with parameter $p \in (-\infty, 1)$.
    \begin{enumerate}
        \item[(5)] Strong Pigou-Dalton principle: if $u_1 > u_2$, then for any $\varepsilon\in (0, (u_1-u_2)/2)$, $f(\bm u) < f(u_1 - \varepsilon, u_2 + \varepsilon, u_3, \cdots, u_n)$. 
    \end{enumerate}
\end{proposition}

For the range $p \in (-\infty, 1)$, the CES welfares are inequality-averse, satisfying the strong Pigou-Dalton principle: they strictly prefer outcomes that reduce disparities when the sum of utilities is on the same level~\citep{moulin2004fair}. We will restrict our attention to this ``fair'' spectrum $p\in(-\infty,1)$. When $p\to -\infty$, the objective
converges to $\min_i u_i$, approaching
the egalitarian criterion, which is fair but overly conservative in terms of efficiency. At the other
extreme, $p=1$ yields the (weighted) utilitarian welfare, which is fully efficiency-oriented but often unfair as it is totally insensitive to how utilities are distributed across $n$ agents. Intermediate values $p\in(-\infty,1)$ interpolate between these
two cases; see \Cref{fig: welfare}. The case with $p=0$
corresponds to the (weighted) Nash welfare, a central objective in the fair allocation
literature~\citep{nash1950bargaining,varian1974equity,cole2017convex,caragiannis2019unreasonable}. In our linear utility setup, the Nash-welfare-maximizing solution gives strong fairness properties of envy-freeness and proportionality, and coincides with the competitive equilibrium of the underlying Fisher market (\citet{moulin2004fair}; interpreting $B_i$ as budgets). 

\paragraph{Performance Metric: Time-averaged Regret.}
We measure the performance of an online algorithm by its regret \textit{w.r.t.} the hindsight optimal value of social welfare $f$. The hindsight optimal welfare value is given by:
\begin{equation}
    \label{eq: hindsight-primal-without-log}
    \mathrm{OPT}(\mathbf v) = \max_{\bm u \geq 0, \bm x_t \geq 0} \left\{ f(\bm{u}) : u_i\leq \frac{1}{T} \sum_{t=1}^T v_{t,i} \cdot x_{t,i}, \ \|\bm x_t\|_1\leq 1 \  \ \  \forall \ t\in[T]  \right\} .
\end{equation}

We refer to the true online arrival as the \textit{online sequence} and denote it as $\mathbf{v}^\actual = (\bm v_{1}^\actual, \cdots, \bm v_T^\actual)$; we add the superscript $\online$ to distinguish it from other sequences in our analysis. We define the (time-averaged) regret as follows:

\begin{definition}\textnormal{(Time-averaged Regret)}
    For an online algorithm $\mathcal{A}$ and online sequence $\mathcal{\mathbf{v}}^\online$, let $\bm u_\mathcal{A}^\dagger(\mathcal{\mathbf{v}}^\online)$ be the resulting time-averaged utility of running the algorithm on the sequence. The time-averaged regret of $\mathcal{A}$ on $\mathcal{\mathbf{v}}^\online$ is defined as $\mathcal{R}_f(\mathcal{A, \mathbf{v}^\online} ) :=  \operatorname{OPT}(\mathbf{v}^\online)-f(\bm u^\dagger_\mathcal 
    A(\mathcal{\mathbf{v}}^\online) )$.
\end{definition}

\subsection{Arrival Models}

We consider a stochastic input model where the $t$'th item value vector $\bm v^{\online}_t$ is \textit{independently }generated from distribution $\mathcal{P}_{t}^\online \in \Delta([0, \vmax]^n)$; the distributions are \textit{unknown} to the decision maker. We use $\mathcal{P}_{s:t}^\online :=(\mathcal{P}_{s}^\online, \cdots, \mathcal P_t^\online)$ to denote the sequence of distributions for time steps $s,\ldots,t$.

\paragraph{Stationary (\textit{i.i.d.}) Arrivals.} We first consider the ideal setting where all $(\mathcal
P_t^\online)_{t\in [T]}$ are identical; the online algorithm has no prior knowledge of the distributions, and does not know any parameters, such as $\vmax$. This input model is studied in \Cref{section: greedy}. 

\paragraph{Nonstationary Arrivals with a Single Sample.} We then consider the case where $(\mathcal
P_t^\online)_{t\in [T]}$ are arbitrarily time-varying (i.e., nonstationary) distributions. This subsumes the deterministic adversarial model where $\bm v_t^\online$ are arbitrary vectors in $[0, \vmax]^n$ (by setting the distributions to singletons), and thus admits the pessimistic result that no algorithm achieves an approximation ratio better than $\mathrm{poly}(1/n)$~\citep{barman2022universal}; this is $\Omega(1)$ in terms of our time-averaged regret metric. 

To achieve meaningful guarantees, we provide the algorithm with 
auxiliary information: a single 
\textit{historical sequence} 
$\mathbf{v}^\historic = (\bm v_1^\historic, \cdots, \bm v_T^\historic)$, which is a sample trace of 
the online sampling procedure. There is 
only a single sample point for each corresponding time step 
$t\in [T]$. We assume that each $\bm v_t^\historic$ in the sample is generated independently from an \textit{unknown} distribution $\mathcal{P}_t^{\historic} \sim \Delta([0, \vmax]^n)$, and is independent of the actual online sequence. We consider this input model in \Cref{section: re-solve}.

While the sample $\bm v^\historic_t$ is meant to provide an estimate of the actual item's value vector, in practice, the sample distribution $\mathcal{P}_t^\historic$ might often be different from the true arrival distribution $\mathcal{P}_t^\online$. We capture this \textit{distribution shift} by the $\ell_1$ Wasserstein metric $\mathcal{W}_1(\mathcal{P}_t^\online, \mathcal{P}_t^\historic)$. We denote the time-averaged Wasserstein discrepancy as $\mathcal W = (1/T) \sum_{t=1}^T \mathcal{W}_1(\mathcal{P}_t^\online, \mathcal{P}_t^\historic).$ We consider the case with distribution shifts in \Cref{sec: robust}.

We summarize the superscript notations and the value sequences in \Cref{tab: superscripts} and \Cref{tab: value-sequences}.

\begin{table}[!htbp]
\centering
\small
\begin{minipage}[t]{0.56\textwidth}
\centering
\caption{Superscripts used in the paper.}
\label{tab: superscripts}
\begin{tabular}{@{}c >{\raggedright\arraybackslash}p{0.41\linewidth} >{\raggedright\arraybackslash}p{0.30\linewidth}@{}}
\toprule
Superscript & Meaning & Example \\
\midrule
$\online$ & True online arrival & $\bm v_t^\online, \ \mathcal{P}_t^\online$ \\
\addlinespace
$\historic$ & Historical sample & $\bm v_t^\historic, \ \mathcal{P}_t^\historic$ \\
\addlinespace
$\dagger$ & Quantity realized by the online algorithm & $\bm x_t^\dagger, \ \bm W_t^\dagger, \ \bm \beta_t^\dagger$ \\
\addlinespace
$\resolving$ & Re-solving quantities & $P_t^\resolving, \ \bm \beta_t^\resolving$ \\
\addlinespace
$\coupling$ & Coupling quantities constructed for the analysis & $D_t^\coupling, \ \bm u_t^\coupling$ \\
\bottomrule
\end{tabular}
\end{minipage}
\hfill
\begin{minipage}[t]{0.42\textwidth}
\centering
\caption{The three value sequences.}
\label{tab: value-sequences}
\begin{tabular}{@{}l >{\raggedright\arraybackslash}p{0.35\linewidth}@{}}
\toprule
Value sequence & Meaning \\
\midrule
$\bv^\online = (\bm v_1^\online, \cdots, \bm v_T^\online)$ & True online arrival\\
\addlinespace
$\bv^\historic = (\bm v_1^\historic, \cdots, \bm v_T^\historic)$ & Historical sample\\
\addlinespace
$\bv^\coupling = (\bm v_1^\coupling, \cdots, \bm v_T^\coupling)$ & Auxiliary coupling sequence used in analysis \\
\bottomrule
\end{tabular}
\end{minipage}
\end{table}

\subsection{Primal and Dual Convex Programs on the Logarithmic Objective}
For subsequent analysis, we set up the primal optimization program with $\log f$ objective and its dual. This enables strong convexity in the objective, which is not satisfied by the generalized-mean welfare. The hindsight log-welfare primal program is given as follows:
\begin{align}
    \label{eq: primal} 
    P^*(\mathbf{v}) = \max_{\bm u \geq 0, \bm x_t \geq 0}  &\left\{ \log f(\bm u) :{u}_i\leq \frac{1}{T} \sum_{t=1}^T v_{t,i} \cdot x_{t,i}, \ \|\bm x_t\|_1\leq 1 \ \  \forall \ t\in[T]  \right\}.
\end{align}

By the strict monotonicity of the $\log (\cdot)$ function, \eqref{eq: primal} and \eqref{eq: hindsight-primal-without-log} share the same optimal solutions. To derive its dual, we let $(-\log f)_+^\star: \mathbb{R}_+^n \to \mathbb{R}$ be the Fenchel conjugate of $\log f$ restricted to the non-negative orthant, i.e., $(-\log f)_+^\star(-\bm \beta) = \max_{\bm u\geq 0}\{\langle \bm u, -\bm \beta \rangle+\log f(\bm u)\}$. By associating each primal utility constraint $u_i \leq (1/T) \sumt v_{t,i} x_{t,i}$ with a dual variable $\beta_i\geq 0$, we can derive the dual as follows,
\begin{align}
    D^*(\mathbf{v}) &=  \min_{\bm \beta\in \mathbb{R}_+^n}  \max_{\bm u\geq 0, \bm x_t\geq 0,  \|\bm x_t\|_1 \leq 1}
    \left\{\log f(\bm u) + \sum_{i=1}^n \beta_i\left(\frac{1}{T} \sum_{t=1}^T v_{t,i} \cdot x_{t,i} - u_i\right)\right\}
    \nonumber 
    \\
    &=  \min_{\bm \beta\in \mathbb{R}_+^n} \left\{  \frac{1}{T} \sum_{t=1}^T\max_{\bm x_t\geq 0, \|\bm x_t\|_1 \leq 1} \left\{{\sum_{i=1}^n} \beta_i \cdot v_{t,i} x_{t,i} \right\} + \max_{\bm u\geq 0} \{\log f(\bm u) +\langle \bm u, -\bm \beta\rangle \}\right\}
    \nonumber \\
    &=\min_{\bm\beta\in \mathbb{R}_+^n} 
        \left\{ \frac{1}{T}\sum_{t=1}^T \max_{i \in [n]}\beta_i v_{t,i} +\left(-\log f\right)_+^\star (-\bm \beta)\right\}
.   \label{eq: dual} 
\end{align}

Let $\bm{u}^*(\mathbf{v}) = (u_i^*(\mathbf{v}))_{i\in[n]}$ be the optimal utility solutions to the primal program, and $\bm{\beta}^*(\mathbf{v}) = (\beta_i^*(\mathbf{v}))_{i\in[n]}$ be the optimal dual solutions; they are unique due to the strict concavity of the $\log f$ objective. With Slater's conditions~\citep{boyd2004convex}, we have strong duality and the corresponding KKT conditions; see \Cref{proof: preliminaries} for details of these derivations.

\begin{proposition}\textnormal{(Strong Duality and KKT Conditions)}
    \label{proposition: kkt}   
    If each agent $i \in [n]$ has at least one item with positive value $v_{t,i}>0$, then strong duality holds for \eqref{eq: primal} and \eqref{eq: dual}, i.e., $P^*(\mathbf{v}) = D^*(\mathbf{v})$. Further, the primal-dual optimal solutions satisfy:
    \begin{enumerate}
        \item $\beta_i^*(\mathbf{v}) = \partial_i  \log f(\bm u^*({\mathbf{v}})), \forall  i\in [n]$. 
        \item $x_{t,i}^*>0 \implies i \in \arg\max_{j\in[n]} {\beta_j^*(\mathbf{v}) \cdot v_{t,j}},\ \forall \ i\in[n], \ \forall \ t\in[T],$ where $\mathbf x^* = (x_{t,i}^*)_{i\in[n], t\in [T]}$ is any primal-optimal allocation solution. 
    \end{enumerate}
\end{proposition}

The second property in the KKT conditions (\Cref{proposition: kkt}) suggests a dual-to-primal decision rule: the $t$'th item should be allocated to the agent who has the \textit{maximum scaled value} $\beta_i^*(\mathbf{v}) \cdot v_{t,i}$. This observation will be useful in our subsequent analyses, even for algorithms that do not explicitly attempt to maintain or estimate anything in the dual space.

We also remark that our target regret $\mathcal{R}_{f}(\mathcal{A}, \mathbf{v}^\online)$ can be bounded via the regret (i.e., additive suboptimality gap) \textit{w.r.t.} the logarithmic-welfare programs by the inequality $e^z\geq z+1$; see \Cref{proposition: regret-to-ratio}. In the rest of the paper, we will refer to \eqref{eq: primal} and \eqref{eq: dual} as the primal and dual programs.  

\begin{proposition} 
\label{proposition: regret-to-ratio}
For any algorithm $\mathcal{A}$ and online sequence $\bv^\online$, denote $\mathcal
R_{\log f}(\mathcal
A, \bv^\online): = P^*(\bv^\online)-\log f(\bm u_{\mathcal{A}}^\dagger)$. Then it holds that
$\mathcal{R}_{f}(\mathcal
A, \bv^\online) \leq \operatorname{OPT}(\mathbf v^\online) \cdot R_{\log f}(\mathcal
A, \bv^\online).$
\end{proposition}

\section{Greedy Algorithm for Stationary Arrivals}
\label{section: greedy}

\begin{algorithm}[!htbp]
\caption{Greedy Algorithm for Stationary Arrivals.}
\label{alg: greedy}
    \SetKwInput{KwInit}{Initialization}
    \KwIn{$n, T$, online sequence $\mathbf{v}^\actual$.}
    \KwInit{
        $\bm{W}_0^\dagger \leftarrow 0$.
    }
    \For{$t = 1, \cdots, T$,}{
        Observe true valuation vector $\bm{v}_t^\actual$. \\

        \If{\textnormal{ {there exists $i\in [n]$ with $W_{t-1,i}^\dagger = 0$ and $v_{t,i}>0$}}}{ Allocate the current item to {such an agent} (arbitrary tie-breaking if there are multiple such agents).}
        \Else{ Greedily choose an integral allocation over the current item that maximizes the welfare so far,
        \begin{equation}
        \label{greedy: rule}
            i_t^\dagger \in \arg \max_{i\in [n]} f\left(\frac{\bm W_{t-1}^\dagger +  \bm v_t^\actual \cdot \bm e_i}{t} \right), \bm x_t^\dagger \gets \bm e_{i_t^\dagger}.
        \end{equation}
        }
        Update the cumulative utilities,
        $
            \bm W_t^\dagger \leftarrow \bm W_{t-1}^\dagger +  \bm v_t^\actual \cdot \bm x_t^\dagger.
        $
    }
\end{algorithm}

We begin with the stationary \textit{i.i.d.}\ arrival model, the canonical stochastic benchmark for online allocation: each value vector is drawn independently from an \emph{unknown} distribution, and the algorithm has no prior knowledge of the distribution. In this section, we establish that a minimalistic greedy algorithm (\Cref{alg: greedy}) is sufficient to achieve {a strong} expected time-averaged regret bound of $\widetilde O(1/T)$. 

\subsection{The Greedy Algorithm}
We now formally describe the greedy algorithm; see details in \Cref{alg: greedy}. The greedy update is defined in the purest sense: at each time step $t\in [T]$, the {algorithm observes} the current arrival and commits to the integral allocation that {myopically maximizes} the generalized-mean welfare so far, taking only the first $t$ steps into account. This is given formally in \eqref{greedy: rule}. For {$p\leq 0$}, the generalized-mean welfare objective is $0$ whenever there are zero-utility agents, causing a tie in \eqref{greedy: rule}. In this case, the algorithm will prefer assigning non-zero values to a zero-utility agent, as a safeguard step to escape zero utilities.

\Cref{alg: greedy} is highly simple and efficient: it only needs $O(n)$ algebraic operations for each time step, and only needs to maintain the past cumulative utilities in its memory. Moreover, it is an \textit{integral} algorithm, while still being competitive against the fractional hindsight solution, as will be shown later. It is also parameter-free and tuning-free. 

\subsection{Analysis of the Greedy Algorithm}

In the rest of the section, we present the main result for the greedy algorithm (\Cref{alg: greedy}), and sketch the key proof insights on how greedy updates lead to no regret. Detailed proofs can be found in \Cref{proof: greedy}.

\begin{restatable}{theorem}{greedy}
\label{thm: greedy-for-iid}
\textnormal{(Regret of the Greedy Algorithm)} Under the stationary arrival model, \Cref{alg: greedy} (denoted by $\mathcal{A}$) achieves
\begin{equation*}
    \mathbb{E} [\mathcal{R}_{f}(\mathcal
    A, \mathbf{v}^{\online})] \leq C_1 \cdot {\frac{n  \log T}{T}},
\end{equation*}
    where $C_1>0$ is a constant {that depends only on $p$, $\mathcal{P}^\online$ and $\vmax$}.
\end{restatable}

\Cref{thm: greedy-for-iid} shows a $\widetilde{O}(1/T)$ rate for the expected time-averaged regret. This no-regret result is strong for the greedy algorithm when viewed against the typical results in online algorithm analysis, where greedy policies are primarily analyzed through competitive ratios, with a constant multiplicative factor apart from optimal even asymptotically---for instance, in classical bipartite matching, the greedy algorithm is only $1/2$-competitive in the worst-case and $(1-1/e)$-competitive under more delicate stationary arrivals~\citep{karp1990optimal}. In contrast, \emph{no-regret} guarantees ask for vanishing per-period loss relative to the hindsight benchmark, and are \textit{not} standard for the pure greedy algorithms especially for a non-separable objective like ours. To this end, we show that for the generalized-mean maximization problem, the purely greedy algorithm with one-step welfare-improving decision each round already achieves $\widetilde{O}(1/T)$ time-averaged regret. Notably, this guarantee is essentially as strong as what is attainable in the\textit{ known} distribution \textit{i.i.d.}\ model with a much more complicated algorithm: the policy of \citet{balseiro2022uniformly} leverages \emph{full} distributional knowledge and repeated solution of high-dimensional fluid optimization programs to obtain $O(1/T)$ time-averaged regret, whereas we match this bound up to a logarithmic factor while dramatically reducing the complexity and informational requirements.

In the rest of the section, we sketch the proof intuition for \Cref{thm: greedy-for-iid}. The proof is based on a \emph{dual charging} argument: we show that each greedy allocation increases a natural primal potential by as much as one would ``pay'' under a certain dual charging rule for that item, up to small errors, and then we sum these per-step comparisons over time. The primal progress is defined \textit{w.r.t.} the following potential $(\Phi_t)_{t\in \{0\}\cup[T]}$, with $\Phi_0 = 0$ and $\Phi_T =T \log f(\bm u_{\mathcal{A}}^\dagger)$:
\[
\Phi_t := t \log f ({\bm W^\dagger_t} / {t}) {, \quad \forall t \in [T].}
\]

To define the charging rule in the dual space, recall the dual program \eqref{eq: dual}. A useful feature of the dual program is that the objective is additively separable across $t\in [T]$. Concretely, we can write the dual objective as
\begin{equation*}
    \frac{1}{T} \sum_{t=1}^T g(\bm \beta; \bm v_t^\actual), \quad  \text{where } \quad g(\bm \beta; \bm v) = \max_{i\in [n]} \beta_i v_i+ (-\log f)_+^\star(-\bm \beta).
\end{equation*}

Here, it is helpful to interpret the dual variable $\bm \beta$ as a \emph{shadow price} vector for per unit of utility. The term $\max_i \beta_i v_{t,i}^\actual$ is then the best price-adjusted value obtainable from the $t$'th item under dual prices $\bm \beta$, while the conjugate term $(-\log f)^{\star}_{+}(-\beta)$ is the dual penalty regularizer ensuring consistency with the target welfare function. To define the dual cost of the greedy algorithm, we need to specify an array of shadow prices that it implicitly generates. This corresponds to the following sequence of \textit{fictitious} dual updates. Intuitively, $\bm \beta_t^\prime$ captures the current marginal welfare weights of agents:
\begin{equation*}
     {\bm{\beta}}^\prime_t  = \nabla \log f ({\bm W^\dagger_{t-1}}/{t}) . 
\end{equation*}
The charging step is to compare the greedy algorithm's progress on the primal potential $\Phi_t$ to the current item's dual cost under the implicit shadow price $\bm \beta_t^\prime$. The greedy nature of the update ensures that the algorithm's myopic primal progress should be no less than any other feasible ones, including the one given by allocating the item among $\arg \max_{i\in [n]} \beta_{t,i}^\prime v_{t,i}^\online$, the maximum-scaled-value rule suggested by the KKT conditions (\Cref{proposition: kkt}). This can be further compared with the dual payment via local convex analysis, using the smoothness properties of the concave function $\log f$. This is formalized in the following lemma.

\begin{restatable}{lemma}{oco}
    \label{lem: greedy-as-oco}
    For any {$t\geq 2$ and $\ulower \in (0, \vmax/n]$}, under \Cref{alg: greedy}, we have
    \begin{equation*}
        \frac{\bm{W}_{t-1}^\dagger}{t-1} \in [\ulower, \vmax]^n \implies 
        \Phi_t - \Phi_{t-1}
        \geq g ({\bm{\beta}_t^\prime}; \bm v_t^\online) -\frac{ \lambda_{\ulower/2}\cdot  \vmax^2}{2 t},
    \end{equation*}
    where 
    $\lambda_{\ulower/2} >0$ is {the maximum of $|\partial_{ii} \log f(\bm u)|$ over $i\in [n]$ and over the vectors $\bm u \in [\ulower/2, \vmax]^n$ with $\sum_{i\in [n]}u_i \leq \vmax$}. 
\end{restatable}

There is an important remark on the \textit{boundedness} of the algorithmic procedure: \Cref{lem: greedy-as-oco} requires the boundedness condition $\bm W_t^\dagger / t \in [\ulower, \vmax]^n$, i.e., a constant lower bound on the current time-averaged cumulative utilities. 
To this end, we show that under the \textit{i.i.d.}\ arrival model, the greedy dynamics \emph{self-stabilize}: 
after a short burn-in period of length $O(\log T)$, the required boundedness will automatically hold with high probability; see \Cref{lem: high-prob-1-greedy}. 

\begin{restatable}{lemma}{highprobability}
    \label{lem: high-prob-1-greedy}
    {Under the \textit{i.i.d.}\ arrival model with $\mathbb{E}[v_{1,i}^\online]>0$ for all $i\in [n]$, there exist constants $\hat c \in (0, \vmax]$ and $\hat d>0$ that depend only on $p$, $\mathcal{P}^\online$ and $\vmax$, such that, with $\hat t := \lceil \hat d\cdot (n+\log (nT))\rceil$, \Cref{alg: greedy} guarantees}
    \begin{equation*}
        \Pr \left( {\bm W_{t}^\dagger}/{t} \in {[\hat c/n, \vmax]^n  ,  \ \forall  t\in [\hat t, T]}\right) \geq 1 - {1/T}.
    \end{equation*}
\end{restatable}

The proof of \Cref{lem: high-prob-1-greedy} critically leverages the greedy update rule. A similar bound is not necessarily true for general online algorithms. 
Also, the bound cannot be enforced via simply adding an initial warm-up phase since it requires lower-bounding $\bm W_t^\dagger$ not just by a constant, 
but by a linear function for all steps $t$ afterwards. Without this implicit lower bound, the objective function will be non-smooth, and one cannot expect regret bounds that are better than $\widetilde{O}(1/\sqrt{T})$ from off-the-shelf online convex optimization results such as \citet{agrawal2014fast}. 

Finally, summing up the inequalities in \Cref{lem: greedy-as-oco} over time steps (treating the initial {$\hat t$} rounds separately), we can lower-bound the algorithm's total primal potential 
by the total dual cost 
up to a logarithmic error term. 
In a stationary environment where values are generated in an \textit{i.i.d.}\ manner, one can show that in expectation, the total dual cost cannot be smaller than the ones given under the hindsight optimal dual prices $\bm \beta^*(\bv^\online)$ up to logarithmic error, with the optimal cost being $T \log \mathrm{OPT}(\mathbf{v^\online})$. Applying \Cref{proposition: regret-to-ratio}, this gives us the desired logarithmic regret bound.  
\section{Re-solving with a Single Sample}
\label{section: re-solve}

We now turn to the model where the arrival distributions are independent but arbitrarily time-varying, and the decision maker has access to a single sample drawn from the $t$'th historical distribution $\mathcal{P}_t^\historic = \mathcal{P}_t^\actual$. With access to historical data, one might naturally hope to ``learn'' the distribution of the future arrivals. However, in the nonstationary setting, there are $T$ potentially different distributions that need to be learned. Estimating such a time-varying process to the accuracy needed for $\widetilde O(1/T)$ regret would generally require $\Omega(T^2)$ item observations, or equivalently, $\Omega(T)$ full-length historical sequences, which is far beyond what is available in most practical scenarios. 

Motivated by this limitation, we adopt the single-sample paradigm and consider data-efficient policies that do not involve explicit estimations of distributions. A useful observation is that even in the more ideal \textit{i.i.d.}\ setting in \Cref{section: greedy}, our greedy algorithm succeeds without explicitly learning the distribution---it simply makes one-step myopic welfare-improving decisions. This motivates us to design the re-solving algorithm, which mirrors the principle of greedy to the data-driven setting---at each step, use the information available so far to form the best predictable proxy for the future, optimize welfare myopically, and then extract the current decision.

\subsection{The Re-solving Algorithms}
\label{section: re-solve-1}

\paragraph{Primal Re-solving.} We first introduce the primal re-solving algorithm, the most natural extension of the greedy algorithm to the single-sample setting; see \Cref{alg: re-solving (primal)}. At each time step $t$, it solves a primal hybrid welfare-maximizing program that contains three types of items: 1) the current $t$'th online item with its true value vector, 2) the existing utility that is derived from the past $t-1$ previous items where allocations are already committed, and
3) $T-t$ forthcoming items, but with values replaced by their counterparts from the historical sequence, since their true values are unknown. See \eqref{eq: primal-re-solving-rule} for the precise formulation of the hybrid program solved in the algorithm. The program optimizes over the allocations of the current and the predicted future items, but only extracts the $t$'th entry and uses it as the decision for the current step. 

\begin{algorithm}[!htbp]
\caption{Online Primal Re-solving}
\label{alg: re-solving (primal)}
    \SetKwInput{KwInit}{Initialization}
    \KwIn{$n, T$, historical sequence ${\mathbf{v}^\historic}$, online sequence $\mathbf{v}^\actual$.}
    \KwInit{
        $\bm{W}_0^\dagger \leftarrow  0$.
    }
    \For{$t = 1, \cdots, T$,}{
        Observe $\bm{v}_t^\actual$, and solve the hybrid welfare-maximization program:
        \begin{equation}
        \label{eq: primal-re-solving-rule}
        \begin{aligned}
               \max_{\bm u \geq 0, \bm{z}_{t}, \cdots, \bm z_{T} \geq0}   \quad  & \log f(\bm u) \\
               \textnormal{s.t.} \quad & u_i\leq \frac{1}{T} \left({W}_{t-1,i}^\dagger+  v_{t,i}^\online  \cdot z_{t,i}+  \sum_{\tau=t+1}^T v_{\tau,i}^\historic \cdot z_{\tau, i}\right), \ \ \forall i\in [n] \\   & \|\bm z_\tau\|_1 \leq 1 \quad \quad  \forall  \tau\in\{t+1, \cdots, T\}  .
        \end{aligned}
        \end{equation}
        
       Set $(\bm z_t^*, \cdots, \bm z_T^*) \gets$ any optimal allocation solution to \eqref{eq: primal-re-solving-rule}. 
       
       Set the current step-$t$ allocation to be the same as the optimizer, i.e., $\bm x_{t}^\dagger \gets \bm z_t^*$. 
       
       Update the cumulative utilities,
        $
            \bm W_t^\dagger \leftarrow \bm W_{t-1}^\dagger +  \bm v_t^\actual \cdot \bm x_t^\dagger.
        $
    }
\end{algorithm}
We remark that the primal re-solving algorithm is a \textit{fractional} algorithm. Different from the previous greedy algorithm, it maximizes over fractional allocations of items $\{t, t+1, \cdots, T\}$. The main reason for allowing fractional allocation in primal re-solving is efficiency: finding the \textit{integral} generalized-mean welfare maximization allocation is $\mathsf{NP}$-hard in general~\citep{caragiannis2019unreasonable}. As a result, finding the optimal integral allocation of \eqref{eq: primal-re-solving-rule} cannot be done in polynomial time unless $\mathsf{P} = \mathsf{NP}$. 
\paragraph{Dual Re-solving.} The primal re-solving algorithm works with a large $n\ (T-t+1)$-dimensional program and involves fractional allocation. 
Given these limitations, we further develop a family of \textit{dual} re-solving algorithms, which admit efficient and integral implementations. We will see that primal re-solving is equivalent to a dual re-solving algorithm under a certain tie-breaking specification.

The dual re-solving algorithm (\Cref{alg: re-solving (dual)}) constructs the hybrid program using the same three item components as we used in the primal re-solving algorithm, but operating in the dual space; see step \eqref{eq: re-solving-rule}. The dual program is $n$-dimensional and is thus usually easier to solve than the $nT$-dimensional primal problem. After obtaining the dual solution $\bm \beta_t^\dagger$ from the hybrid re-solving program, the algorithm allocates the item arbitrarily among the agents with maximum paced values $\beta_{t,i}^\dagger v_{t,i}^\online$, as suggested by the KKT conditions. Arbitrary tie-breaking is allowed here---by allocating to any of the winning agents or dividing between them, \Cref{alg: re-solving (dual)} is actually a family of algorithms with different specifications of tie-breaking. The easiest and most efficient way is to give the whole item to a single agent, which ensures integral allocation. In \Cref{section: infrequent-re-solving-and-other-accelerations} we discuss more practical approaches to accelerate the dual re-solving algorithm.

We explain why primal re-solving (\Cref{alg: re-solving (primal)}) is subsumed by \Cref{alg: re-solving (dual)}. By the KKT conditions (\Cref{proposition: kkt}), we know that the primal re-solving algorithm's allocation is supported on the maximum-paced-value agents under the re-solved dual solution $\bm \beta_t^\dagger$. Therefore, the primal re-solving algorithm can be explained as an instance of dual re-solving that breaks ties optimally (in terms of the welfare objective). Since dual re-solving is strictly more general, in subsequent analysis we will focus on \Cref{alg: re-solving (dual)}; all results and analyses can be automatically applied to the primal re-solving algorithm.
\begin{algorithm}[!htpb]
\caption{Online Dual Re-solving.}
\label{alg: re-solving (dual)}
    \SetKwInput{KwInit}{Initialization}
    \KwIn{$n, T$, historical sequence ${\mathbf{v}^\historic}$, online sequence $\mathbf{v}^\actual$.}
    \KwInit{
        $\bm{W}_0^\dagger \leftarrow  0$.
    }
    \For{$t = 1, \cdots, T$,}{
        Observe $\bm{v}_t^\actual$, and solve the hybrid welfare-maximization program:{\footnotemark}
        \begin{equation}
        \label{eq: re-solving-rule}
            \bm{\beta}_{t}^\dagger \leftarrow 
            \arg \min_{\bm{\beta}\in \mathbb{R}_+^n} \left\{ \frac{1}{T} \left(\langle \bm{\beta}, \bm{W}_{t-1}^\dagger \rangle + \max_{i\in[n]} \beta_i v_{t, i}^\actual + \sum_{\tau=t+1}^{T} \max_{i\in[n]}\beta_i v_{\tau,i}^\historic\right)   +\left(-\log f\right)_+^\star(-\bm \beta)
            \right\}.
        \end{equation}
       Select an agent with the highest multiplied value, with arbitrary tie-breaking:
       \begin{equation}
       \label{eq: re-solving-dual-to-primal}
           I_t^\dagger \gets \arg \max_{i\in[n]} \beta_{t,i}^\dagger v_{t, i}^\actual, \ \ \bm x_{t}^\dagger \ \in \  \operatorname{conv}\{\bm e_i: i\in I_t^\dagger\}. 
       \end{equation}
        Update the cumulative utilities,
        $
            \bm W_t^\dagger \leftarrow \bm W_{t-1}^\dagger +  \bm v_t^\actual \cdot \bm x_t^\dagger.
        $
    }
\end{algorithm}
\footnotetext{{On the exceptional event that the dual solution does not exist, the algorithm allocates the item to the smallest-indexed agent. Our analysis bounds the probability of this event by $O(1/T +\mathcal{W})$, so this does not affect the regret bounds.}}
\paragraph{Hybrid Primal-Dual Programs.} We introduce compact notations for the hybrid programs such as \eqref{eq: primal-re-solving-rule} and \eqref{eq: re-solving-rule} in the re-solving algorithms. These programs involve solving the future allocations over incoming items given the committed past allocations, and thus can be defined over general past utilities $\bm W = (W_1, \cdots, W_n)$ and future item values $\bm v_{t+1:T}$.
\begin{equation}
\label{eq: primal-re-solve-t}
    \begin{aligned}
        P^*_t(\bm{W};\bm{v}_{t+1:T}) = \max_{\bm{u} \geq 0, \bm x_{t},\cdots, x_{T}\geq 0 } \quad & \log f(\bm u)  \\
               \textnormal{s.t.} \quad & u_i\leq \frac{1}{T} \left({W}_{i}+  \sum_{\tau=t+1}^T v_{\tau,i}\cdot x_{\tau, i}\right), \forall i\in [n]
        \\& \|\bm{x}_\tau\|_1\leq 1,   \quad \quad \forall \tau\in\{ t+1, \cdots, T\}. \\ 
    \end{aligned}
\end{equation}
\begin{equation}
\label{eq: dual-re-solve-t}
    D_t^*(\bm{W};\bm{v}_{t+1:T}) = \min_{\bm{\beta}\in \mathbb{R}_{+}^n} \ \ 
    \left\{  \frac{1}{T} \left( \langle \bm{\beta}, \bm{W} \rangle + \sum_{\tau=t+1}^{T} \max_{i \in[n]}\beta_i v_{\tau,i} \right)  +(-\log f)_+^\star (-\bm \beta)\right\}.
\end{equation}

We denote the optimal primal utility solution and the dual solutions as $\bm{u}_{t}^{*}(\bm{W};\bm{v}_{t+1:T})$ and $\bm{\beta}_{t}^{*}(\bm{W};\bm{v}_{t+1:T})$, respectively. Notice that a past item (value vector $\bm v$) allocated to $i$ can be interpreted as an item with value vector $v_i \bm e_i$. Therefore, hybrid programs can be interpreted as standard log-welfare optimization programs, and share the properties such as solution uniqueness, strong duality, and KKT conditions.

For simplicity, we denote the optimization programs solved in \eqref{eq: primal-re-solving-rule} and \eqref{eq: re-solving-rule} as $P_t^\resolving, D_t^\resolving$, and refer to them as \textit{re-solving programs}; denote their optimal solutions as $\bm u_t^\resolving, \bm \beta_t^\resolving$. 

\subsection{Analysis of the Re-solving Algorithms}
\label{section: re-solve-3}

In this section, we introduce the assumption that with probability $1$, the optimal time-averaged utilities are bounded from below by a constant $\ulower^*>0$, for both $\mathcal{P}^\online$ and $\mathcal{P}^\historic$. This rules out pessimistic adversarial constructions where optimal time-average utility diminishes for some agents. 

\begin{assumption}\textnormal{(Hindsight Solutions are Lower-Bounded)}
    \label{assumption: lower-bound}
    There exists a constant $\ulower^* \in (0, \vmax]$ such that with probability
$1$, $\bm u^*(\bv^\online), \bm u^*(\bv^\historic) \in [\ulower^*, \vmax]^n.$
\end{assumption}

Our first main result for the re-solving algorithms is the following $\widetilde{O}(1/T)$ regret bound (with $O(n)$ dependence on the number of agents; \Cref{thm: adaptive-with-density}) when the historical sample distributions match the true arrival distributions exactly, under a density upper bound assumption on the distributions. Asymptotically, this recovers the exact dependence on both $T$ and $n$ of the greedy algorithm's regret in the ideal \textit{i.i.d.}\ case.
\begin{restatable}{theorem}{densityA}
\label{thm: adaptive-with-density}
    When $\mathcal{P}_t^\actual =\mathcal{P}_t^\historic$ and the distribution has a density which is at least $\rho$ times the uniform distribution's density, under \Cref{assumption: lower-bound}, \Cref{alg: re-solving (dual)} (denoted by $\mathcal{A}$) achieves
    \begin{equation*}
    \mathbb{E}[\mathcal{R}_{f}(\mathcal{A}, \mathbf{v}^\actual)] \leq C_2 \cdot \frac{\rho \cdot n \cdot \log T}{T},
    \end{equation*}
    where $C_2>0$ is a constant that does not depend on $\rho,$ $n$ or $T$.
\end{restatable}

The rest of the section is devoted to highlighting the important techniques from our analysis. We first decompose the regret into the sum of differences of consecutive \textit{coupling programs}. Then, we use concentration analysis to show that up to $\widetilde{O}(1/T)$ consecutive coupling programs are different in expectation.

\subsubsection{A Coupling Analysis Framework}

To track the algorithm's performance dynamically over time, we introduce a sequence of auxiliary hybrid programs, with their existing utility and tail parts specified as follows,  
\begin{equation}
    \label{eq: coupling-program-def-for-no-error}
    P_t^\coupling := P_t^*(\bm W_{t}^\dagger ; \bm v_{t+1:T}^\online), \ \ D_t^\coupling := D_t^*(\bm W_{t}^\dagger ; \bm v_{t+1:T}^\online).
\end{equation}
We call these \emph{coupling programs}, as they play a role similar to the coupling programs from \citet{vera2019bayesian}.
Denote their optimal primal and dual solutions as $\bm u_t^\coupling , \bm \beta_t^\coupling$, respectively. These coupling programs are designed only for theoretical analysis; they are not maintained in the algorithm and need not be solved computationally. 
See \Cref{tab: hybrid-programs} for a comparison between actual re-solving programs and coupling programs.
\begin{table}[!htbp]
\centering
\small
\caption{ Re-solving and coupling programs.}
\label{tab: hybrid-programs}
\begin{tabular}{@{}l >{\raggedright\arraybackslash}p{0.203\textwidth} >{\raggedright\arraybackslash}p{0.108\textwidth} >{\raggedright\arraybackslash}p{0.492\textwidth}@{}}
\toprule
\makecell[l]{Primal, dual\\ programs} & Meaning & \makecell[l]{Past utility} & Tail items \\
\midrule
$P_t^\resolving,\ D_t^\resolving$
& Re-solving program, solved by the algorithm at step $t$.
& $\bm W_{t-1}^\dagger$
& $(\bm v_t^\online, \bm v_{t+1:T}^\historic)$: the true item $t$, followed by historical sample \\
\addlinespace
$P_t^\coupling,\ D_t^\coupling$
& Coupling program constructed for analysis.
& $\bm W_{t}^\dagger$
& 
$\bm v_{t+1:T}^\coupling \!=\!
\begin{cases}
\bm v_{t+1:T}^\online, & \text{\Cref{section: re-solve-3}},\\
\text{jointly sampled with } v_{t+1:T}^\online, & \text{\Cref{sec: robust}}.
\end{cases}$ \\
\bottomrule
\end{tabular}
\end{table}

We explain how the coupling programs track the online algorithm's decisions. They capture the dynamic ``best possible'' objective given the committed decisions 
if one optimizes the future allocations optimally.   
First, the coupling programs are forced to take the existing past utilities $\bm W_t^\dagger$ that the online algorithm has allocated.
Meanwhile, on the future part, the coupling programs are \textit{clairvoyant}, since they have access to the true incoming sequence $\bm v_{t+1:T}^\online$,
which is unknown to the algorithm at that timestep, and optimize the decision over the future. 
It is thus clear that the coupling program at time zero, $D_0^\coupling$, is the hindsight optimal program, and the last coupling program $D_T^\coupling$ is the outcome of the sequence of decisions taken by our algorithm. 
\begin{equation*}
    \underbrace{\log \mathrm{OPT}^*(\mathbf{v}^\actual) = D_0^\coupling}_{\text{hindsight optimal}} \geq D_1^\coupling \geq \cdots \geq D_{T-1}^\coupling \geq \underbrace{D_T^\coupling  = \log f(\bm u_{\mathcal{A}}^\dagger(\mathbf{v}^\online ))}_{\text{algorithm's outcome}}.
\end{equation*}

Using this coupling framework, our main proof strategy is to bound the differences between adjacent coupling programs; 
summing up these bounds yields a bound on the total sub-optimality for the logarithmic welfare. 
To bound the difference, we observe that if the online decision coincides with the coupling program on the current item's allocation, then in the next step the coupling benchmark stays unchanged; see \Cref{prop: coupling-with-no-error}. 

\begin{proposition}
    \label{prop: coupling-with-no-error}
    For the coupling programs given in \eqref{eq: coupling-program-def-for-no-error}, if the online algorithm and $P_{t-1}^\coupling$ agree on the allocation of the item $t$, then $\bm u_t^\coupling = \bm u_{t-1}^\coupling$ .
\end{proposition}

With \Cref{prop: coupling-with-no-error}, it suffices to bound the number of total disagreements 
between the re-solving and coupling programs, and the fluctuation caused by each disagreement. This critically relies on the welfare maximizer's \textit{stability} and \textit{concentration}, which we introduce next. 
\subsubsection{Stability and Concentration}
\label{section: re-solve-2}

We derive stability and concentration properties for the generalized-mean welfare maximization convex programs. In particular, we investigate the hindsight welfare-maximizing utilities $\bm{u}^*(\mathbf{v})$ as a function of the input value sequence $\mathbf{v}$. As offline properties, these bounds may be of independent interest. To the best of our knowledge, they are new to the literature even for the well-studied case $p=0$, providing a new aspect for characterizing the structure of the welfare maximizers (which is also the \textit{competitive equilibria} of the market in the $p=0$ case).

\paragraph{Monotonicity and Stability when Adding/Dropping Items.} We characterize how $\bm{u}^*(\mathbf{v})$ changes when an extra item is added to the value instance.\footnote{While these are offline results, we still use the same $T$ to normalize the time-averaged utility for consistency.} We show that adding an item weakly increases every agent's utility in the optimal solution, with an upper bound on the increment. This characterizes how sensitive the offline welfare-maximization program is to input changes. We prove a more general result with $K$ added items.

\begin{restatable}{lemma}{stability}\textnormal{(Optimizer Monotonicity and
Stability)}
\label{lem: stability}
\label{lem: monotonicity}
For any instance $\mathbf v^-$ that admits a strictly positive utility solution, let
$\mathbf v^+$ be obtained by adding $K>0$ items to $\mathbf v^-$. Then for every $i\in[n]$,
\begin{equation*}
0\leq u_i^*(\mathbf v^+)-u_i^*(\mathbf v^-)
\leq \frac{K\vmax}{T}
\frac{\max_{j\in[n]}\beta_j^*(\mathbf v^-)}{\beta_i^*(\mathbf v^-)}.
\end{equation*}
\end{restatable}

We sketch the intuition of the stability result. Adding $K$ items in the allocation as endowment changes each agent's time-averaged utility by at most $K\vmax / T$. \Cref{lem: stability} thus shows that re-optimizing generalized-mean welfare over the expanded market does not amplify this increment by more than a multiplicative factor $\max_j \beta_j^*(\bv^-)/\beta_i^*(\bv^-)$, which quantifies heterogeneity in agent utilities through the ratio between the differentials (by the KKT conditions, the $i$'th dual variable is the $i$'th derivative at optimum). This can be understood as characterizing how easily the welfare objective would trade off between different agents' utilities. If agent $j\in [n]$ has a tiny utility level and thus a large dual multiplier, then increasing/decreasing agent $j$'s utility will have a high marginal gain/loss for the welfare, causing high sensitivity on all agents $i\in [n]$ due to potential competition with agent $j$. Conversely, if all utilities lie in a fixed range $[\ulower, \uupper]$, the sensitivity multiplier will also be bounded by $\kappa = \max_{i, j\in [n]} \max_{\bm u\in [\ulower, \uupper]^n}
\partial_i f(\bm u)/\partial_j f(\bm u)$.

\paragraph{Optimal Solution Concentration under Independent Inputs.} We now investigate the concentration behavior of $\bm u^*$ when the items at different steps are independently drawn from distributions $(\mathcal{P}_t)_{t\in [T]}$. Since our subsequent analysis involves hybrid programs, we investigate the more general case with existing past utility $\bm W$, and say that an incoming sequence $\bm v_{t+1:T} \sim \mathcal{P}_{t+1:T}$ is a \textit{good tail} for it when $\bm u_t^*(\bm W; \bm v_{t+1:T} )$ is close to a deterministic underlying center.\footnote{The center might not be the mean. We instead use the mean of a Lipschitz extension of the optimal utility function; see details in the proof.}

\begin{definition}\textnormal{(Good Sequence Tails)} Let $\bm v_{t+1:T} \sim
\mathcal{P}_{t+1:T}$ be a randomized sequence tail. For given bounds $0<\ulower<\uupper \leq \vmax$, we say the tail is $\varepsilon$-good for $\bm W$ iff for each $i\in [n]$,
\begin{equation*}
\bm u_t^*(\bm W; \bm v_{t+1:T}) \in [\ulower, \uupper]^n \implies
\left| u_{t,i}^*(\bm W; \bm v_{t+1:T})
- m_{t,i}(\bm W; \mathcal{P}_{t+1:T}) \right| \leq \varepsilon,
\end{equation*}
where $m_{t,i}(\cdot\,; \mathcal{P}_{t+1:T})$ is a deterministic center.
Further, we say $\bm v_{t+1:T}$ is \emph{uniformly} $\varepsilon$-good iff the
above holds for all $\bm W \in [0, t\vmax]^n$.
\end{definition}

In the concentration analysis of the re-solving program's iterate 
$\bm u_t^\resolving$, we will show that the tails are good with high probability not just for the past $\bm W_{t-1}^\dagger$, but \textit{uniformly} for all possible cumulative utility levels $\bm W$. This allows us to effectively treat the tail $\bm v_{t+1:T}^\historic$ as fresh samples from the distributions in every step $t$, despite the correlation between $\bm v_{t+1:T}^\historic$ and $\bm W_{t-1}^\dagger$. 

\begin{restatable}{lemma}{uc}
    \label{lem: uc}
    \ifappendixlemma
    Let $\mathbf{v} \sim \mathcal{P}_{1:T}$ be a randomized sequence.  For given $0<\ulower<\uupper \leq \vmax$ and $\eta\in(0,1)$, there are deterministic functions $m_{t,i}(\bm W;\mathcal P_{t+1:T})$ such that with probability at least $1-\eta$, for every $t\in\{0,\ldots,T\}$, $\bm W\in[0,t\vmax]^n$, and $i\in[n]$,
    \begin{equation*}
        \bm u_t^*(\bm W;\bm v_{t+1:T})\in[\ulower,\uupper]^n
        \implies
        \left|u_{t,i}^*(\bm W;\bm v_{t+1:T})
        -m_{t,i}(\bm W;\mathcal P_{t+1:T})\right|\leq\varepsilon_0,
    \end{equation*}
    where
    \begin{equation*}
        \varepsilon_0 = \kappa \vmax \cdot 
            \left(
            \sqrt{\frac{n\log (T+1)+\log\left(2nT/\eta\right)}{2T}}
            + \frac{n}{T}\right),
    \end{equation*}
    where parameter $\kappa = \max_{i, j \in [n]} \max_{\bm u\in [\ulower, \uupper]^n} \partial_i f(\bm u) / \partial_j f(\bm u)$ is the largest derivative ratio of $f$ across entries on $\bm u \in [\ulower, \uupper]^n.$
    
    \else
    \textnormal{(Informal)} 
    For any constant bounds $0<\ulower<\uupper \leq \vmax$, with probability at least $1-1/(2T)$, all tails are uniformly $\varepsilon$-good for $\varepsilon$ = O($\sqrt{n\log T/T}$)
    \fi
\end{restatable}

Here, uniform convergence is \textit{conditional}: it only applies to the utility solutions that lie in $[\ulower, \uupper]^n$, with a convergence rate that depends on these bounds. This is aligned with the observation from \Cref{lem: stability} that only bounded solutions are stable.

\subsubsection{Bounding the Disagreements} 

We now bound the disagreements between the re-solving and coupling programs $D_{t-1}^\coupling$ and $D_{t}^\resolving$. Recall that their tails share the same distribution, so by \Cref{lem: uc} we have informally,\footnote{Since uniform convergence in \Cref{lem: uc} is conditional, in the formal proof one has to carefully check that all intermediate solutions are bounded; the hidden constants in \eqref{eq: proof-sketch-with-density} depend on these implicit bounds.} with high probability,
\begin{equation}
\label{eq: proof-sketch-with-density}
\left.
\begin{aligned}
& \left\|\bm u_t^\resolving - \bm m_{t-1}^*(\bm W_{t-1}^\dagger; \bm v_{t:T})\right\|_\infty  = \widetilde{O}(1/\sqrt{T})\\
& \left\|\bm u_{t-1}^\coupling -\bm m_{t-1}^*(\bm W_{t-1}^\dagger; \bm v_{t:T}) \right\|_\infty  = \widetilde{O}(1/\sqrt{T})
\end{aligned}
\right\}
\ \implies\|\bm u_t^\resolving  - \bm u_{t-1}^\coupling\|_\infty =  \widetilde{O}(1/\sqrt{T}).
\end{equation}

By the stability result (\Cref{lem: stability}) both $\bm u_t^\resolving , \bm u_{t-1}^\coupling$ are also close to $\bm u_{t-1}^\resolving$, so by the KKT conditions we know that $\bm \beta_t^\dagger, \bm \beta_{t-1}^\coupling$ lie in the vicinity of $\bm \beta_{t-1}^\dagger$ up to $\widetilde{O}(1/\sqrt{T})$ error, with high probability.\footnote{This is also informal due to various boundedness requirements.}

Next, we need to argue that close dual multipliers can indeed induce the same primal decision for most value vectors, under the maximum-scaled-value decision rule. We say an item value vector is \textit{safe} for a dual multiplier $\bm \beta$ if all dual multipliers in the vicinity of $\bm \beta$ allocate the item to the same winner under the maximum-scaled-value rule $\arg \max_{i} \beta_i v_i$. Formally, for each dual multiplier $\bm \beta \in \mathbb{R}_+^n$, define its \textit{$\iota$-safe region} as follows,

\begin{definition}
    \label{definition: safe-region}
    For $\iota>0$, the $\iota$-safe region of values for $\bm \beta$, denoted by $\mathcal{S}(\bm \beta; \iota)$, is defined as
    \begin{equation*}
    \mathcal{S}(\bm \beta; \iota):= \{\bm v \in [0, \vmax]^n: \arg\max_{i\in [n]} \beta_i^\prime v_{i} = \arg \max_{i\in [n]} \beta_i v_i, \forall\bm \beta^\prime \in \mathcal{B}_\infty(\bm \beta; \iota)\cap [0, +\infty)^n \}.
\end{equation*}
\end{definition}

We argue that for any dual multiplier $\bm \beta$, only an $O(\iota)$ fraction of the value space $[0,\vmax]^n$ is not $\iota$-safe. When the value distributions have bounded density, the probability of a value being unsafe is then $O(\iota)$; see the following lemma. 
\begin{restatable}{lemma}{safety} \textnormal{(Lower Bound of Safe Volume)}
    \label{lem: safe-volume} \label{corollary: unsafe-counts}
    For each $\bm \beta\in [\betalower, +\infty)^n$ and $\iota>0$, the volume of its $\iota$-safe region is at least
    \begin{equation*}
        \mathrm{Vol}(\mathcal{S}(\bm \beta; \iota)) \geq \vmax^n (1 - {2n\iota}/{ 
        \betalower}).
    \end{equation*}
\end{restatable}

We can then bound the total number of disagreements by \Cref{lem: uc} and \eqref{lem: safe-volume}. To get our target regret bound, we further leverage the local smoothness and stability properties of the primal-dual programs, where we also inductively maintain the high-probability boundedness conditions over time. The missing proofs can be found in \Cref{proof: re-solve}.

\section{Robustness to Distribution Shifts}
\label{sec: robust}
\label{section: re-solve-4}
\label{section: adaptive-1}

In real-world platforms, there can often be distribution drifts caused by sample bias and systematic environment changes. Motivated by this reality, we next consider the case with different historical and online arrival distributions, i.e., $\mathcal{P}_t^\online \neq \mathcal{P}_t^\historic$, with the shifts measured in the time-averaged $\ell_1$ Wasserstein distance $\mathcal{W}$. 

In this setting,
our main result is that the re-solving algorithm achieves $\widetilde{O}(1/\sqrt{nT} + \mathcal{W})$ expected time-averaged regret. The bound degrades smoothly as $\mathcal{W}$ increases, suggesting that re-solving is robust to small distribution discrepancies in the historical data.

\begin{restatable}{theorem}{adaptiveA}
\label{thm: adaptive}
    Under \Cref{assumption: lower-bound}, \Cref{alg: re-solving (dual)} (denoted by $\mathcal{A}$) achieves
    \begin{equation*}
       \mathbb{E}[\mathcal{R}_f(\mathcal{A}, \mathbf{v}^\actual)] \leq  C_3 \cdot \sqrt{\frac{\log T}{nT}} + C_4 \cdot \mathcal{W}, \quad \forall \, T \geq c'n\log T, 
    \end{equation*}
    where $c', C_3, C_4 >0$ are constants independent of $n$ and $T$.
\end{restatable}

In the rest of the section, we analyze the case with distribution shifts using a novel and more general coupling scheme, which generalizes the framework in the previous section.

\paragraph{A More General Coupling Framework.} 
We consider a sequence of coupling programs similar to the previous section, but allow the tail to be chosen from any general auxiliary \textit{coupling sequence} 
$\mathbf{v}^\coupling = (\bm{v}_1^\coupling, \cdots, \bm{v}_T^\coupling)$; in previous \Cref{section: re-solve-1} we only considered $\bv^\coupling = \bv^\online.$
The $t$'th coupling program, defined as in \eqref{eq: def-coupling}, is forced to accept the online algorithm's decision in the first $t$ rounds, but is able to optimize over the future arrivals as if they are $\bm  v_{t+1:T}^\coupling$, the tail of the coupling sequence. We denote the coupling program's optimal solutions as $\bm u_{t}^\coupling, \bm \beta_{t}^\coupling$.

\begin{equation}
\label{eq: def-coupling}
    P_t^\coupling:= P_t^* (\bm{W}_t^\dagger; \bm{v}_{t+1:T}^\coupling), \ D_t^\coupling:= D_t^* (\bm{W}_t^\dagger; \bm{v}_{t+1:T}^\coupling). 
\end{equation}

Similar to the last section, the key step is to bound the difference between adjacent coupling programs. This is provided by \Cref{lem: coupling-main}, which holds deterministically for general choice coupling sequences and any sequence of dual multipliers.

\begin{restatable}{lemma}{couplingmain}
    \label{lem: coupling-main}
    Let $\mathbf{v}^\coupling$ be any coupling sequence with $\|\bv^\coupling\|_\infty\leq \vmax$.
    Then
    \begin{equation*}
        D_{t-1}^\coupling -D_t^\coupling \leq \frac{2}{T}\vmax \| \bm{\beta}_t^\coupling - \bm{\beta}_t^\dagger \|_\infty +\frac{1}{T}\|\bm{\beta}_t^\coupling \|_\infty \|\bm{v}_t^\coupling - \bm{v}_{t}^\actual\|_\infty 
            .
        \end{equation*}
\end{restatable}
\paragraph{Proof.}
Since $\bm{\beta}_{t}^\coupling$ is the optimal solution to the $t$'th coupling program $D_{t}^\coupling$, we have
\begin{equation}
\label{eq: coupling-eq1}
    D_{t}^\coupling =  {\frac{1}{T}\left(\langle \bm{\beta}_{t}^\coupling, \bm{W}_{t}^\dagger \rangle + \sum_{\tau=t+1}^{T} \max_{i \in[n]}\beta_{t,i}^\coupling  v_{\tau,i}^\coupling\right)  +(-\log f)_+^\star (-\bm \beta_t^\coupling)}.
\end{equation}
On the $(t-1)$'th coupling program, the same multiplier $\bm{\beta}_{t}^\coupling$ is sub-optimal: 
\begin{equation}
\label{eq: coupling-eq2}
    D_{t-1}^\coupling \leq {\frac{1}{T}\left(\langle \bm{\beta}_{t}^\coupling, \bm{W}_{t-1}^\dagger \rangle + \sum_{\tau=t}^{T} \max_{i \in[n]}\beta_{t,i}^\coupling  v_{\tau,i}^\coupling\right)   +(-\log f)_+^\star (-\bm \beta_t^\coupling)}. 
\end{equation}
Denote the instantaneous utility given by the algorithm at time step $t$ as $\bm{y}_t = (v_{t, 1}^\actual x_{t, 1}^\dagger, \cdots,v_{t, n}^\actual x_{t, n}^\dagger)$, then $\bm{y}_t = \bm{W}_t^\dagger - \bm{W}_{t-1}^\dagger$. Taking difference of the above \eqref{eq: coupling-eq1} and \eqref{eq: coupling-eq2},
\begin{equation}
    \label{eq: coupling-proof}
    \quad D_{t-1}^\coupling - D_t^\coupling \leq \frac{1}{T}
        \left( 
            \max_{i\in[n]} \beta_{t,i}^\coupling  v_{t,i}^\coupling 
                - 
            \langle \bm{\beta}_{t}^\coupling  ,\bm{y}_t \rangle \right).
\end{equation}
{Since we allow any tie-breaking rule that fully allocates the item, we have
$\max_{i\in[n]} \beta_{t,i}^\dagger v_{t, i}^\actual = \langle \bm{\beta}_t^\dagger, \bm{y}_t\rangle$ and $\|\bm y_t\|_1\leq\vmax$.}
By adding and subtracting terms on the right-hand side of \eqref{eq: coupling-proof},
\begin{align*}
    & \quad D_{t-1}^\coupling - D_t^\coupling \\
    &\leq \frac{1}{T}\left( 
        \left(
            \max_{i\in[n]} \beta_{t,i}^\coupling v_{t,i}^\coupling - \max_{i\in[n]} \beta_{t,i}^\dagger v_{t,i}^\actual
        \right)
            +
        \left(
            \langle \bm{\beta}_t^\dagger, \bm{y}_t
            \rangle
            -
             \langle \bm{\beta}_{t}^\coupling  ,\bm{y}_t \rangle
        \right)
           \right)\\
    &=\frac{1}{T}\left( 
        \left(
            \max_{i\in[n]} \beta_{t,i}^\coupling v_{t,i}^\coupling - \max_{i\in[n]} \beta_{t,i}^\coupling v_{t,i}^\actual
        \right)
            +
        \left(
            \max_{i\in[n]} \beta_{t,i}^\coupling v_{t,i}^\actual - \max_{i\in[n]} \beta_{t,i}^\dagger v_{t,i}^\actual
        \right)
            +
        \left(
            \langle \bm{\beta}_t^\dagger, \bm{y}_t
            \rangle
            -
             \langle \bm{\beta}_{t}^\coupling  ,\bm{y}_{t} \rangle
        \right)
           \right)\\
       &\leq \frac{1}{T}
        \left(
            \|\bm{\beta}_t^\coupling \|_\infty \|\bm{v}_t^\coupling - \bm{v}_{t}^\actual\|_\infty 
            +
        2\vmax \| \bm{\beta}_t^\coupling - \bm{\beta}_t^\dagger \|_\infty
        \right). 
\end{align*}
In the last step, we used the Lipschitzness of $\max_{i\in[n]} \beta_i v_i$ as a function of $(v_1, \cdots, v_n)$ and as a function of $(\beta_1, \cdots, \beta_n)$, respectively. The bound holds for integral and fractional tie-breaking alike.

$\hfill \square$

\Cref{lem: coupling-main} helps us to decompose the regret on the logarithmic welfare into the sum of differences between the adjacent coupling programs, plus an 
additional drift term since $\bv^\coupling \neq \bv^\online$ in general.

\begin{align}
    \mathcal{R}_{\log}(\mathcal{A}, \bv^\online) &= \left\{\sum_{t=1}^T(D_{t-1}^\coupling - D_t^\coupling) \right\}+ P^*(\mathbf{v}^\actual) - P^*(\mathbf{v}^\coupling) \nonumber \\
    &\leq 
     \underbrace{   \frac{2}{T}\vmax \sum_{t=1}^T \| \bm{\beta}_t^\coupling - \bm{\beta}_t^\dagger \|_\infty}_{R_1}+
    \underbrace{\frac{1}{T} \sum_{t=1}^T\|\bm{\beta}_t^\coupling \|_\infty \|\bm{v}_t^\coupling - \bm{v}_{t}^\actual\|_\infty}_{R_2}
             +
     \underbrace{P^*(\mathbf{v}^\actual) - P^*(\mathbf{v}^\coupling)}_{R_3}.
    \label{corollary: regret-decomposition}
\end{align}

Critically, this decomposition separates the \textit{statistical learning error} of the algorithm's decision and a \text{coupling drift} component that is (mostly) algorithm-independent. Concretely, 
\begin{itemize}
    \item $R_1$ is the statistical learning error. It measures how close the algorithm's dual decisions are to the coupling program's choices. It decreases when we set the coupling sequence closer to the algorithm's dual decision trace. 
    \item $R_2$ and $R_3$ are coupling drift terms. They decrease when we choose the auxiliary coupling sequence to be closer to the actual online arrival.  
\end{itemize}

\paragraph{Choosing the Coupling Sequence in the Face of Distribution Shift.} 
There is a natural tension between the statistical learning error ($R_1$) and the drift terms ($R_2$ and $R_3$): controlling $R_1$ requires maintaining coupling solutions that are close to the algorithm's decisions, while bounding $R_2$ and $R_3$ requires the coupling sequence to be close to the actual sequence as well. 
Our next step is then to carefully construct a coupling sequence $\bv^\coupling$ to control all terms simultaneously under distribution shifts. We start with two naive and natural choices and learn from their failure.

(1) \textit{Setting the coupling sequence to be the true online arrival, i.e., $\mathbf{v}^\coupling = \mathbf{v}^\actual$.} This recovers what we did in \Cref{section: re-solve-1}. As an immediate benefit, the drift terms $R_2$ and $R_3$ become zero. However, due to the discrepancies 
between $\mathcal{P}_{t:T}^\historic$ and $\mathcal{P}_{t:T}^\actual$, the two dual multipliers $\bm \beta_t^\coupling$ and $\bm \beta_t^\dagger$ have different means, 
since they are learned from tails with different distributions. Each summand in the term $R_1$ will suffer from a constant deviation, and eventually $R_1 =  \Omega(1)$. Intuitively, the online algorithm is unable to approach  $\bm \beta_t^\coupling = \bm{\beta}_t^*(\bm{W}_t^\dagger;\bm{v}_{t+1:T}^\actual)$ with its actual dual multipliers since its information of the tail is limited and incorrect.
We note that this is similar to the Compensated Coupling scheme by \citet{vera2019bayesian}, which operates only in the primal space and cannot handle distribution shifts. 

(2) \textit{Setting the coupling sequence to be the historical samples, i.e., $\mathbf{v}^\coupling = \mathbf{v}^\historic$.} This immediately bounds $R_1$ to be $\widetilde O (1/T)$ in expectation, because the coupling program $D_t^\coupling$ and the re-solving program $D_t^\resolving$ would almost be the same. 
However, since $\mathbf{v}^\historic$ and $\mathbf{v}^\online$ are independently drawn from different distributions, $R_2$ and $R_3$ cannot be controlled. Intuitively, in this case, the coupling benchmark fails to include useful hindsight information and potentially diverges from the true hindsight optimal with large discrepancies.

Although these simple constructions fail to give us the desired regret bound, they help us to understand the intrinsic tension between the stochastic estimation term and the drift terms. On the one hand, to get an $\widetilde{O}(1/\sqrt
    T)$ statistical estimation error in $R_1$, the historical and coupling sequences should have the same distribution so it is possible for the re-solving algorithm to learn the coupling programs using historical data. On the other hand, $\bv^\coupling$ should be correlated with $\bv^\online$ so the coupling benchmark meaningfully tracks the hindsight optimum, and thus controls the drift terms $R_2$ and $R_3$. We show that such a $\bv^\coupling$ could be constructed, by sampling jointly with $\bv^\online$ from the optimal distribution in the definition of the Wasserstein discrepancy; see \Cref{proposition: wasserstein}. 

\begin{proposition} \textnormal{(The Choice of the Coupling Sequence)}
    \label{proposition: wasserstein}
    Let $\delta_{1:T} =(1/T)\sum_{t=1}^T \|\bm v_t^\coupling - \bm{v}_t^\actual\|_1$ be the time-averaged $\ell_1$ distance between $\bv^\coupling$ and $\bv^\online$. By the definition of Wasserstein discrepancy, there exists a joint distribution $\Gamma^*\in \Delta([0,\vmax]^{n\times T}\times [0, \vmax]^{n\times T})$ such that for $(\bv^\actual, \bv^\coupling) \sim \Gamma^*$, 
    \begin{enumerate}
        \item The marginal distribution of $\mathbf{v}^\actual$ is $\mathcal{P}^\actual$.
        \item The marginal distribution of $\mathbf{v}^\coupling$ is $\mathcal{P}^\historic$.
        \item The expected $\ell_1$ distance is equal to the Wasserstein metric, ${\mathbb{E}}_{(\mathbf{v}^\actual,\mathbf{v}^\coupling)\sim \Gamma^* }[\delta_{1:T}] = \mathcal{W}.$
    \end{enumerate}
\end{proposition}

\paragraph{Bounding the Regret Terms.} We briefly sketch how to bound each regret term in \eqref{corollary: regret-decomposition} under the coupling choice in \Cref{proposition: wasserstein}. To satisfy various boundedness requirements, we will show that all intermediate utility solutions from the re-solving and coupling programs are entry-wise lower-bounded by some constant $\ulower>0$ with high probability. For $R_1$, with $\mathcal{P}_{t+1:T}^\coupling = \mathcal{P}_{t+1:T}^\historic$ we know that the optimal re-solving and coupling programs share close solutions. Applying the uniform convergence results in \Cref{lem: uc}, we can show $\mathbb{E}[R_1] =\widetilde{O}(1/\sqrt{T})$. For $R_2$, we note that all dual iterates $\bm \beta_t^\coupling$ are upper-bounded by the KKT conditions once we secure that $\bm u_t^\coupling \in [\ulower, \vmax]^n$. Therefore, $R_2$ is of the same order as the average $\ell_1$ discrepancy, which is $\mathcal{W}$ in expectation. For $R_3$, we observe that it measures how the coupling benchmark deviates from the true hindsight optimum, with an $O(\mathcal{W})$ bound that is independent of the algorithmic process; see \Cref{lemma: r2-and-r3}.

\section{Experiments}

\label{section: experiments}

We evaluate the performance of our algorithms on the task of \textit{online fair notification and recommendation}. In these scenarios, the online items are arriving users, and the task is to fairly distribute users' impressions, via feeds or recommendations, across different content creators or departments of interest that can benefit from user's attention. These ``agents'' have heterogeneous preferences over users with different features. The balance of fairness and efficiency is an important concern here~\citep{kroer2023fair}, since the platform's revenue connects closely to the total engagement from these impressions, while the impressions themselves are consumed by different channels or groups of the platform, which the platform needs to balance for its long-term growth and sustainability.

For agents' valuations for the items, we use the real-world preference data from the Instagram notification dataset~\citep{kroer2023fair} and the MovieLens dataset~\citep{harper2015movielens}; see \Cref{proof-appendix-experiments} for dataset and preprocessing details.

\paragraph{Arrival Models.} We run the greedy algorithm (\Cref{alg: greedy}) and the dual re-solving algorithm (\cref{alg: re-solving (dual)}) under the following three arrival models:

\begin{itemize}
    \item [(1)] The standard \textit{i.i.d.}\ model, where at each time step a platform user is drawn uniformly from the dataset with replacement.
    \item [(2)] A nonstationary model, where the horizon and the agents are uniformly partitioned into $Q$ periods or groups ($Q = 4, 5$ for the two datasets, respectively); in the $q$'th period the item is still uniformly sampled from the dataset, but the agents from the $q$'th agent group double their values.
    \item [(3)] True temporal model, where the arrivals are ordered according to the timestamps in the dataset.
\end{itemize}

For the first two arrival models, the historical sample trace is generated independently using the same manner as the true arrival. For the true temporal case where the true arrival sequence is deterministic, we simulate the historical sample by perturbing each true valuation $v_{t,i}^\online$ by an independent Gaussian noise drawn from $\mathcal{\mathcal{N}}(0, v_{t,i}^\online/2)$. 

\paragraph{Metric and Benchmarks.} For each time step $t$, we consider the time-averaged regret \textit{w.r.t.}\ the optimal generalized-mean welfare over the first $t$ items; for $t=T$ this becomes the eventual time-averaged regret \textit{w.r.t.} the true hindsight optimum. To better align the results from different experiments, we scale the outcome of each simulation so the hindsight optimal welfare value is $1$, and plot all results on a \textit{logarithmic} scale. For instance, $10^{-1}$ time-averaged regret in the plot means $90\%$ approximation ratio of the hindsight objective. To compare our algorithms with existing candidates for online generalized-mean maximization, we also run the algorithms proposed by \citet{barman2022universal} and \citet{huang2025long}. 

\begin{figure*}[!htbp] 
    \centering
    \begin{subfigure}
	\centering
\includegraphics[width=0.48\linewidth]{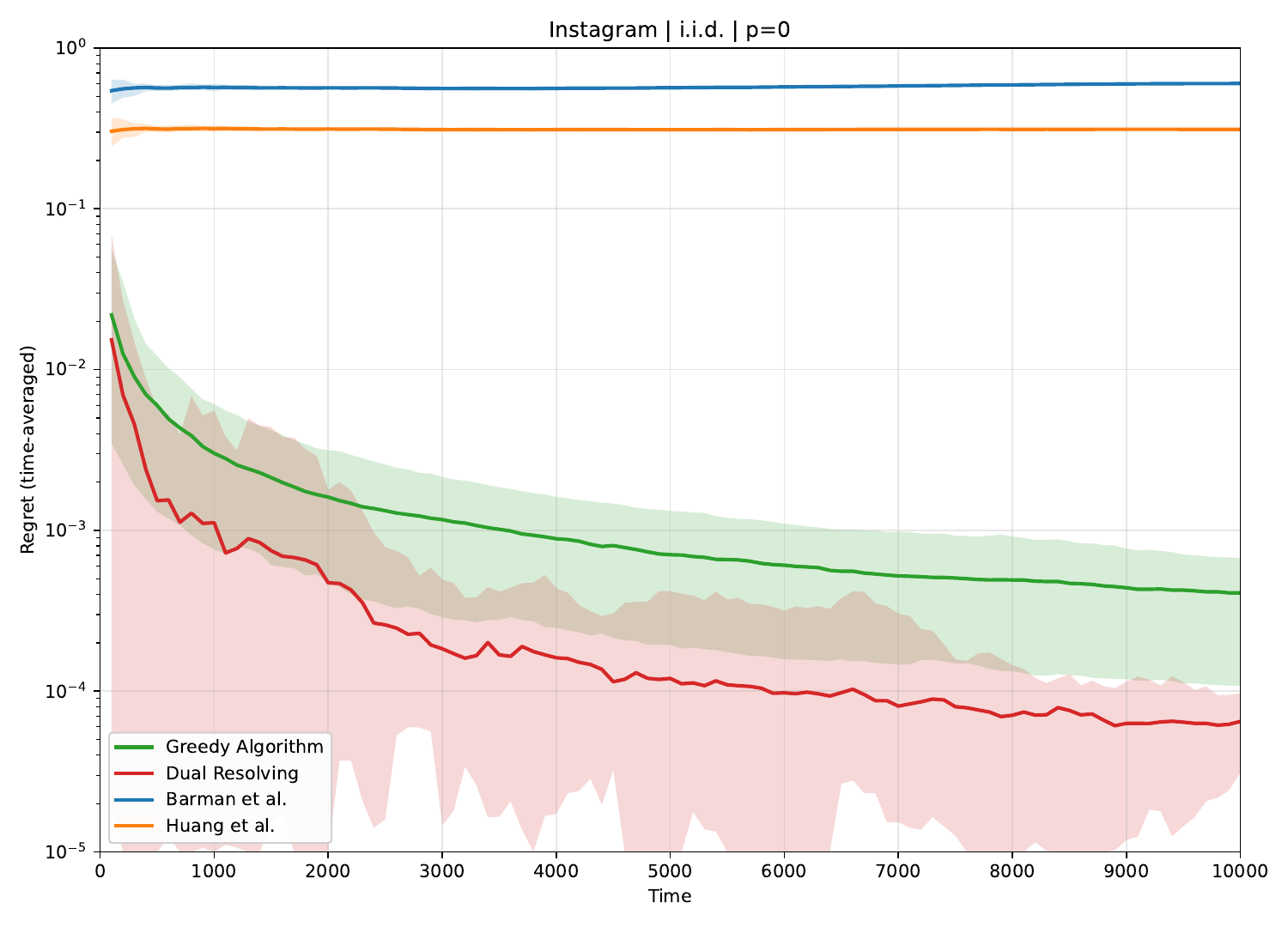}
	\end{subfigure}
     \begin{subfigure}
	\centering
\includegraphics[width=0.48\linewidth]{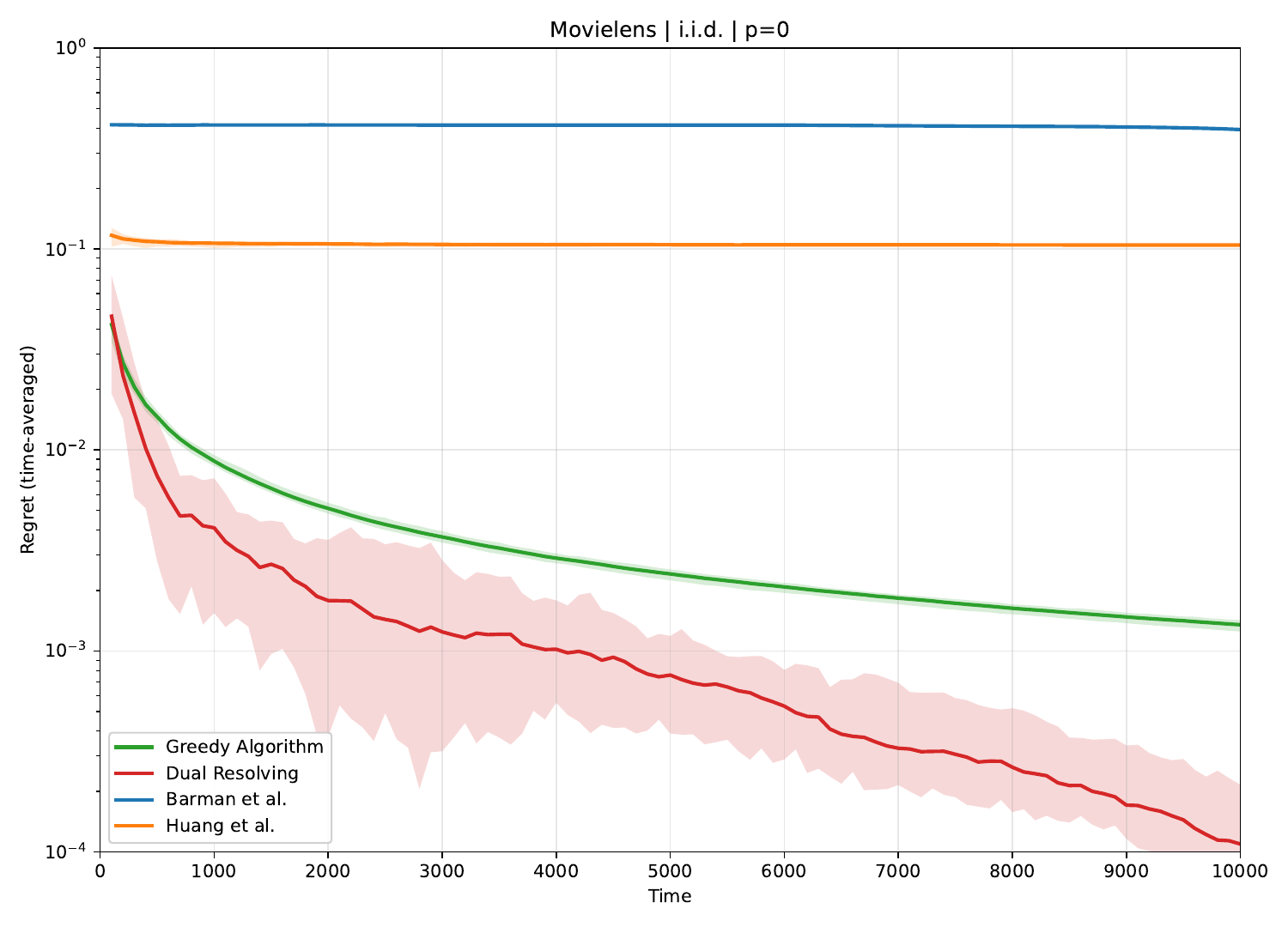}
	\end{subfigure}
     \begin{subfigure}
	\centering
\includegraphics[width=0.48\linewidth]{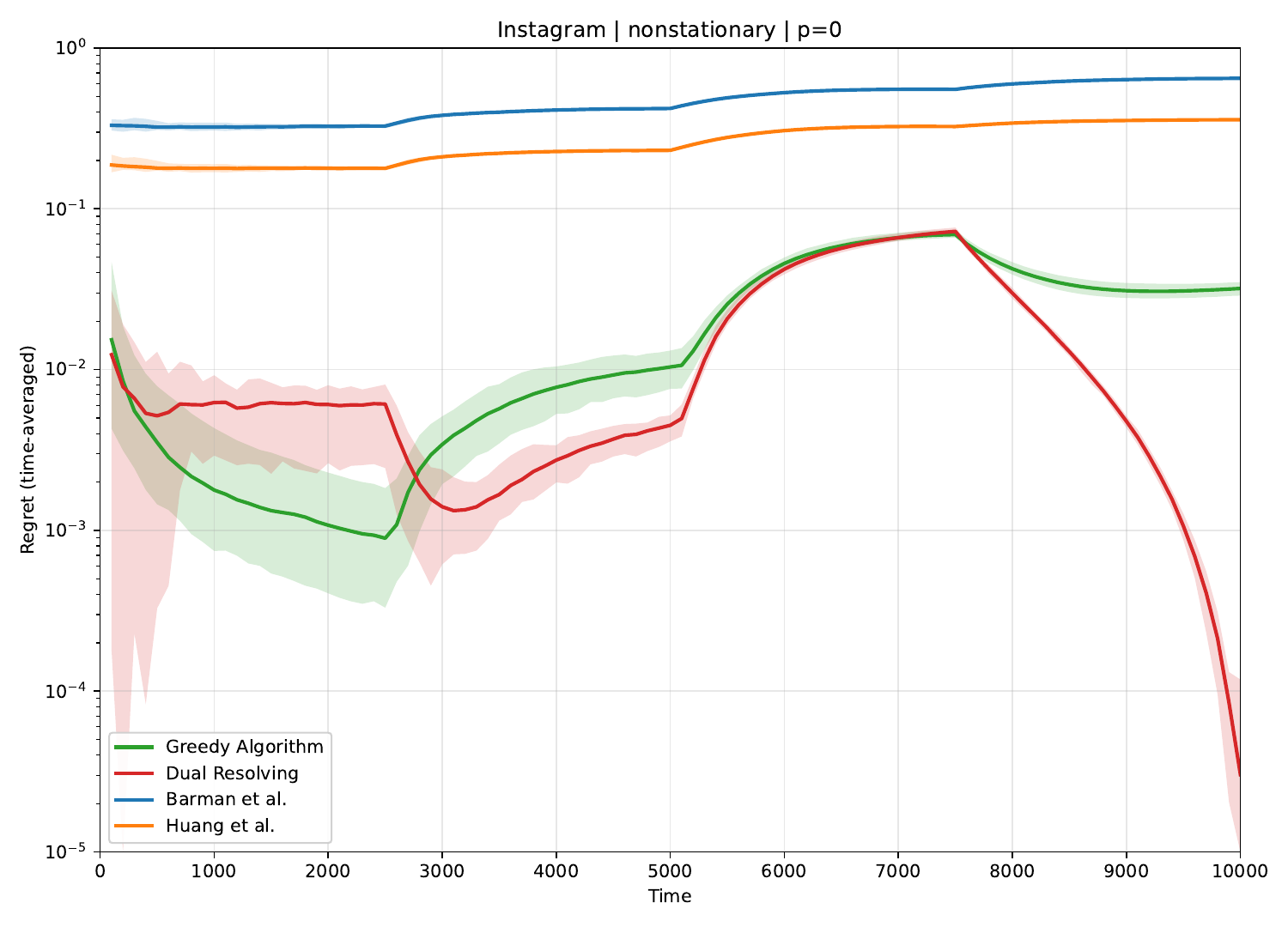}
	\end{subfigure}
     \begin{subfigure}
	\centering
\includegraphics[width=0.48\linewidth]{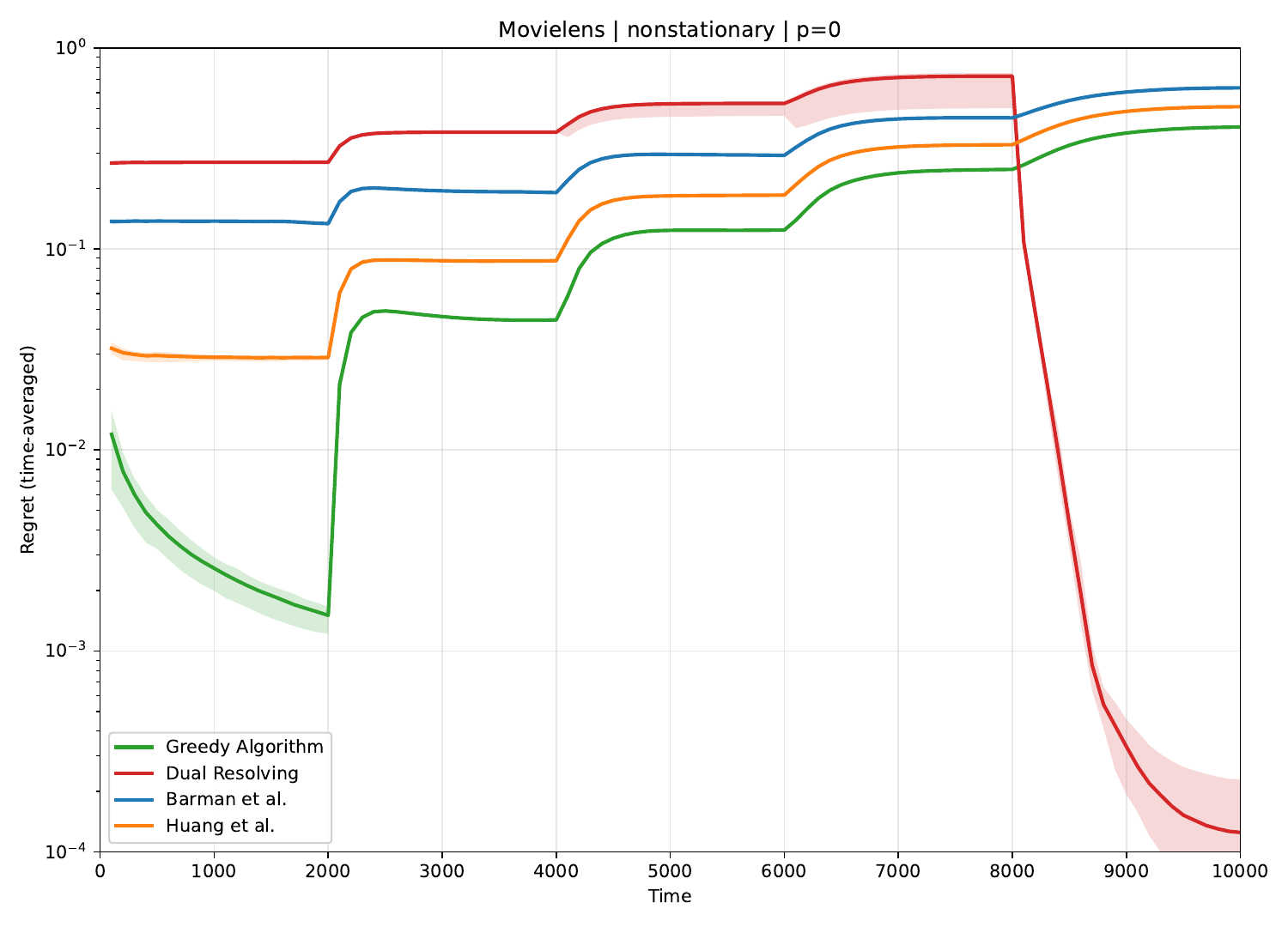}
	\end{subfigure}
     \begin{subfigure}
	\centering
\includegraphics[width=0.48\linewidth]{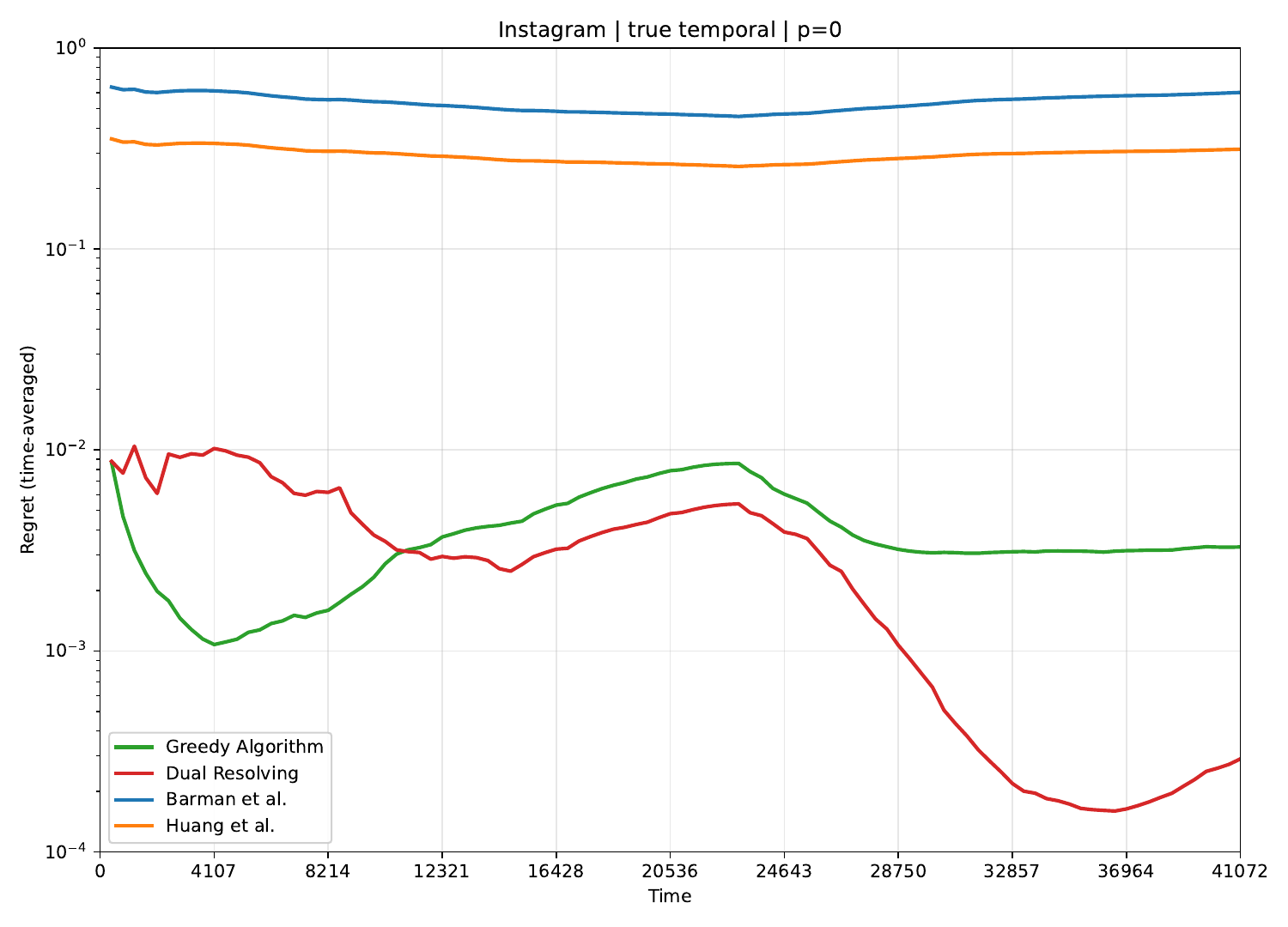}
	\end{subfigure}
     \begin{subfigure}
	\centering
\includegraphics[width=0.48\linewidth]{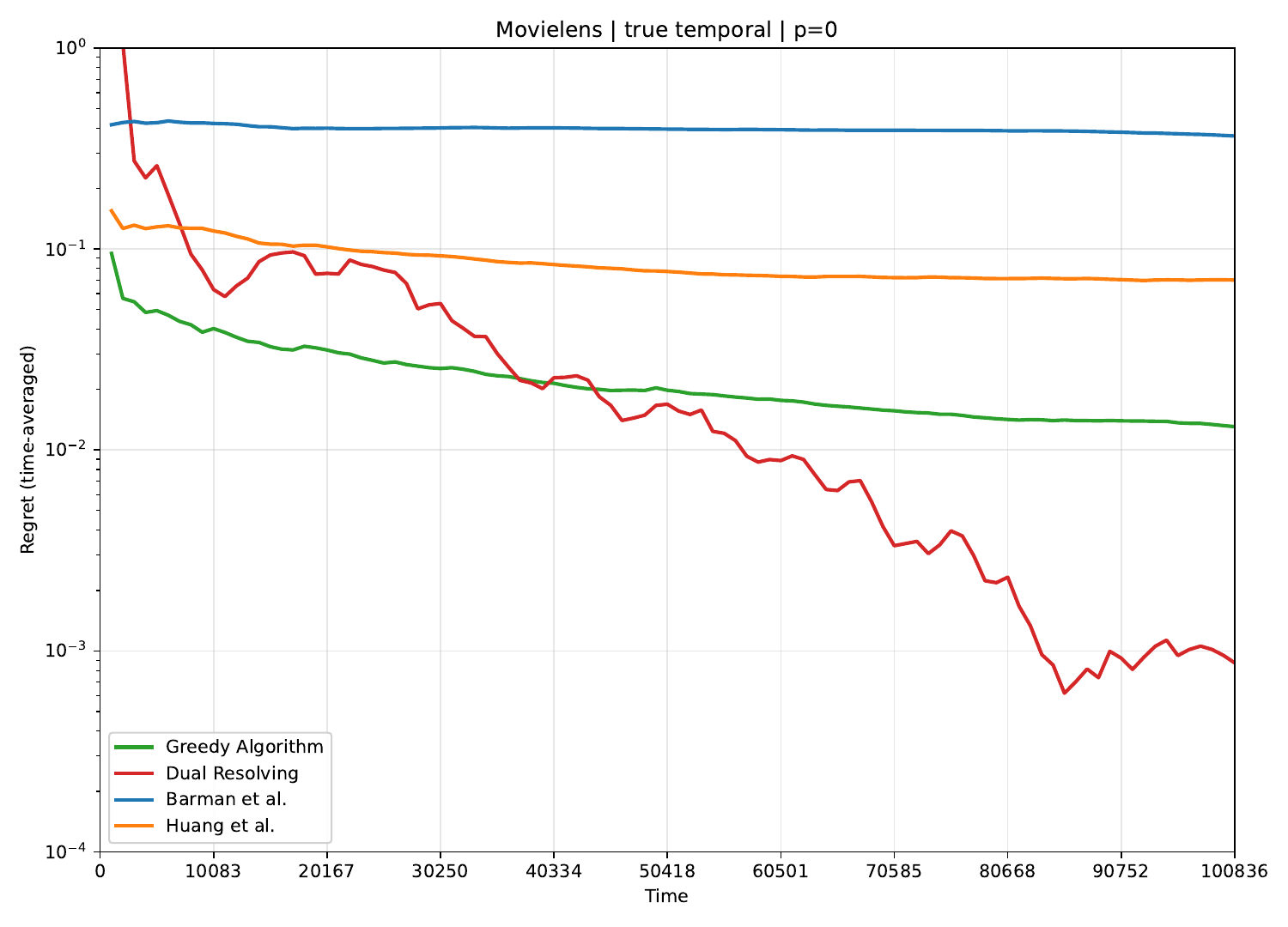}
	\end{subfigure}
    \vspace{1em}
    \caption{\small Simulations of the greedy algorithm and the re-solving algorithm on the Instagram notification dataset \textit{(left)} and the MovieLens dataset \textit{(right)} with the Nash welfare objective $(p=0)$, under three inputs models: \textit{i.i.d.}\ \textit{(top)}, periodic \textit{(middle)}, and true temporal \textit{(bottom)}.}
    \label{fig: p=0}
\end{figure*}

\paragraph{Results.} We report the simulation results for $p =0$ (Nash welfare) in \Cref{fig: p=0}; the results for $p = -1$ and $p = 0.5$ are deferred to \Cref{proof-appendix-experiments}. For the \textit{i.i.d.}\ and nonstationary arrival models, we plot the average outcome over $K=20$ independent repetitions. 

The greedy algorithm displays strong convergence behavior under \textit{i.i.d.}\ arrivals, achieving around $99.9\%$ approximation of the hindsight welfare after $10,000$ rounds for both datasets. It outperforms the algorithms by \citet{barman2022universal} and \citet{huang2025long} in all experiments. Their algorithms are designed for the worst-case inputs and achieve no better than a $0.9$ approximation ratio in the \textit{i.i.d.} setting. For true temporal arrivals, the greedy algorithm also achieves good numerical performance, suggesting its robustness to a broader range of real-world data. 

The dual re-solving algorithm consistently outperforms other candidates in all experiments. For the $Q$-period nonstationary simulations, the re-solving algorithm may deviate from the prefix optimal solution in the first $Q-1$ periods, but it quickly converges to the hindsight optimal in the final period. The results suggest that a single sample per period can be sufficient to provide meaningful structural information about the entire horizon, despite the high degree of nonstationarity. 

\subsection{Efficient Implementation of Dual Re-solving}

\label{section: infrequent-re-solving-and-other-accelerations}

We now discuss the computational cost of \Cref{alg: re-solving (dual)} and how \eqref{eq: re-solving-rule} can be solved more efficiently. The re-solving program \eqref{eq: re-solving-rule} is an $n$-dimensional convex program. Its objective consists of a linear term $\langle \bm \beta, \bm W_{t-1}^\dagger\rangle/T$, a sum of $T-t+1$ piecewise-linear terms $\max_{i\in[n]}\beta_i v_{\tau, i}$, and the conjugate term $(-\log f)_+^\star(-\bm \beta)$; the conjugate term can be explicitly computed in closed form for our generalized-mean objectives. 
Therefore, the re-solving program is an unconstrained convex minimization over the positive orthant, and can be solved using standard first-order methods. Moreover, we leverage the program structure to significantly accelerate the algorithm. We introduce three useful accelerations.

\paragraph{Stochastic minibatches over the future items.} Since the re-solving program \eqref{eq: re-solving-rule} has a sum over $T-t+1$ sampled future items, evaluating its exact subgradient takes $O(n(T-t+1))$ on each $\bm \beta$. However, this sum has a natural unbiased estimate using random item minibatches. For the $t$'th re-solving program, we sample $b$ items from the historical tail $\bm v_{t+1}^{(h)}$, evaluate the subgradients on them and scale accordingly. Each iteration then costs $O(nb)$, which is independent of $t$ or $T$. In our experiments we use stochastic mirror descent for each re-solving program, with minibatch size $b=64$, stepsize $1/\sqrt{k}$, and AdaGrad-style normalization on each co-ordinate.

\paragraph{Warm Start.} From step $t$ to step $t+1$, the re-solving programs changes only in two items. Our stability result (\Cref{lem: stability}) then suggests that the optimal dual solutions $\bm \beta_t^\resolving$ and $\bm \beta_{t+1}^\resolving$ are $O(1/T)$ apart. We can then warm-start each re-solving program by setting the initial iterate in the first-order solver at the previous dual solution. In our implementation, cold-solving the program at $t=1$ requires $50, 000$ iterations, while each subsequent programs only requires about $2,000$ iterations.

\paragraph{Running Time.} With the above accelerations, we report the actual running time of the algorithm per item (i.e., per re-solve), with $T = 10{,}000$; see \Cref{tab: runtime}.

\begin{table}[!htbp]
    \centering
    \begin{tabular}{lcc}
    \toprule
     & Instagram $(n = 4)$ & MovieLens $(n=10)$ \\
    \midrule
    Time per item (ms) & $27.6$ & $29.6$ \\
    \bottomrule
    \end{tabular}
    \caption{\small Time per allocated item of
     the re-solving algorithm 
     when $T = 10, 000$. 
     Single-thread execution. 
     Device: Intel Core i5-14600KF CPU ($5.3$ GHz) with $32$ GB of RAM, under Ubuntu 24.04 with Python 3.12 and NumPy 2.4.}
    \label{tab: runtime}
\end{table}

\paragraph{Infrequent Re-Solving and its Performance.} We can further accelerate the algorithm by decreasing the frequency of re-solving. Since stability suggests that dual solution changes slowly over time, it makes sense to only re-solve the program infrequently: concretely, re-solve every $\tau$ steps, and re-use the dual variable in the next $\tau$ allocation decisions. This reduces the number of re-solved programs from $T$ to $\lceil T/\tau\rceil$, and speeds up \Cref{alg: re-solving (dual)} for approximately $\tau$ times. 

\begin{figure*}[!htbp] 
    \centering
    \begin{subfigure}
	\centering
\includegraphics[width=0.48\linewidth]{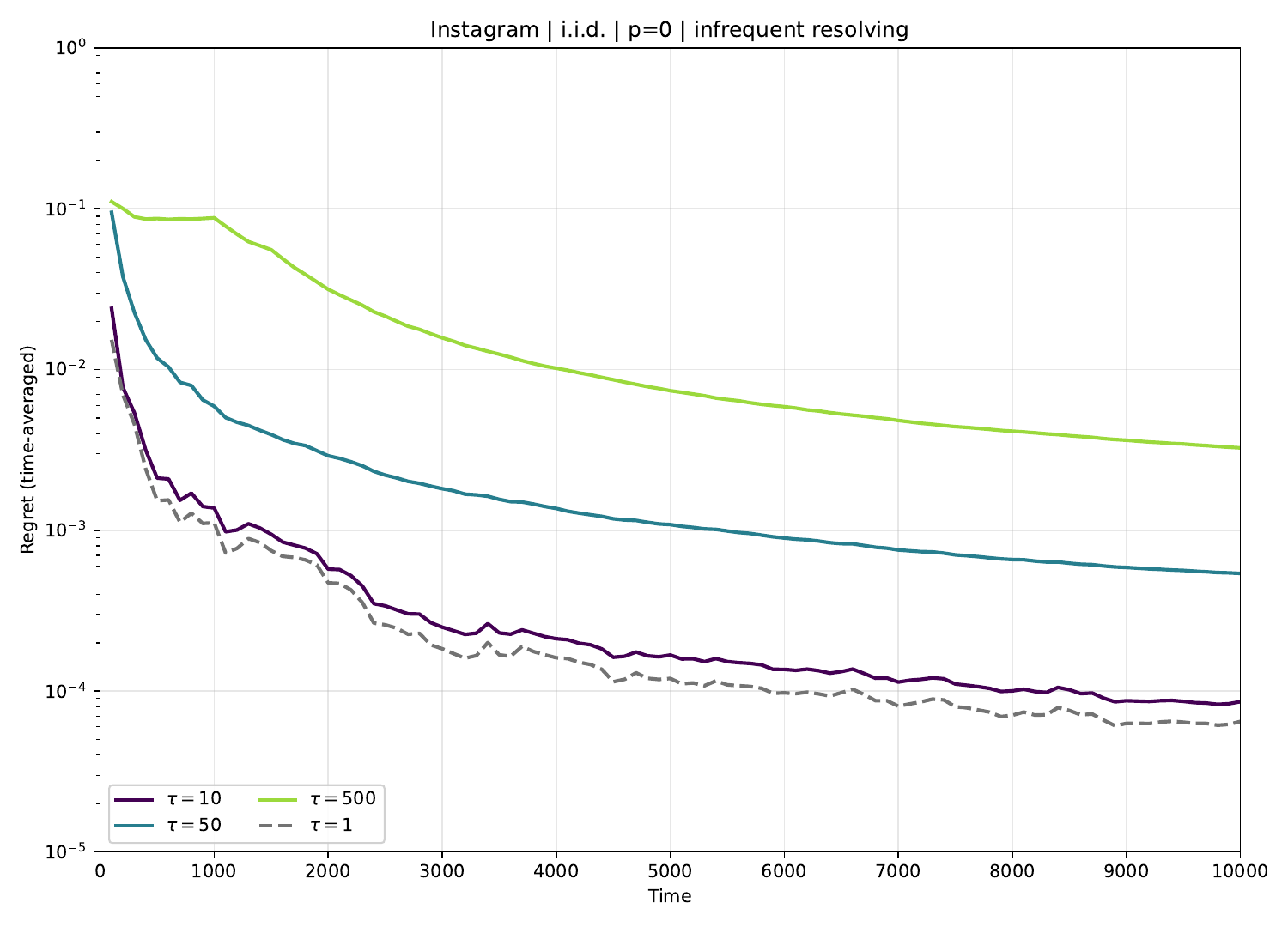}
	\end{subfigure}
     \begin{subfigure}
	\centering
\includegraphics[width=0.48\linewidth]{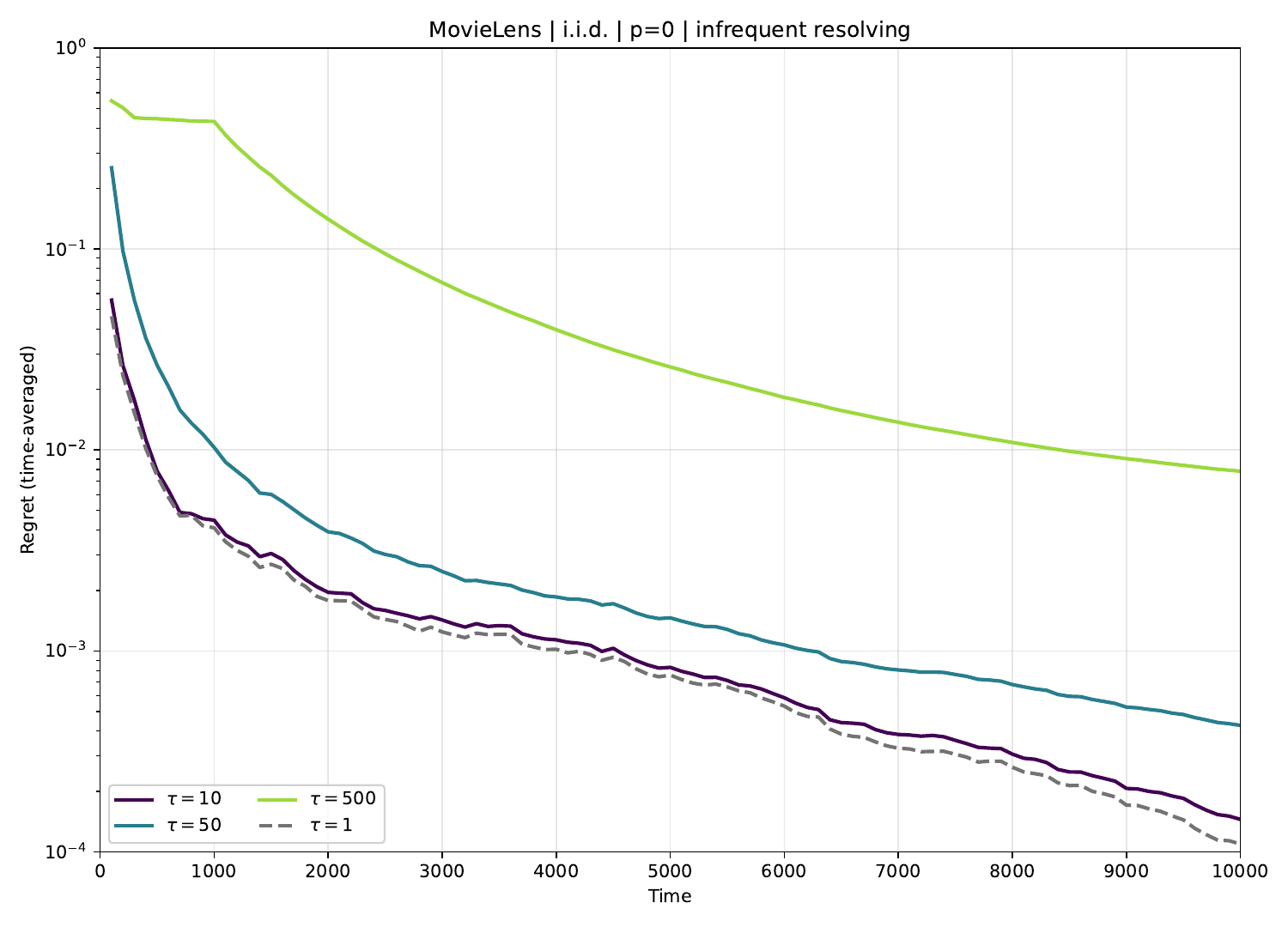}
	\end{subfigure}
     \begin{subfigure}
	\centering
\includegraphics[width=0.48\linewidth]{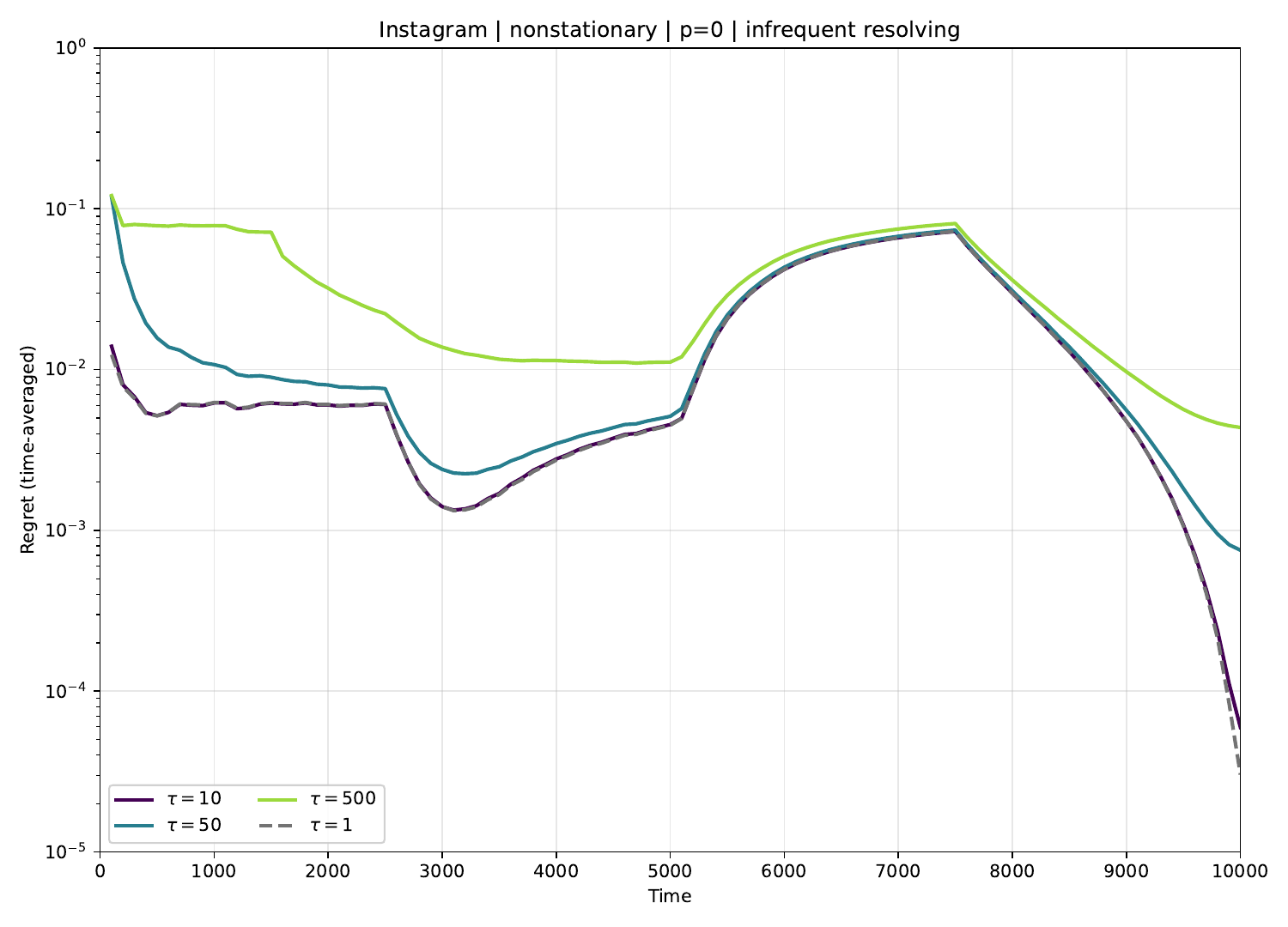}
	\end{subfigure}
     \begin{subfigure}
	\centering
\includegraphics[width=0.48\linewidth]{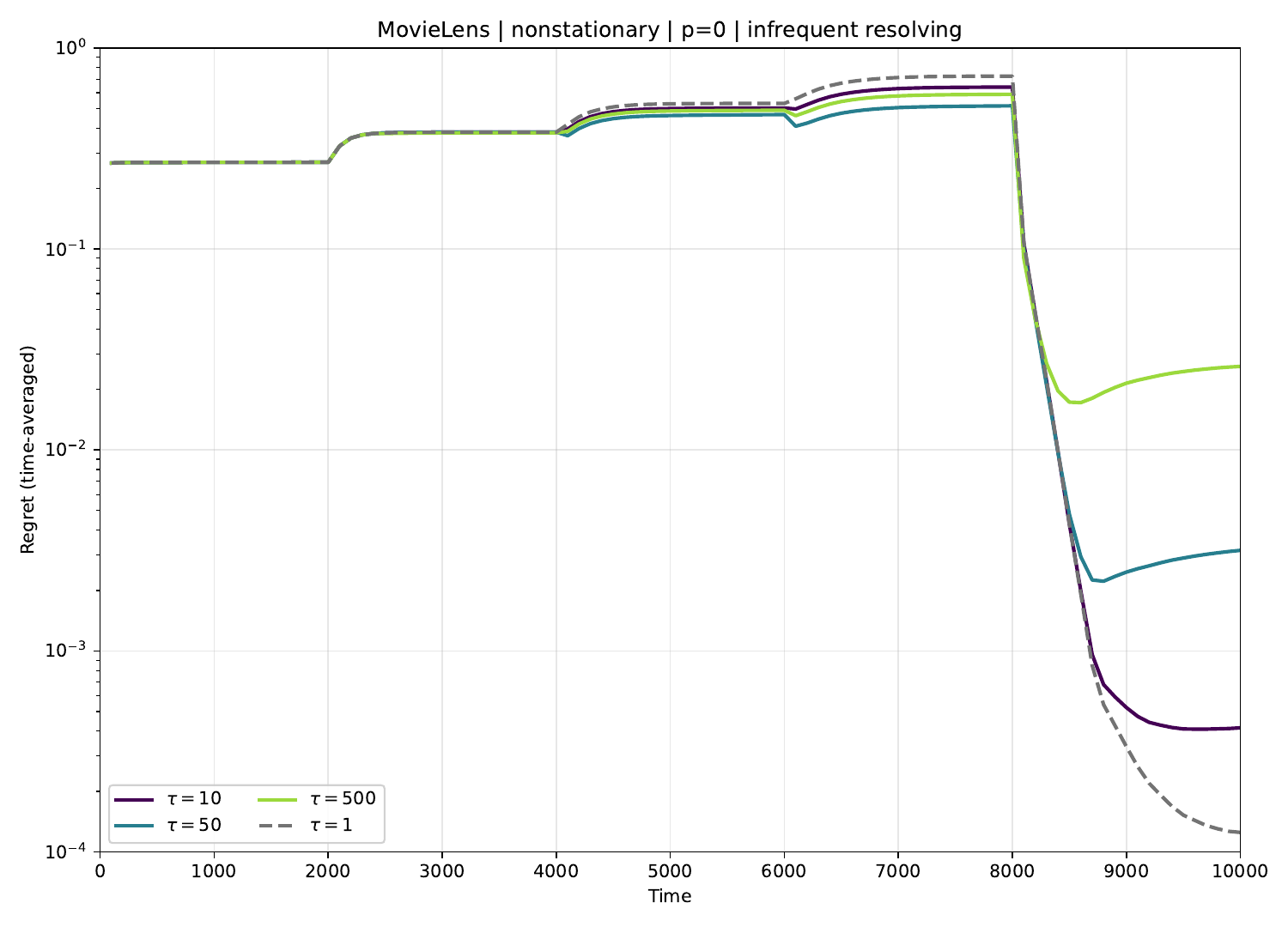}
	\end{subfigure}
     \begin{subfigure}
	\centering
\includegraphics[width=0.48\linewidth]{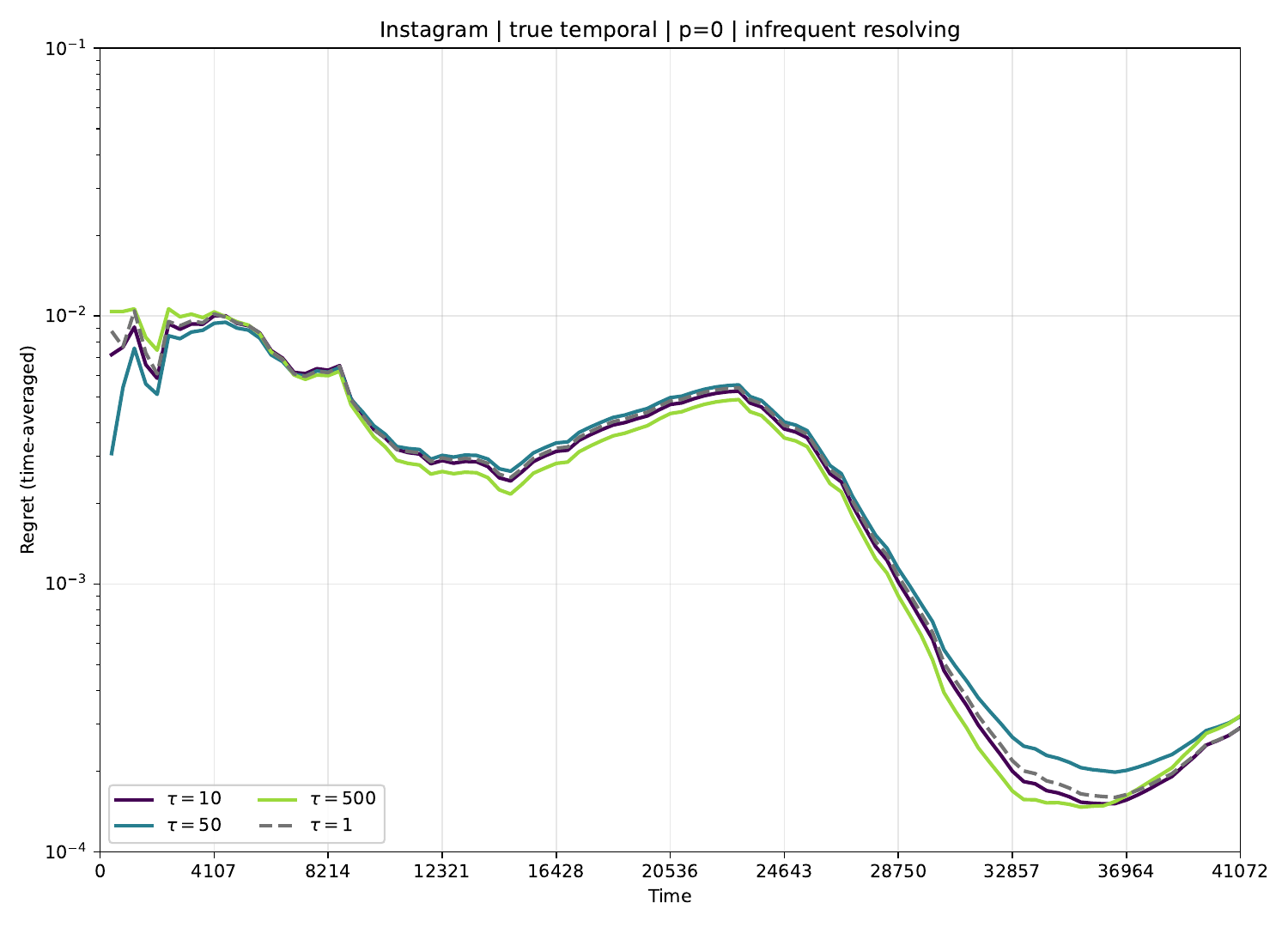}
	\end{subfigure}
     \begin{subfigure}
	\centering
\includegraphics[width=0.48\linewidth]{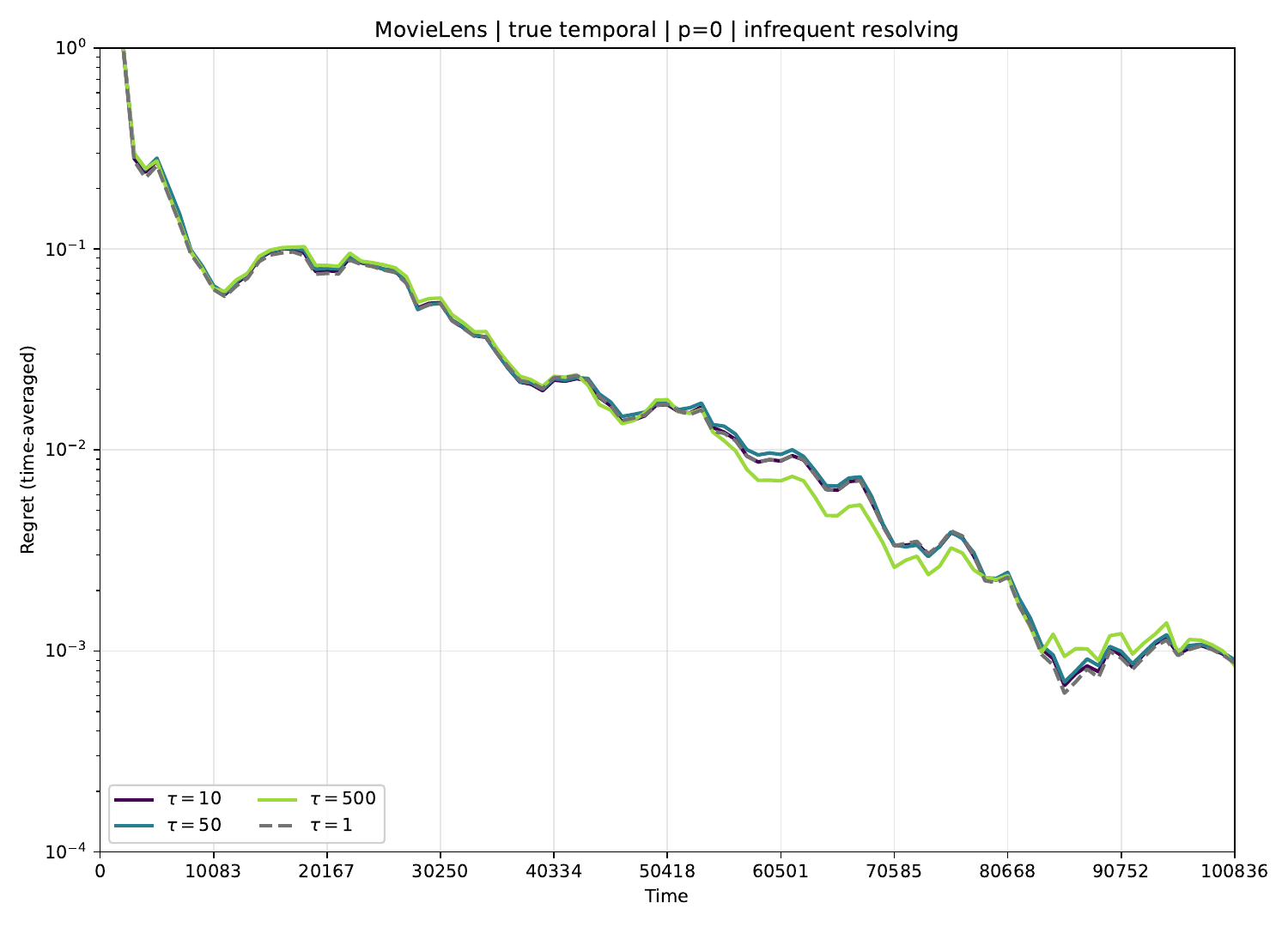}
	\end{subfigure}
    \vspace{1em}
    \caption{\small Simulations of infrequent dual re-solving with re-solving interval $\tau\in\{1, 10, 50, 500\}$ on the Instagram notification dataset \textit{(left)} and the MovieLens dataset \textit{(right)} with the Nash welfare objective $(p=0)$, under three inputs models: \textit{i.i.d.}\ \textit{(top)}, periodic \textit{(middle)}, and true temporal \textit{(bottom)}. The dashed curve $\tau = 1$ is the standard dual re-solving algorithm.}
    \label{fig: tau}
\end{figure*}

While our analysis in \Cref{sec: robust} cannot be exactly applied to infrequent dual re-solving, we provide numerical evidences that solving infrequently trades off little on the performance when $\tau$ is relatively small. We show the time-averaged regret of infrequent re-solving for $\tau = 10, 50, 500$, and compare the results with the $\tau = 1$ case (i.e., the standard re-solving algorithm), for the $p=0$ case. As shown in \Cref{fig: tau}, the regret degrades smoothly as $\tau$ increases. For infrequent re-solving with $\tau = 10$, the algorithm still achieves $99.9\%$ of the hindsight welfare in all cases, tracking the $\tau = 1$ case closely while being approximately $10$ times faster. 

\section{Conclusion}
\label{section: conclusion}

For \textit{i.i.d.}\ arrivals, we show that the simplest greedy algorithm attains $\widetilde{O}(1/T)$ time-averaged regret for our generalized-mean objective, with no knowledge of the arrival distribution. 
Whether the optimality of greedy is restricted to our objective is unknown, and we raise the open question: \textit{Are there broader classes of objectives in online convex optimization where greedy is near-optimal?}

For nonstationary arrivals, where $\Omega(1)$ regret is unavoidable without additional data, our re-solving algorithm recovers the $\widetilde{O}(1/T)$ rate from a single historical sample per distribution, and is robust to distribution shift. This contrasts with the results in online resource allocation with a single sample~\citep{balseiro2023robust,ghuge2025single}, where plain re-solving is not robust and one needs to update the dual variable using a specific online policy.
An open question is: \textit{What are the conditions that enable robustness of the re-solving algorithm for general convex objectives?} Our coupling framework decomposes the regret into a learning error plus drift errors, and could be useful in understanding this question.

\newpage

\bibliographystyle{plainnat}
\bibliography{ref}
\newpage
\appendix

\section{Derivations of the Dual Program and the KKT conditions}
\label{proof: preliminaries}
We show how to derive the dual program \eqref{eq: dual} to the primal program \eqref{eq: primal}. We associate a Lagrange multiplier $\beta_i\geq 0$ to each utility constraint $u_i \leq (1/T) \sumt v_{t,i} x_{t,i}$, and a Lagrange multiplier $p_t \geq 0$ to each supply constraint $\sumi x_{t,i}\leq 1$. We then construct the Lagrangian,
\begin{align*}
    L(\bm u, \bm x_1, \cdots, \bm x_T, \bm \beta, p_1, \cdots, p_T) = \log f(\bm u) - \sumi \ \beta_i \left(u_i - \frac{1}{T}\sumt x_{t,i}v_{t,i}\right) - \sumt \ p_t \left(1- \sumi  x_{t,i}\right).
\end{align*}

The Lagrange dual function is then given by
\begin{align*}
    \tilde{L} (\bm \beta, p_1, \cdots, p_T) &= \max_{\bm u\geq 0, \bm x_t\geq 0}  L(\bm u, \bm x_1, \cdots, \bm x_T, \bm \beta, p_1, \cdots, p_T) \\
    &= \sum_{t=1}^T p_t + \max_{\bm u\geq 0} (\log f(\bm u) - \langle \bm \beta, \bm u\rangle) + \sum_{i=1}^n\sum_{t=1}^T  \max_{x_{t,i}\geq 0} \left(\frac{1}{T} \beta_i v_{t,i} - p_t\right) x_{t,i} \\
    &= \sum_{t=1}^T p_t + (-\log f)_+^\star(-\bm \beta) + \sum_{i=1}^n\sum_{t=1}^T\begin{cases}
        0 & \text{($p_t \geq \beta_i v_{t,i}/T$)}\\
        +\infty & (o.w.)
    \end{cases}.
\end{align*}

The indicator functions can be written as part of a dual constraint. We then get the dual programs
\begin{equation}
\label{eq: dual-without-simplification}
    \begin{aligned}
        \min_{\bm \beta \geq 0, p_1 \cdots, p_T \geq 0} \quad &  \sum_{t=1}^T p_t + (-\log f)_+^\star(-\bm \beta) \\
        \text{s.t.} \quad \ \  \quad & \ \ p_{t} \geq \beta_i v_{t,i}/T, \ \ \ \ \ \ \ \ \ \ \ \ \ \ \ \ \ \ \ \ \forall \ t\in[T], i\in[n].
    \end{aligned}
\end{equation}
The above program \eqref{eq: dual-without-simplification} is a minimization program with the objective increasing in each $p_t$. Therefore, its optimal solution must satisfies $p_t = \max_{i\in[n]} \beta_i v_{t,i}/T$. Bringing this back to the program, we get the equivalent formulation without $p_1, \cdots, p_T.$
\begin{equation*}
    \begin{aligned}
        \min_{\bm \beta \geq 0} \quad &  \frac{1}{T}\sum_{t=1}^T \max_{i\in[n]} \beta_i v_{t,i} + (-\log f)_+^\star(-\bm \beta) .
    \end{aligned}
\end{equation*}
This is exactly what we have in \eqref{eq: dual}. To derive the KKT conditions (\Cref{proposition: kkt}), we first check strong duality via Slater's conditions: the feasible utility space is an $n$-dimensional polyhedron, and there is always an inner point as long as each agent $i$ has at least one nonzero valuation $v_{t,i}>0$. To see the first statement in \Cref{proposition: kkt}, notice that Lagrangian optimality implies
\begin{equation*}
    \bm u^* \in \arg \max_{\bm u\geq 0} (\log f(\bm u) - \langle \bm \beta, \bm u\rangle) \ \implies \ \bm \beta  = \nabla (\log f)(\bm u^*). 
\end{equation*}
To see the second statement, notice that complementary slackness from the dual \eqref{eq: dual-without-simplification} requires $x_{t} >0 \implies p_t = \beta_{i} v_{t,i}/T$, and the fact that $p_t = \max_{i\in[n]} \beta_i v_{t,i}/T$.
\section{Missing Proofs in \Cref{section: greedy}}
\label{proof: greedy}

\subsection{Proof of \Cref{lem: greedy-as-oco}}
\oco*

\paragraph{Proof.}

We write $g(\cdot ;\bm v_t) = \phi(\cdot; \bm v_t) + \psi(\cdot)$, and $\phi(\bm \beta; \bm v_t) = \max_{i\in [n]} \beta_i v_{t, i}, \psi (\bm \beta) = (-\log f)_+^\star(- \bm \beta)$. For each step $t\in [T]$, we define the space of feasible utility vectors that could be derived from item $t$ as 
\begin{equation*}
    \mathcal{U}_{t}^\online = \left\{ \bm u: u_i \in [0, v_{t,i}^\online x_i] {\ \forall i \in [n], \text{ for some } \bm x \geq 0 \text{ with }} \|\bm x\|_1 \leq 1 \right\}.
\end{equation*}

{
We take $\bm u_t^\prime$ to be the vertex of the subdifferential $\partial \phi(\bm \beta_t^\prime; \bm v_t^\actual)$ that is supported on a single agent, i.e., $\bm u_t^\prime = v_{t,i}^\online \cdot \bm e_i$ for an agent $i \in \arg\max_{k\in [n]} \beta_{t,k}^\prime v_{t,k}^\online$. Since $\phi(\cdot; \bm v_t^\actual)$ is a maximization of linear functions, this choice satisfies $\langle \bm \beta_t^\prime, \bm u_t^\prime\rangle = \phi(\bm \beta_t^\prime; \bm v_t^\actual)$, and it is the instantaneous utility vector produced by the integral allocation $\bm x_t = \bm e_i$, so that $\bm u_t^\prime \in \mathcal{U}_t^\online$.

Under the hypothesis of the lemma, no agent has zero cumulative utility at the end of step $t-1$, so \Cref{alg: greedy} allocates item $t$ according to \eqref{greedy: rule}. Since $\bm u_t^\prime \in \mathcal{U}_t^\online$ is a feasible instantaneous utility vector for item $t$, the allocation of \Cref{alg: greedy} at step $t$ has a larger increment in the objective than $\bm u_t^\prime$, i.e.,
}
\begin{equation}
    \label{eq: greedy-as-oco-1}
    \log f\left( \frac{\bm W_{t}^\dagger}{t}\right)  \geq \log f\left(\frac{\bm W_{t-1}^\dagger +\bm u_{t}^\prime }{t}\right).
\end{equation}

{
Since $\bm W_{t-1}^\dagger/(t-1) \in [\ulower, \vmax]^n$ and $t \geq 2$, all coordinates of $\bm W_{t-1}^\dagger/t$ and of $(\bm W_{t-1}^\dagger + \bm u_t^\prime)/t$ are at least $\ulower (t-1)/t \geq \ulower/2$. At most one unit of supply is allocated at each step, so $\sum_{i\in [n]}W_{t-1,i}^\dagger \leq (t-1)\vmax$ and $\|\bm u_t^\prime\|_1 \leq \vmax$. Therefore, both vectors lie in the set $\{\bm u \in [\ulower/2,\vmax]^n: \sum_{i\in [n]}u_i \leq \vmax\}$, which is convex, so the whole segment between them lies in it as well. This segment moves along the single agent on which $\bm u_t^\prime$ is supported, and its length is at most $\vmax/t$. We expand $\log f$ to the second order along this segment, bound the second-order term through $\lambda_{\ulower/2}$, and obtain
\begin{align}
    \log f\left(\frac{\bm W_{t-1}^\dagger +\bm u_{t}^\prime }{t}\right) &\geq \log f\left(\frac{\bm W_{t-1}^\dagger }{t}\right)  + \frac{1}{t}\cdot \left\langle \bm \beta_{t}^\prime, \bm u_{t}^\prime\right\rangle - \frac{\lambda_{\ulower/2} \cdot \vmax^2}{2t^2}\nonumber  \\
    &= \log f\left(\frac{\bm W_{t-1}^\dagger }{t}\right)  + \frac{ \phi(\bm \beta_{t}^\prime;\bm v_t^\online)}{t}- \frac{\lambda_{\ulower/2} \cdot \vmax^2}{2t^2}
    \label{eq: greedy-as-oco-2}
\end{align}
}

\begin{lemma}[Lemma 11 from \citet{xiao2009dual}] 
    \label{lem: conjugate-auxiliary}
    For convex function $\psi$, if $\bm z = \nabla \psi^\star(-\bm s/t)$, then it holds that
    \begin{equation*}
        t\psi^\star (-\bm s/t) + \psi(\bm z) \leq (t-1)\psi^\star (-\bm s/(t-1)).
    \end{equation*}
\end{lemma}
Since $\psi(\bm \beta) = (-\log f)_+^\star (-\bm \beta)$, by concavity of $\log f$ we know that $\log f(\bm u) = -\psi^\star(-\bm u)$. {Taking $\bm s = \bm W_{t-1}^\dagger$ in \Cref{lem: conjugate-auxiliary}, we have $\nabla \psi^\star (-\bm W_{t-1}^\dagger/t) = \nabla \log f (\bm W_{t-1}^\dagger/t) = \bm \beta_t^\prime$.} Hence, we can apply \Cref{lem: conjugate-auxiliary} to get
\begin{equation}
    \label{eq: greedy-as-oco-3}
    t\log f\left( \frac{\bm W^\dagger_{t-1}}{t}\right) - {(t-1) \log f\left( \frac{\bm W_{t-1}^\dagger}{t-1}\right)} \geq \psi (\bm \beta_{t}^\prime). 
\end{equation}
Combining \eqref{eq: greedy-as-oco-1}, \eqref{eq: greedy-as-oco-2} and \eqref{eq: greedy-as-oco-3}, we have
\begin{align*}
& \quad \  t \log f\left( \frac{\bm W_{t}^\dagger}{t}\right) - (t-1)\log f\left(\frac{\bm W
^\dagger_{t-1}}{t-1}\right) \\
    &\geq t \log f\left(\frac{\bm W_{t-1}^\dagger +\bm u_{t}^\prime }{t}\right) - (t-1)\log f\left(\frac{\bm W_{t-1}^\dagger}{t-1}\right) \\
    &\geq t \log f\left(\frac{\bm W_{t-1}^\dagger  }{t}\right) - (t-1)\log f\left(\frac{\bm W_{t-1}^\dagger}{t-1}\right) + {\phi(\bm \beta_{t}^\prime;\bm v_{t}^\online)} -\frac{\lambda_{\ulower/2} \cdot \vmax^2}{2t} \\
    &\geq \psi(\bm \beta_{t}^\prime)+ \phi(\bm \beta_{t}^\prime;\bm v_t^\online) -\frac{\lambda_{\ulower/2} \cdot \vmax^2}{2t} .
\end{align*}
$\hfill \square$

\subsection{Proof of \Cref{lem: high-prob-1-greedy}}

\highprobability*

\paragraph{Proof.}
{
We first fix the parameters associated with the value distribution that determine the constants in the lemma. First, let $\mu_i: = \mathbb{E}[v_{1,i}^\actual] >0$ be the per-item value expectation of agent $i$. By the boundedness of values and a simple conditional probability argument, we have
\begin{equation*}
    \Pr(v_{1,i}^\online \geq \mu_i /2) \geq \frac{\mu_i}{2\vmax}, \ \forall \, i\in[n].
\end{equation*}

For $\delta \in (0,\vmax]$ and $i \in [n]$, consider the sequence of events
\begin{equation*}
    \left\{v_{1, i}^\actual \geq \frac{\mu_i}{2}, \ \bm v_1^\actual  \notin (0, \delta)\right\}.
\end{equation*}

As $\delta \to 0$, the event sequences approaches $\{v_{1,i}^\actual \geq \mu_i/2\}$ since the value vector's positive coordinates are bounded away from zero. By continuity from below of the probability measure, there exists a sufficiently small $\delta \in (0, \vmax]$ such that 
\begin{equation*}
    \Pr \left(v_{1, i}^\actual \geq \frac{\mu_i}{2}, \ \bm v_1^\actual  \notin (0, \delta)\right) \geq q:= \min_{k\in [n]} \frac{\mu_k}{4\vmax}.
\end{equation*}

For notation simplicity, we set $\hat c := (\delta q/8)\cdot \min_{i \in [n]}(\mu_i/(2\vmax))^{1/(1-p)}$ and $\hat d := \max\{128/q^2, 2\vmax/\hat c\}$; these parameters only depend on $\vmax$, the value distribution, and $\vmax$. Since $q \leq 1/4$ and $\delta \leq \vmax$, we have $\hat c \in (0,\vmax]$.
We take $\hat t = \lceil \hat d \cdot (n+\log(nT))\rceil$.

For $t\in [T]$ and $i \in [n]$, consider the set of timesteps $\mathcal{M}_{t,i}=\{\tau\leq t: v_{\tau, i}^\online \geq \mu_i/2, v_{\tau, j} \notin (0, \delta) \ \forall j \in [n]\}$. The indicators of these steps are determined by the value inputs alone, and they are \textit{i.i.d.}\ across $\tau$ with success probability at least $q$. Applying the multiplicative Chernoff bound to the sum of the indicators, we have $$\Pr(|\mathcal{M}_{t,i}|< qt/2)\leq \exp\left(-\frac{qt}{8}\right), \ \ \forall \, t \in [T].$$  We then apply a union bound over $i \in [n]$ and over the steps $t \in [\lceil q \hat t/8\rceil, T]$, and obtain
\begin{equation}
    \label{eq: high-prob-greedy-count}
    \Pr\left( |\mathcal{M}_{t,i}|\geq \frac{qt}{2}, \ \forall i \in [n], \ \forall t \in \left[\left\lceil q\hat t/8\right\rceil, T\right]\right) \geq 1 - nT \exp\left( -\frac{q^2 \hat t}{64}\right) \geq 1 - \frac 1T,
\end{equation}
where in the last step we use the facts $\hat d \geq 128/q^2$ and $\log (nT^2)\leq 2 \log (nT)$. The rest of the proof is deterministic and works on the event in \eqref{eq: high-prob-greedy-count}.

We argue that all agents have positive cumulative utility from step $\lceil q\hat t/8\rceil$ onward. At any step $\tau$ at which some agent has zero cumulative utility and positive value for item $\tau$, \Cref{alg: greedy} allocates item $\tau$ to such an agent by its safeguard rule. Meanwhile, the number of agents with zero cumulative utility strictly decreases, so there are at most $n$ such steps over the whole horizon. Consequently, if agent $i$ has zero cumulative utility at the end of step $t$, then at most $n$ of the steps $\tau \leq t$ have $v_{\tau,i}^\actual>0$, and in particular $|\mathcal{M}_{t,i}|\leq n$. On the event in \eqref{eq: high-prob-greedy-count} and for $t \geq \lceil q \hat t /8\rceil$, we have $|\mathcal{M}_{t,i}| \geq qt/2 \geq q^2\hat t/16 \geq 8(n+\log (nT))>n$, so $W_{t,i}^\dagger>0$ for all $i \in [n]$.

Now fix an agent $i \in [n]$ and a step $t \geq \hat t$. Since $q\hat t/8 \geq 1$, we have $\lceil q\hat t /8\rceil \leq q\hat t/4 \leq qt/4$, so the number of steps in $\mathcal{M}_{t,i}$ that are larger than $\lceil q \hat t/8\rceil$ is at least $qt/4$. Since \Cref{alg: greedy} allocates all items fully, there exists an agent $j \in [n]$ that receives at least $qt/(4n)$ of these steps; let $\tau^\prime$ be the latest received by $j$ among such items.

Consider any step $\tau$ received by $j$ among these steps. Since $\tau > \lceil q\hat t/8\rceil$, no agent has zero cumulative utility at the end of step $\tau -1$, so \Cref{alg: greedy} allocates item $\tau$ according to \eqref{greedy: rule} exactly, i.e., without the zero-utility concern.
By the definition of $\mathcal{M}_{t,i}$ we have $v_{\tau,j}^\actual \geq \delta$. Summing over the steps received by $j$, all of which are at most $\tau^\prime$, 

\begin{equation*}
    W_{\tau^\prime,j}^\dagger \geq \delta \cdot \frac{qt}{4n}.
\end{equation*}

It remains to transfer this lower bound from agent $j$ to agent $i$. If $j = i$, then $$W_{t,i}^\dagger \geq W_{\tau^\prime,i}^\dagger \geq \delta q t/(4n) \geq 2\hat c t/n,$$ where in the last step we use $\hat c \leq \delta q/8$. 

If $j \neq i$, notice that all cumulative utilities at the end of step $\tau^\prime -1$ are positive and $v_{\tau^\prime,j}^\actual>0$. We can then apply \Cref{lem: greedy-rule-implication} at step $\tau^\prime$ and obtain
\begin{align*}
    W_{t,i}^\dagger \geq W_{\tau^\prime -1,i}^\dagger &\geq \left( \frac{v_{\tau^\prime,i}^\actual}{v_{\tau^\prime,j}^\actual}\right)^{\frac{1}{1-p}}\cdot W_{\tau^\prime -1,j}^\dagger - v_{\tau^\prime,i}^\actual \geq \left( \frac{\mu_i}{2\vmax}\right)^{\frac{1}{1-p}}\cdot \left( W_{\tau^\prime,j}^\dagger - \vmax\right) - \vmax \\
    &\geq \left( \frac{\mu_i}{2\vmax}\right)^{\frac{1}{1-p}}\cdot \frac{\delta q t}{4n} - 2\vmax \geq \frac{2 \hat c t}{n} - 2 \vmax,
\end{align*}
where we use $v_{\tau^\prime,i}^\actual \geq \mu_i/2$, $v_{\tau^\prime,j}^\actual \leq \vmax$, $W_{\tau^\prime-1,j}^\dagger \geq W_{\tau^\prime,j}^\dagger - \vmax$, and $(\mu_i/(2\vmax))^{1/(1-p)}\leq 1$. 

Note that in both cases, we have $$W_{t,i}^\dagger \geq 2\hat c t/n - 2\vmax.$$ 

Since $\hat d \geq 2\vmax/\hat c$, we have $t \geq \hat t \geq 2n\vmax/\hat c$ and hence $2\vmax \leq \hat c t /n$, so that $W_{t,i}^\dagger \geq \hat c t/n$. 
This proves the lemma.
}

$\hfill \square$

\subsection{Proof of \Cref{thm: greedy-for-iid}}

\greedy*
\paragraph{Proof.}

{
Without loss of generality, we consider the case with $n\geq 2$ and every agent has positive expected value; the agents with all zero values can be dropped without affecting the allocation and regret results. 
Suppose that $\mathbb{E}[v_{1,i}^\actual] = 0$ for some agent $i$; 
then $v_{t,i}^\actual = 0$ for all $t\in [T]$ with probability one, and agent $i$ has zero utility under every allocation. 

Let $\hat c$, $\hat d$ and $\hat t$ be the constants of \Cref{lem: high-prob-1-greedy}, 
and let $\hat A := \{W_{t,i}^\dagger \geq \hat c \cdot t/n, \ \forall t \in [\hat t, T], \forall i \in [n]\}$ be the good event specified there, so that $\Pr(\hat A)\geq 1-1/T$. 

If $T \leq \hat t$, 
notice that we have $n \log T/T \geq 1/(4\hat d+1)$. Our targets then holds after enlarging the constant $C_1$. In the rest of the proof we assume $T > \hat t$.
}

Applying \Cref{lem: greedy-as-oco}, we know that 
\begin{align}
    \hat A \implies T \log f\left(\frac{\bm W_T^\dagger}{T}\right) &\geq \hat t \log f \left(\frac{\bm W_{\hat t}^\dagger}{\hat t}\right) + \sum_{\tau = \hat t+1}^T g(\bm \beta_\tau^\prime ; \bm v_{\tau}^\online) - \sum_{\tau = \hat t+1}^T  \frac{\lambda^\prime \cdot \vmax^2}{2\tau} \nonumber \\
    & \geq  {\hat t \log \frac{\hat c}{n}} +\sum_{\tau = \hat t+1}^T g(\bm \beta_\tau^\prime ; \bm v_{\tau}^\online) - \frac{\lambda^\prime \cdot \vmax^2}{2} \cdot \log T \label{eq: greedy-dual-regret-0-1},
\end{align}
{where we use $\lambda^\prime$ to denote the constant $\lambda_{\hat c/(2n)}$ of \Cref{lem: greedy-as-oco} for $\ulower = \hat c/n$. On $\hat A$ we have $\bm W_{\tau-1}^\dagger/(\tau-1)\in [\hat c/n, \vmax]^n$ for every $\tau \in [\hat t+1, T]$, so \Cref{lem: greedy-as-oco} applies to these timesteps. 
By applying \Cref{lem: curvature} with $\ulower = \hat c/n$, we also have
\begin{equation}
    \label{eq: greedy-dual-curvature}
    \lambda^\prime \leq \frac{4\left( 1 - 2\min\{p,0\}\right)}{\hat c^2}\cdot \left( \frac{\hat c}{2\vmax}\right)^{2 \min\{p, 0\}}\cdot n .
\end{equation}
}

Next, we show that
\begin{equation}
\label{eq: greedy-dual-regret-0}
    \mathbb{E}\left[\sum_{\tau = \hat t+1}^T g(\bm \beta_\tau^\prime ; \bm v_{\tau}^\online)   \ | \ \hat A \right]\geq T\cdot \mathbb{E}[ \log \mathrm{OPT}(\mathbf{v}^\actual) \ |\ \hat A] { -  g^\fluid \hat t - 2 (\bar g + g^\fluid)},
\end{equation}
{where the finite constants $\bar g, g^\fluid$ will be specified below.}

Let $G(\bm \beta) = \mathbb{E}_{\bm v \sim \mathcal P_t^\online}[g (\bm \beta; \bm v)]$ be the ex-ante (one-step) dual loss function and $\bm \beta^\fluid$ = $(\beta_i^\fluid)_{i\in [n]}$ be its minimizer. {The function $G$ is convex and lower semicontinuous on $\mathbb{R}_+^n$. 
Then $g^\fluid := \max_{\bm v \in [0,\vmax]^n}|g(\bm \beta^\fluid; \bm v)| \leq \vmax \|\bm \beta^\fluid\|_\infty + |\psi(\bm \beta^\fluid)|$ is finite and is determined by $p$, 
 $\mathcal{P}^\online$ and $\vmax$. We investigate the difference
\begin{align}
    &\quad \ \nonumber g(\bm \beta_\tau^\prime;\bm v_\tau^\online)-g(\bm \beta^\fluid;\bm v_\tau^\online) \\&= \left\{g(\bm \beta_\tau^\prime;\bm v_\tau^\online) -G(\bm \beta_\tau^\prime)\right\} - \left\{g(\bm \beta^\fluid;\bm v_\tau^\online) -G(\bm \beta^\fluid)\right\} + \{G(\bm \beta_\tau^\prime) - G(\bm \beta^\fluid) \}  \nonumber \\
     &\geq \underbrace{\left\{g(\bm \beta_\tau^\prime;\bm v_\tau^\online) -G(\bm \beta_\tau^\prime)\right\}}_{\text{(a)}} - \underbrace{\left\{g(\bm \beta^\fluid;\bm v_\tau^\online) -G(\bm \beta^\fluid)\right\}}_{\text{(b)}}. \label{eq: greedy-dual-regret-1}
\end{align}

The last inequality is by the optimality of $\bm \beta^\fluid$. We bound the conditional expectations of (a) and (b) as follows. {We consider the filtration generated by the online value sequence $\mathcal{F}_{\tau-1} = \sigma(\bm v_{1:\tau-1}^\actual)$.\footnote{In the case where the algorithm uses randomized tie-breaking, this additionally includes the random bits used by the algorithm, which are independent of the item sequence.} The cumulative utility $\bm W_{\tau-1}^\dagger$ and the shadow price $\bm \beta_\tau^\prime = \nabla \log f (\bm W_{\tau-1}^\dagger/\tau)$ are $\mathcal{F}_{\tau-1}$-measurable, while $\bm v_\tau^\actual$ is independent of $\mathcal{F}_{\tau-1}$. By the tower rule, $\mathbb{E}[(\text{a})|\hat A] = \mathbb{E}[\mathbb{E}[(\text{a}) \ | \ \mathcal{F}_{\tau-1}, \hat A] \ | \ \hat A]$, and we bound the inner conditional expectation as follows,}
\begin{align*}
    &\quad \mathbb{E}\left[
        { g(\bm \beta_\tau^\prime;\bm v_\tau^\online) -G(\bm \beta_\tau^\prime)} \ | \ \mathcal{F}_{\tau-1}, \hat A
    \right] \\
    & = \int_{[0,\vmax]^n} g(\bm \beta_\tau^\prime; \bm v)  {\mathcal{P}}_\tau^\actual(\mathrm{d} \bm v \ | \ \mathcal{F}_{\tau-1}, \hat A)- \int _{[0, \vmax]^n}g(\bm \beta_\tau^\prime; \bm v) \mathcal{P}_\tau^\actual(\mathrm{d}\bm v) \\
    &\geq -2 \bar g \cdot \|{\mathcal{P}}_\tau^\actual(\cdot \ | \ \mathcal{F}_{\tau-1}, \hat A) - \mathcal{P}_\tau^\actual(\cdot )\|_{\mathrm{TV}}, 
\end{align*}
where the second step uses the $\mathcal{F}_{\tau-1}$-measurability of $\bm \beta_\tau^\prime$ together with the definition of $G$, and $\bar g$ is the maximum absolute value of $g(\bm \beta_\tau^\prime; \bm v)$ over $\bm v \in [0,\vmax]^n$ and over the histories with $\Pr(\hat A \ | \ \mathcal{F}_{\tau-1})>0$. 

We next show the upper bound on $\bar g$. 
Since $f$ satisfies positive $1$-homogeneity, we have that $\langle \bm u, \nabla \log f(\bm u)\rangle = 1$ for every $\bm u \in \mathbb{R}_{++}^n$; since $\bm \beta_\tau^\prime = \nabla \log f (\bm W_{\tau-1}^\dagger/\tau)$ is non-negative, each of its coordinates is at most $2n/\hat c$. Therefore, $\max_{i\in [n]}\beta_{\tau,i}^\prime v_i \leq 2n \vmax/\hat c$ for all $\bm v \in [0,\vmax]^n$. Meanwhile, notice that $$\psi(\bm \beta_\tau^\prime) = \log f (\bm W_{\tau-1}^\dagger/\tau)-1.$$ By the boundedness of utility vector we have $\bar g \leq 2n\vmax/\hat c + 1 + \log \max\{2n/\hat c, \vmax\}$.

For the outer expectation, we apply the bound $\|\mathcal{P}(\cdot|A) - \mathcal{P}(\cdot)\|_{\mathrm{TV}} \leq 1 -\Pr(A)$, valid for any $\mathcal{P}$-measurable event $A$, {conditionally on $\mathcal{F}_{\tau-1}$, and we use the independence of $\bm v_\tau^\actual$ from $\mathcal{F}_{\tau-1}$ in the form $\mathcal{P}_\tau^\actual(\cdot \ | \ \mathcal{F}_{\tau-1}) = \mathcal{P}_\tau^\actual(\cdot)$. We obtain}
\begin{align*}
    \mathbb{E}[(\text{a})|\hat A] &\geq - 2\bar g \cdot \mathbb{E}\left[ \|\mathcal{P}_\tau^\actual(\cdot \ | \ \mathcal{F}_{\tau-1}, \hat A) - \mathcal{P}_\tau^\actual(\cdot )\|_{\mathrm{TV}} \ | \ \hat A\right] \\
    &\geq - 2\bar g \cdot \mathbb{E}\left[ 1-\Pr(\hat A \ |\ \mathcal{F}_{\tau-1}) \ | \ \hat A\right]\\
    & \geq -2\bar g (1-\Pr(\hat A)),
\end{align*}
{where the last step uses the tower rule and Jensen's inequality,
\begin{equation*}
    \mathbb{E}\left[ \Pr(\hat A \ |\ \mathcal{F}_{\tau-1}) \ | \ \hat A \right] = \frac{\mathbb{E}\left[ \Pr(\hat A\ |\ \mathcal{F}_{\tau-1})^2\right]}{\Pr(\hat A)} \geq \frac{\mathbb{E}\left[ \Pr(\hat A \ |\ \mathcal{F}_{\tau-1})\right]^2}{\Pr(\hat A)} = \Pr(\hat A).
\end{equation*}
By the same argument applied to the fixed price $\bm \beta^\fluid$ and without the inner conditional expectation, we get $\mathbb{E}[\text{(b)}|\hat A] \leq 2 g^\fluid (1-\Pr (\hat A))$.} Bringing this back to \eqref{eq: greedy-dual-regret-1}, and summing up over $\tau = \hat t+1, \cdots, T$ gives us 
\begin{align*}
    \mathbb{E}\left[\sum_{\tau = \hat t+1}^T g(\bm \beta_\tau^\prime ; \bm v_{\tau}^\online)   \ | \ \hat A \right] &\geq \mathbb{E}\left[\sum_{\tau = \hat t+1}^T g(\bm \beta^\fluid ; \bm v_{\tau}^\online)   \ | \ \hat A \right] - 2 (T-\hat t)(\bar g + g^\fluid ) (1- \Pr (\hat A)) \\
    & \geq \mathbb{E}\left[\sum_{\tau = 1}^T g(\bm \beta^\fluid ; \bm v_{\tau}^\online)   \ | \ \hat A \right] - g^\fluid\hat t - 2(\bar g + g^\fluid ) \\
    &= T\cdot \mathbb{E}\left[ D(\bm \beta^\fluid ; \mathbf{v}^\online)   \ | \ \hat A \right] - g^\fluid\hat t - 2(\bar g + g^\fluid )\\
    & \overset{\text{(I)}}{\geq}  T \cdot \mathbb{E}\left[D^*(\mathbf v^\online)  \ | \ \hat A \right]- g^\fluid\hat t - 2(\bar g + g^\fluid )\\
    &\overset{\text{(II)}}{=}  T \cdot  \mathbb{E}\left[ \log \mathrm{OPT}(\mathbf v^\online)  \ | \ \hat A \right] - g^\fluid\hat t - 2(\bar g + g^\fluid ),
\end{align*}
where { $D(\bm \beta; \mathbf{v}^\online) := (1/T)\sum_{t=1}^T g(\bm \beta; \bm v_t^\online)$ is the objective of the dual program \eqref{eq: dual}, the second inequality uses $(T-\hat t)(1-\Pr(\hat A))\leq 1$ and $g(\bm \beta^\fluid;\bm v)\leq g^\fluid$ for the first $\hat t$ terms, and} {step} (I) is by the instance-wise optimality {of $D^*(\mathbf v^\online)$ over $\bm \beta \in \mathbb{R}_+^n$}, and {step} (II) is by strong duality (\Cref{proposition: kkt}). This proves \eqref{eq: greedy-dual-regret-0}. 

{
Combining \eqref{eq: greedy-dual-regret-0-1} and \eqref{eq: greedy-dual-regret-0} with the definition $\mathcal{R}_{\log f}(\mathcal{A},\bv^\online) = \log \mathrm{OPT}(\mathbf{v}^\online) - \log f(\bm u_{\mathcal{A}}^\dagger)$, we have
\begin{equation}
    \label{eq: greedy-dual-regret-2}
    \mathbb{E}[\mathcal{R}_{\log f}(\mathcal{A}, \bv^\online)|\hat A] \leq \frac{1}{T} \left( \hat t \log \frac{n}{\hat c} + g^\fluid \hat t + 2 (\bar g + g^\fluid) + \frac{\lambda^\prime \vmax^2 \log T}{2}\right).
\end{equation}

Recall the definition and bounds on $\bar g, \hat t, \hat c$, and $\lambda'.$ We can show that each of the four terms in the bracket in \eqref{eq: greedy-dual-regret-2} can be by constant multiples of $n \log T$, with the constants determined by $p$, $\mathcal{P}^\online$ and $\vmax$. 

}

Finally, using a conditional expectation argument, we have 
\begin{align*}
     \mathbb{E}[\mathcal{R}_f(\mathcal{A}, \mathbf v^\actual)] &\leq {f(\mathbf{1}\vmax)}\cdot  (1-\Pr(\hat A))+ {f(\mathbf{1}\vmax)} \cdot \mathbb{E}[\mathcal{R}_{\log f}(\mathcal{A, \bv^\actual}) \ | \ \hat A],
\end{align*}
{where the first step uses \Cref{proposition: regret-to-ratio} on $\hat A$ and $\mathcal{R}_f (\mathcal{A}, \mathbf{v}^\actual)\leq \mathrm{OPT}(\bv^\online)$ on its complement, together with the bound $\mathrm{OPT}(\bv^\online)\leq f(\mathbf 1 \vmax) = \vmax$.
This proves \Cref{thm: greedy-for-iid}. 

$\hfill \square$

\subsection{\Cref{lem: greedy-rule-implication} and its Proof}
\begin{lemma}
    \label{lem: greedy-rule-implication}
    For step $t\in [T-1]$ with $\bm W_t^\dagger \in \mathbb{R}_{++}^n$, if item $t+1$ is allocated to agent $j$ by \Cref{alg: greedy} with $v_{t+1,j}^\online>0$. Then for any $i\neq j$, we have 
    \begin{equation*}
        {W_{t, i}^\dagger} \geq \left(\frac{v_{t+1, i}^\online }{v_{t+1, j}^\online}\right)^{\frac{1}{1-p}} \cdot {W_{t,j}^\dagger} - v_{t+1,i}^\online. 
    \end{equation*}
\end{lemma}
\paragraph{Proof.} For simplicity, in this proof we use shorthands $a = v_{t+1, i}^\online, b = v_{t+1, j}^\online$. We also drop the $\dagger$ superscripts on existing utilities; there will be no ambiguity in this proof. 

We consider the cases with different $p$ in the objective welfare function. 
\begin{itemize}
    \item Case 1: $p \in (0,1)$. Since the greedy algorithm prefers $j$ to $i$ on step $t+1$, we have
    \begin{equation*}
        (W_{t,i}+a)^p - W_{t,i}^p \leq (W_{t,j}+b)^p - W_{t,j}^p. 
    \end{equation*}
    By the identity $(x+y)^p - x^p = p \int_{x}^{x+y} z^{p-1} \mathrm{d}z$, this implies 
    \begin{equation*}
        p\int_{W_{t,i}}^{W_{t,i}+a} z^{p-1} \mathrm{d}z \leq p\int_{W_{t,j}}^{W_{t ,j}+b} z^{p-1} \mathrm{d}z.
    \end{equation*}
    Since $p>0$ and $z^{p-1}$ is decreasing on $\mathbb{R}_+$, this further implies
    \begin{equation*}
        a (W_{t,i}+a)^{p-1} \leq bW_{t,j}^{p-1}. 
    \end{equation*}
    Solving this inequality gives us the desired result.
    \item Case 2: $p = 0$. Notice that the objective is equivalent to the product of agent utilities. Since the greedy algorithm prefers $j$ to $i$ on step $t+1$, we have
    \begin{equation*}
        W_{t,j} (W_{t,i}+a) \leq W_{t,i} (W_{t,j} + b).
    \end{equation*}
    The desired inequality follows from rearranging terms.
    \item Case 3: $p\in (-\infty, 0)$. Since the greedy algorithm prefers $j$ to $i$ on step $t+1$, we have 
    \begin{equation*}
        (W_{t,i}+a)^p - W_{t,i}^p \geq (W_{t,j}+b)^p - W_{t,j}^p. 
    \end{equation*}
    By the identity $(x+y)^p - x^p = p \int_{x}^{x+y} z^{p-1} \mathrm{d}z$, this implies 
    \begin{equation*}
        p\int_{W_{t,i}}^{W_{t,i}+a} z^{p-1} \mathrm{d}z \geq p\int_{W_{t,j}}^{W_{t ,j}+b} z^{p-1} \mathrm{d}z.
    \end{equation*}
{We divide both sides by $p<0$, which reverses the inequality, and then bound the left integral from below by $a(W_{t,i}+a)^{p-1}$ and the right integral from above by $bW_{t,j}^{p-1}$, using the fact that $z^{p-1}$ is decreasing on $\mathbb{R}_+$. This further implies}
    \begin{equation*}
        a (W_{t,i}+a)^{p-1} \leq bW_{t,j}^{p-1}. 
    \end{equation*}
    This is same as the inequality we get in Case $1$. Solving this inequality gives us the desired result. 
\end{itemize}

$\hfill \square$

{
\subsection{\Cref{lem: curvature} and its Proof}
\begin{lemma}
    \label{lem: curvature}
    For any $\ulower \in (0, \vmax/n]$, the constant $\lambda_{\ulower/2}$ in \Cref{lem: greedy-as-oco} satisfies
    \begin{equation*}
        \lambda_{\ulower/2} \leq \frac{4\left( 1-2\min\{p,0\}\right)}{n \ulower^2}\cdot \left(\frac{n \ulower}{2 \vmax}\right)^{2 \min\{p, 0\}}.
    \end{equation*}
\end{lemma}
\paragraph{Proof.} Fix an agent $i \in [n]$ and a feasible utility vector $\bm u \in [\ulower/2, \vmax]^n$ with $\sum_{k \in [n]}u_k \leq \vmax$. For $p = 0$ we have $\partial_{ii}\log f(\bm u) = -1/(n u_i^2)$, and the claim follows from $u_i \geq \ulower/2$. We next consider $p \in (-\infty, 1]\backslash\{0\}.$ 

For $p \neq 0$, one can compute that
\begin{equation*}
    \partial_{ii}\log f(\bm u) = \frac{(p-1)u_i^{p-2}}{\sum_{k\in [n]}u_k^p} - \frac{p \cdot u_i^{2p-2}}{\left(\sum_{k \in [n]}u_k^p\right)^2}, \quad  \quad \left| \partial_{ii}\log f(\bm u)\right| \leq \frac{(1-p)u_i^{p-2}}{\sum_{k\in [n]}u_k^p} + \frac{|p| \cdot u_i^{2p-2}}{\left(\sum_{k \in [n]}u_k^p\right)^2}.
\end{equation*}

Notice that both exponents $p-2$ and $2p-2$ are negative for $p<1$. For $p \in(0,1)$, applying the utility upper and lower bounds directly gives us 
\begin{equation*}
    |\partial_{ii}\log f(\bm u)| \leq 4(1-p)/(n\ulower^2) + 4p/(n^2 \ulower^2)\leq 4/(n \ulower^2).
\end{equation*}

For $p<0$, the function $x \mapsto x^p$ is convex and decreasing on $(0,\infty)$, so we apply Jensen's inequality and $\sum_{k\in [n]}u_k \leq \vmax$ to obtain $(1/n)\sum_{k\in [n]}u_k^p \geq \left((1/n)\sum_{k \in [n]}u_k\right)^p \geq (\vmax/n)^p$, i.e., $\sum_{k \in [n]}u_k^p \geq n^{1-p}\vmax^p$. Substituting the three bounds, we obtain
\begin{equation*}
    \left| \partial_{ii}\log f(\bm u)\right| \leq \frac{4(1-p)}{n \ulower^2}\cdot \left( \frac{n \ulower}{2 \vmax}\right)^p + \frac{4|p|}{n^2 \ulower^2}\cdot \left( \frac{n \ulower}{2\vmax}\right)^{2p} \leq \frac{4(1-2p)}{n \ulower^2}\cdot \left( \frac{n\ulower}{2\vmax}\right)^{2p},
\end{equation*}
where in the last step we use $n \ulower/(2\vmax)\leq 1/2$ and $p<0$ to bound $(n \ulower/(2\vmax))^p \leq (n\ulower/(2\vmax))^{2p}$, together with $n \geq 1$. Taking the maximum over $i \in [n]$ and over $\bm u$ proves the lemma.

$\hfill \square$
}

\section{Missing Proofs in \Cref{section: re-solve}}
\label{proof: re-solve}

\subsection{Proof of \Cref{lem: stability}}
\stability*
\paragraph{Proof of Monotonicity. }
We first prove $u_i^*(\bv^+) - u_i^*(\bv^-) \geq 0$. We let $(\bm u^-,\bm\beta^-)$ and $(\bm u^+,\bm\beta^+)$ denote the optimal primal and dual solutions for the two instances. Fix any optimal allocations $\mathbf x^-$ and $\mathbf x^+$ that do not allocate items to an agent with zero value. 

We prove by contradiction. For the sake of contradiction, assume $I^\star:=\{i\in[n]:u_i^+<u_i^-\}$ is nonempty. Then, on the common items, every agent in $I^\star$ obtains strictly more utility under $\mathbf x^-$ than under $\mathbf x^+$. If not, substituting the former $\mathbf x^-$ for $\mathbf{x}^+$ on the common items will strictly improve every agent in $I^\star$ while weakly improve other agents, contradicting the Pareto optimality of $\mathbf x^+$.

Consequently, there exists a common item $t^\star$ on which the aggregate share of $I^\star$ is larger under $\mathbf x^-$ than under $\mathbf x^+$. Therefore, there exist agents $i\in I^\star$ and $j\notin I^\star$ such that
$x_{t^\star,i}^->x_{t^\star,i}^+$ and
$x_{t^\star,j}^+>x_{t^\star,j}^-$. By the KKT conditions,
\begin{equation}
\label{eq: proof-monotonicity}
    \frac{\beta_i^+}{\beta_j^+}
    \leq \frac{v_{t^\star,j}}{v_{t^\star,i}}
    \leq \frac{\beta_i^-}{\beta_j^-}.
\end{equation}
By the stationarity condition of the optimal solutions, we have 
\begin{equation}
\label{eq: proof-monotonicity-contradiction}
    \frac{\beta_i^+}{\beta_j^+}
    =\frac{B_i}{B_j}\left(\frac{u_i^+}{u_j^+}\right)^{p-1}, \ \  \frac{\beta_i^-}{\beta_j^-}
    =\frac{B_i}{B_j}\left(\frac{u_i^-}{u_j^-}\right)^{p-1}.
\end{equation}
Since $u_i^+<u_i^- , u_j^+\geq u_j^-$, and $p<1$, \eqref{eq: proof-monotonicity-contradiction} implies $\beta_i^+/\beta_j^+>\beta_i^-/\beta_j^-$.  This contradicts \eqref{eq: proof-monotonicity}.

\paragraph{Proof of Stability.}
We next show the upper bound on $u_i^*(\bv^+) - u_i^*(\bv^-)$. Let $\bm u^M$ be the utility that $\bm x^-$ obtains on the $K$ added items. The first-order optimality condition at $\bm u^-$ gives
\begin{align}
    \sum_{i=1}^n\beta_i^-(u_i^+-u_i^-)
    &\leq \sum_{i=1}^n\beta_i^-u_i^M
    \leq \max_{j\in[n]}\beta_j^-\sum_{i=1}^n u_i^M
    \leq \frac{K\vmax}{T}\max_{j\in[n]}\beta_j^-.
    \label{eq: proof-stability-weighted}
\end{align}
Since the monotonicity argument gives $u_i^+-u_i^-\geq0$ for every $i$, isolating the $i$th term in \eqref{eq: proof-stability-weighted} proves the result. 
\begin{equation*}
    0\leq u_i^+-u_i^-
    \leq \frac{K\vmax}{T}\frac{\max_{j\in[n]}\beta_j^-}{\beta_i^-}.
\end{equation*}
The same argument applies to hybrid programs: an increase of $W_i$ is an added one-hot utility endowment, with its magnitude divided by $T$ in the resulting utility.

$\hfill\square$

\subsection{Stability under a Single Modification}
\label{subsec: single-modification}

We show how \Cref{lem: monotonicity} could be used to analyze the changes in the hybrid program in one iteration of our online procedure, after the algorithm decides on an allocation $\bm x$ of item $t$. Specifically, we compare the hybrid program  $P_{t-1}^*(\bm W;\bm v_{t:T})$ with $P_t^*(\bm W+\bm y;\bm v_{t+1:T})$, where $\bm y \in \mathbb{R}_+^n$ is an arbitrary endowment with $\|\bm y\|_1\leq\vmax$ . We fix any $\bm W\in[0,(t-1)\vmax]^n$, constant bounds $0<\ulower<\uupper\leq\vmax$, use the same value $\kappa=\max_{r,s\in[n]}\max_{\bm u\in[\ulower,\uupper]^n}\partial_r f(\bm u)/\partial_s f(\bm u)$.

\begin{lemma}\textnormal{(Stability after fixing an allocation)}
\label{lemma: fixed-allocation}
If the optimizers of $P_{t-1}^*(\bm W;\bm v_{t:T})$ and $P_t^*(\bm W+{\bm y};\bm v_{t+1:T})$ both lie in $[\ulower,\uupper]^n$, then $\|\bm u_{t-1}^*(\bm W;\bm v_{t:T})-\bm u_t^*(\bm W+{\bm y};\bm v_{t+1:T})\|_\infty\leq\kappa\vmax/T$.
\end{lemma}

\paragraph{Proof.}
{Consider the hybrid program with existing utility $\bm W+\bm y$ and tail $\bm v_{t:T}$, which carries both the endowment and the item. It is obtained from $P_{t-1}^*(\bm W;\bm v_{t:T})$ by adding the endowment $\bm y$, of total raw magnitude $\|\bm y\|_1\leq\vmax$, and from $P_t^*(\bm W+\bm y;\bm v_{t+1:T})$ by adding the item $\bm v_t$, of raw magnitude at most $\vmax$.} Each of the two programs therefore satisfies the hypothesis of \Cref{lem: stability} as the smaller instance, since its optimizer is strictly positive.

Next, we claim that the multiplier ratio that \Cref{lem: stability} attaches to the smaller instance is at most $\kappa$. To see this, notice that by \Cref{proposition: kkt} we have $\beta_i^*=\partial_i\log f(\bm u^*)=\partial_i f(\bm u^*)/f(\bm u^*)$, so $\max_{j\in[n]}\beta_j^*/\beta_i^*=\max_{j\in[n]}\partial_j f(\bm u^*)/\partial_i f(\bm u^*)$. Applying \Cref{lem: stability} from each of the two programs to the enlarged one gives, for every $i\in[n]$, that the $i$th coordinate of the enlarged program's optimizer exceeds $u_{t-1,i}^*(\bm W;\bm v_{t:T})$ by at least $0$ and at most $\kappa\vmax/T$, and exceeds $u_{t,i}^*(\bm W+{\bm y};\bm v_{t+1:T})$ by at least $0$ and at most $\kappa\vmax/T$. Subtracting these two estimates coordinate-wise bounds $|u_{t-1,i}^*(\bm W;\bm v_{t:T})-u_{t,i}^*(\bm W+{\bm y};\bm v_{t+1:T})|$ by $\kappa\vmax/T$, and taking the maximum over $i\in[n]$ proves the claim.

$\hfill\square$

\begin{lemma}\textnormal{(Boundedness after fixing an allocation)}
\label{lemma: one-sided-bound}
For any $a \in [0, \vmax/2]$, if $\bm u_{t-1}^*(\bm W;\bm v_{t:T})\geq 2a$ coordinate-wise and $T \geq \vmax/a$, then the optimizers of $P_{t-1}^*(\bm W;(\bm v_t',\bm v_{t+1:T}))$ and $P_t^*(\bm W+{\bm y};\bm v_{t+1:T})$ lies in the bounded area $[a,\vmax]^n$.
\end{lemma}

\paragraph{Proof.}
We first consider removing the item $\bm v_t$ from the tail, and thus the hybrid program becomes $P_t^*(\bm W;\bm v_{t+1:T})$. For the same optimal allocation of $P_{t-1}^*(\bm W;\bm v_{t:T})$ on the rest of the items, the resulting time-averaged utility decreases by at most $\vmax/T$ for an agent. Therefore, $P_t^*(\bm W;\bm v_{t+1:T})$ has a feasible utility solution with every coordinate at least $2a-\vmax/T\geq a>0$.

Let $k\in[n]$ be the minimum-utility agent in the program $P_{t-1}^*(\bm W;(\bm v_t',\bm v_{t+1:T}))$, i.e., $k \in \arg \min_{i\in [n]} u_{t,i}^*(\bm W;\bm v_{t+1:T})$. 
Since $\partial_i \log f(\bm u)$ is decreasing in $u_i$, by \Cref{proposition: kkt}, $k$ also maximizes $\beta_k^*(\bm W;\bm v_{t+1:T})$. Therefore, the multiplier ratio in \Cref{lem: stability} equals $1$ at the coordinate $k$. Since $P_{t-1}^*(\bm W;\bm v_{t:T})$ is obtained from $P_t^*(\bm W;\bm v_{t+1:T})$ by adding back the single item $\bm v_t$, applying \Cref{lem: stability} gives us $$0\leq u_{t-1,k}^*(\bm W;\bm v_{t:T})-u_{t,k}^*(\bm W;\bm v_{t+1:T})\leq\vmax/T.$$ Therefore, $u_{t,k}^*(\bm W;\bm v_{t+1:T})\geq2a-\vmax/T\geq a$. Since we choose $k$ to be the minimum-utility agent, we then have that every coordinate of $\bm u_t^*(\bm W;\bm v_{t+1:T})$ is at least $a$.

Finally, notice that each of the two modified programs can be obtained from $P_t^*(\bm W;\bm v_{t+1:T})$ by adding an item. By the monotonicity result in \Cref{lem: monotonicity}, we prove the lower bound. The upper bound simply follows from the boundedness of value vectors.

$\hfill\square$

\subsection{Proof of \Cref{lem: uc}}
We first state \Cref{lem: uc} formally.
\appendixlemmatrue
\uc*

\paragraph{Proof.}
We denote \(\mathcal V_t=[0,t\vmax]^n\) for simplicity. We will apply concentration analysis to a globally Lipschitz extension of the optimal utility function; the extension exactly matches the optimal utility when the optimal utility is in $[\ulower , \uupper]^n$. The proof of \Cref{lemma: concentration-single} is deferred to \Cref{proof: concentration-single}.

\begin{restatable}{lemma}{nonuc}[Lipschitz extension of the bounded optimizer]
\label{lemma: concentration-single}
For every \(t\in\{0,\ldots,T\}\) and \(i\in[n]\) there exists a Borel function \(\widehat u_{t,i}\) on \(\mathcal V_t\times([0,\vmax]^n)^{T-t}\) that agrees with \(u_{t,i}^*\) when \( u_{t,i}^* \in [\ulower,\uupper]^n\), and satisfies, throughout its domain, \(|\widehat u_{t,i}(\bm W;\bm v_{t+1:T})-\widehat u_{t,i}(\bm W';\bm v'_{t+1:T})|
\leq (\kappa/T)\cdot({\|\bm W-\bm W'\|_1}+{\vmax\cdot|\{\tau>t:\bm v_\tau\neq\bm v'_\tau\}|})\).
\end{restatable}

We then prove \Cref{lem: uc} with \(m_{t,i}(\bm W;\mathcal P_{t+1:T})=\mathbb E_{\bm V_{t+1:T}\sim\mathcal P_{t+1:T}}[\widehat u_{t,i}(\bm W;\bm V_{t+1:T})]\). By taking expectations on the Lipschitz bound of \Cref{lemma: concentration-single}, we have \(|m_{t,i}(\bm W;\mathcal P_{t+1:T})-m_{t,i}(\bm W';\mathcal P_{t+1:T})|\leq\kappa\|\bm W-\bm W'\|_1/T\). Notice that for \(t<T\) the tail consists of \(T-t\) independent item vectors, and changing one of them moves \(\widehat u_{t,i}\) by at most \(\kappa\vmax/T\). Applying McDiarmid's inequality, for every fixed \(\bm W\in\mathcal V_t\) and every \(\varepsilon>0\),

\begin{equation}
\label{eq: concentration-single-correct}
 \Pr\left(
 \left|\widehat u_{t,i}(\bm W;\bm V_{t+1:T})
 -m_{t,i}(\bm W;\mathcal P_{t+1:T})\right|>\varepsilon
 \right)
 \leq
 2\exp\left(-\frac{2\varepsilon^2T^2}
 {\kappa^2\vmax^2(T-t)}\right).
\end{equation}

To get a uniform convergence result over all $\bm W \in \mathcal{V}_t$, we consider the lattice discretization of the space $\mathcal V_t$ as follows, 
\begin{equation*}
    \mathcal V_t^{\discretized}
    =\{\bm W\in\mathbb R_+^n:W_i=k_i\vmax,\ k_i\in\{0,\ldots,t\}\}.
\end{equation*}

We set the target error threshold as $ a_t=\frac{\kappa\vmax}{T}
    \sqrt{\frac{T-t}{2}
    \left(n\log(t+1)+\log\frac{2nT}{\eta}\right)}.$ By \eqref{eq: concentration-single-correct} and union bounds over the lattice points and $i\in [n]$, 
\begin{align}
 &\Pr\left(
 \exists i\in[n],\ \bm W^{\discretized}\in\mathcal V_t^{\discretized}:
 \left|\widehat u_{t,i}(\bm W^{\discretized};\bm V_{t+1:T})
 -m_{t,i}(\bm W^{\discretized};\mathcal P_{t+1:T})\right|>a_t
 \right)\nonumber\\
 &\qquad\leq
 2n(t+1)^n
 \exp\left(-\frac{2a_t^2T^2}
 {\kappa^2\vmax^2(T-t)}\right)
 =\frac{\eta}{T}.
\label{eq: uc-discretized}
\end{align}

For an arbitrary \(\bm W\in\mathcal V_t\), let
\(\bm W^{\discretized}\) be a coordinate-wise nearest lattice point.  Then
\(\|\bm W-\bm W^{\discretized}\|_1\leq n\vmax/2\).  On the complement of the event in
\eqref{eq: uc-discretized}, whenever
\(\bm u_t^*(\bm W;\bm V_{t+1:T})\in[\ulower,\uupper]^n\), by \Cref{lemma: concentration-single} and the Lipschitzness of $m_{t,i}$, we have
\begin{align*}
 &\left|u_{t,i}^*(\bm W;\bm V_{t+1:T})
 -m_{t,i}(\bm W;\mathcal P_{t+1:T})\right|\\
 &\quad\leq
 \left|\widehat u_{t,i}(\bm W;\bm V_{t+1:T})
 -\widehat u_{t,i}(\bm W^{\discretized};\bm V_{t+1:T})\right|+
 \left|\widehat u_{t,i}(\bm W^{\discretized};\bm V_{t+1:T})
 -m_{t,i}(\bm W^{\discretized};\mathcal P_{t+1:T})\right|\\
 &\qquad+
 \left|m_{t,i}(\bm W^{\discretized};\mathcal P_{t+1:T})
 -m_{t,i}(\bm W;\mathcal P_{t+1:T})\right|\\
 &\quad\leq a_t+\frac{n\kappa\vmax}{T}.
\end{align*}
Taking a union bound over \(t=0,\ldots,T-1\), and observing that the assertion for
\(t=T\) is deterministic, proves the simultaneous claim with probability at least
\(1-\eta\).  The eventual error bound is then 
\[
 a_t+\frac{n\kappa\vmax}{T}
 \leq\kappa\vmax\left(
 \sqrt{\frac{n\log(T+1)+\log(2nT/\eta)}{2T}}
 +\frac{n}{T}\right)=\varepsilon_0.
\]

This proves \Cref{lem: uc}.

\(\hfill\square\)

\subsection{Proof of \Cref{lemma: concentration-single}}

\nonuc*
\label{proof: concentration-single}
\paragraph{Proof.}
Fix \(t\in\{0,\ldots,T\}\) and \(i\in[n]\), with \(\mathcal V_t\), \(\kappa\), \(\ulower\) and \(\uupper\) as in the proof of \Cref{lem: uc}.  Equip \(\mathcal V_t\times([0,\vmax]^n)^{T-t}\) with the metric \(d((\bm W,\bm z),(\bm W',\bm z'))=\|\bm W-\bm W'\|_1+\vmax\cdot|\{\tau>t:\bm z_\tau\neq\bm z'_\tau\}|\),
and let \(\mathcal G_t\) be the set of pairs \((\bm W,\bm z)\) for which the
hybrid program admits a strictly positive feasible utility and its optimizer
lies in \([\ulower,\uupper]^n\).

We first show that the function
\((\bm W,\bm z)\mapsto u_{t,i}^*(\bm W;\bm z)\), restricted to
\(\mathcal G_t\), is \(\kappa/T\)-Lipschitz in the metric \(d\).  Take two
points \((\bm W,\bm z)\) and \((\bm W',\bm z')\) of \(\mathcal G_t\), and let \(J\) be the set of all indices in the tail where
their item vectors differ. We then construct a common enlarged hybrid instance with
existing utility \(\bm W\vee\bm W'\), one copy of every common tail item, and
both copies of every item in \(J\).  Relative to either original program, the
total magnitude of the added one-hot endowments and items is at most
\(\|\bm W-\bm W'\|_1+\vmax|J|=d((\bm W,\bm z),(\bm W',\bm z'))\).

Applying the stability result in \Cref{lem: stability}, we have
\begin{equation*}
    |u_{t,i}^*(\bm W;\bm z)-u_{t,i}^*(\bm W';\bm z')|\leq(\kappa/T)\cdot d((\bm W,\bm z),(\bm W',\bm z')).
\end{equation*}

The feasible utility correspondence has a compact value space, and is continuous in
\((\bm W,\bm z)\).  By the uniqueness of the utility maximizer and Berge's maximum theorem, the utility continuous on \(\mathcal G_t\).  Moreover,
\(\mathcal G_t\) is compact. To see compactness, notice that along a convergent sequence in this set, the lower
bound \(\ulower\) keeps the limiting optimizer strictly positive. Therefore, \(d\) is lower semicontinuous.  Consequently the infimum below, taken over a
compact set, is lower semicontinuous and hence Borel measurable.

If \(\mathcal G_t\neq\varnothing\), take the McShane extension of this
restricted function and project it to \([0,\vmax]\):
\begin{align*}
 \widehat u_{t,i}(\bm W;\bm z)
 =\min\bigg\{\vmax,\max\bigg\{0,\ 
 \inf_{(\bm W',\bm z')\in\mathcal G_t}
 \left[u_{t,i}^*(\bm W';\bm z')
 +\frac{\kappa}{T}d((\bm W,\bm z),(\bm W',\bm z'))\right]
 \bigg\}\bigg\}.
\end{align*}
If \(\mathcal G_t=\varnothing\), set \(\widehat u_{t,i}=0\).  By the Lipschitz bound on \(\mathcal G_t\), the McShane extension matches 
\(u_{t,i}^*\) and is globally \(\kappa/T\)-Lipschitz in \(d\); since
\(u_{t,i}^*\in[0,\vmax]\) on \(\mathcal G_t\) and the projection onto
\([0,\vmax]\) is \(1\)-Lipschitz, we have the desired properties.

\(\hfill\square\)

\subsection{Proof of \Cref{lem: safe-volume}}
\safety*
\paragraph{Proof.}
The result is trivial for $n=1$. We next consider $n \geq 2$. Let
\(\bm v\sim\mathrm{Uniform}[0,\vmax]^n\).  Since \(\beta_i\) is strictly positive, the
maximizer of \(\beta_i v_i\) is unique with probability $1$.  For a fixed realization $\bm v$, if \( v\notin \mathcal S(\bm\beta;\iota)\), then there exist a
\(\bm\beta'\in\mathcal B_\infty(\bm\beta;\iota)\cap\mathbb R_+^n\) and an index
\(j\neq i\) that maximizes \(\beta'_r v_r\).  Hence
\begin{align}
 0\leq\beta_i v_i-\beta_jv_j
 &= (\beta_i-\beta'_i)v_i
   +(\beta'_iv_i-\beta'_jv_j)
   +(\beta'_j-\beta_j)v_j
 \leq 2\vmax\iota.
\label{eq: safe-volume-gap}
\end{align}

For each \(i\), let \(Y_{-i}=\max_{j\neq i}\beta_jv_j\) be the second highest multiplied value. By \eqref{eq: safe-volume-gap} and a union bound, we have
\begin{align*}
 \Pr(\bm v \notin\mathcal S(\bm\beta;\iota))
 &\leq\sum_{i=1}^n
 \Pr\left(Y_{-i}\leq\beta_iv_i\leq Y_{-i}+2\vmax\iota\right).
\end{align*}
Notice that the variable \(v_i\) is uniform on
\([0,\vmax]\) after conditioning on \(Y_{-i}\). Therefore,
\begin{align*}
 \Pr(\bm v\notin\mathcal S(\bm\beta;\iota))
 &\leq\sum_{i=1}^n\min\left\{1,\frac{2\iota}{\beta_i}\right\}
 \leq\frac{2n\iota}{\betalower}.
\end{align*}
Multiplying the complementary probability by the volume \(\vmax^n\), we get the desired volume bound. 

\(\hfill\square\)
\subsection{Proof of \Cref{thm: adaptive-with-density}}
\densityA*
\paragraph{Proof.}
Let \(a=\ulower^*/4\), and set \(\eta=1/(4T)\). We will apply \Cref{lem: uc} with the bounds $\ulower = a$ and $\uupper = \vmax$; let $\kappa$ and $\varepsilon_0$ be specified as in \Cref{lem: uc}. For \(s\in\{0,\ldots,T\}\), define the
auxiliary historical-tail program and its solutions by
\begin{equation}
\label{eq: theorem-density-auxiliary}
 P_s' =P_s^*(\bm W_s^\dagger;\bm v_{s+1:T}^\historic),
 \qquad
 (\bm u_s',\bm\beta_s')
 =(\bm u_s^*(\bm W_s^\dagger;\bm v_{s+1:T}^\historic),
   \bm\beta_s^*(\bm W_s^\dagger;\bm v_{s+1:T}^\historic)).
\end{equation}
Thus \(P_0'\) is the hindsight program on the historical instance.  At round \(t\), the
re-solving program differs from \(P_{t-1}'\) only by replacing
\(\bm v_t^\historic\) with \(\bm v_t^\actual\), and \(P_t'\) is obtained after
fixing the allocation selected at round \(t\).

\paragraph{Local constants.}
On the compact box \([a,\vmax]^n\), there are constants \(b,L,\sigma>0\),
independent of \(n,T,\rho\), such that
\begin{equation}
\label{eq: theorem-density-local-constants}
 \min_{i\in[n]}\partial_i\log f(\bm u)\geq\frac{b}{n},\qquad
 \|\nabla\log f(\bm u)-\nabla\log f(\bm z)\|_\infty
 \leq\frac{L}{n}\|\bm u-\bm z\|_\infty,
\end{equation}
and \(\log f\) is \(\sigma/n\)-strongly concave with respect to
\(\|\cdot\|_\infty\).  We next prove this rigorously. 

First consider the
case \(p\neq0\).  Since every coordinate of \(\bm u\) lies in \([a,\vmax]\), we
have \(\min\{a^q,\vmax^q\}\leq u_j^q\leq\max\{a^q,\vmax^q\}\) for each of the three
exponents \(q\in\{p,p-1,p-2\}\). Sum the case \(q=p\) over \(j\), 

\begin{equation*}
    n \min \{a^p,\vmax^p\} \leq \sum_{j=1}^nu_j^p \leq n \max \{a^p,\vmax^p\}.
\end{equation*}

We differentiate the generalized-mean objective and
obtain \(\partial_i\log f(\bm u)=u_i^{p-1}\big/\sum_{j=1}^nu_j^p\), which is at
least \(b/n\) for \(b=\min\{a^{p-1},\vmax^{p-1}\}/\max\{a^p,\vmax^p\}\); this is
the first inequality in \eqref{eq: theorem-density-local-constants}.  We
differentiate once more and bound each row of the Hessian by
\begin{equation}
 \sum_{j=1}^n\left|\partial_j\partial_i\log f(\bm u)\right|
 \leq \frac{|p-1|\,u_i^{p-2}}{\sum_{j=1}^nu_j^p}
 +\frac{|p|\,u_i^{p-1}}{\left(\sum_{j=1}^nu_j^p\right)^2}\sum_{j=1}^n u_j^{p-1}
 \leq\frac{L}{n},
\label{eq: theorem-density-Hessian-row}
\end{equation}
where \(L\) depends only on \(a,\vmax\) and \(p\).  We then apply the mean-value
theorem together with \eqref{eq: theorem-density-Hessian-row} and obtain the
second inequality in \eqref{eq: theorem-density-local-constants}.

For the curvature bound we compute, for every \(\bm z\in\mathbb R^n\),
\begin{equation}
 \bm z^\top\nabla^2\log f(\bm u)\bm z
 =\frac{p-1}{\sum_{j=1}^nu_j^p}\sum_{i=1}^n u_i^{p-2}z_i^2
 -\frac{p}{\left(\sum_{j=1}^nu_j^p\right)^2}
 \left(\sum_{i=1}^n u_i^{p-1}z_i\right)^2.
\label{eq: theorem-density-Hessian-quadratic}
\end{equation}
If \(0<p<1\), we discard the second term of
\eqref{eq: theorem-density-Hessian-quadratic}, which is nonpositive, and obtain
\(-\bm z^\top\nabla^2\log f(\bm u)\bm z\geq(1-p)\sum_iu_i^{p-2}z_i^2/\sum_ju_j^p\).
If \(p<0\), we apply Cauchy--Schwarz 
and obtain
\(\left(\sum_iu_i^{p-1}z_i\right)^2\leq\sum_ju_j^p\cdot\sum_iu_i^{p-2}z_i^2\).
We multiply this inequality by \(p<0\) and substitute it into the negative of
\eqref{eq: theorem-density-Hessian-quadratic}, which gives
\begin{equation*}
 -\bm z^\top\nabla^2\log f(\bm u)\bm z
 \geq\frac{1-p}{\sum_{j=1}^nu_j^p}\sum_{i=1}^n u_i^{p-2}z_i^2
 +\frac{p}{\sum_{j=1}^nu_j^p}\sum_{i=1}^n u_i^{p-2}z_i^2
 =\frac{\sum_{i=1}^n u_i^{p-2}z_i^2}{\sum_{j=1}^nu_j^p}.
\end{equation*}
In either case we bound \(\sum_iu_i^{p-2}z_i^2\) from below by
\(\min\{a^{p-2},\vmax^{p-2}\}\|\bm z\|_2^2\) and \(\sum_ju_j^p\) from above, and
we use \(\|\bm z\|_2\geq\|\bm z\|_\infty\) to reach the asserted
\(\sigma/n\)-strong concavity.  

For \(p=0\), we have
\(\partial_i\log f(\bm u)=1/(nu_i)\) and
\(\bm z^\top\nabla^2\log f(\bm u)\bm z=-(1/n)\sum_i z_i^2/u_i^2\), and the same
three conclusions follow.

We set \(\iota=L(2\varepsilon_0+\kappa\vmax/T)/n\) and prove the result first
under the conditions \(2\varepsilon_0+2\kappa\vmax/T\leq a\) and
\(4n\vmax\iota\leq\sigma a^2\).  The remaining case is treated at the end of
the proof.

\paragraph{Uniform convergence and boundedness.}
We apply \Cref{lem: uc} separately to the historical the online sequences.  Since
\(\mathcal P_t^\historic=\mathcal P_t^\actual\), the two hybrid programs converge to the same center.  Let \(\mathcal E_{\mathrm{uc}}\) be the intersection of
the two events on which the conclusion of \Cref{lem: uc} holds.  Each of these
two events fails with probability at most \(\eta\), so a union bound gives
\(\Pr(\mathcal E_{\mathrm{uc}})\geq1-1/(2T)\).  On \(\mathcal E_{\mathrm{uc}}\),
each of the two optimizers stays within \(\varepsilon_0\) of the common
center, so we apply the triangle inequality and obtain, for every
\(s\in\{0,\ldots,T\}\),
\begin{equation}
\label{eq: theorem-density-two-sample}
 \bm u_s',\bm u_s^\coupling\in[a,\vmax]^n
 \implies
 \|\bm u_s'-\bm u_s^\coupling\|_\infty\leq2\varepsilon_0.
\end{equation}
Here, we note that \Cref{lem: uc} is a uniform convergence result over the existing utility space,
so we may evaluate it at the random \(\bm W_s^\dagger\).

We now show by induction that all auxiliary, re-solving, and coupling solutions
that occur in the argument lie in \([a,\vmax]^n\).  At \(s=0\), the claim follows
from \Cref{assumption: lower-bound}, since both hindsight utility vectors have
every coordinate at least \(\ulower^*=4a\).  Suppose the claim has been verified
through round \(t-1\).  The coupling programs are successive restrictions of the
full online hindsight program.  We apply the dual comparison proved below to the
rounds already verified, and obtain
\begin{equation}
\label{eq: theorem-density-coarse-gap}
 0\leq D_0^\coupling-D_{t-1}^\coupling
 \leq\frac{2\vmax L}{n}\left(2\varepsilon_0+\frac{\kappa\vmax}{T}\right)
 =2\vmax\iota.
\end{equation}
The vector \(\bm u_{t-1}^\coupling\) is feasible for the full online program, so
we combine the first-order optimality condition at \(\bm u_0^\coupling\) with the
strong concavity in \eqref{eq: theorem-density-local-constants} and obtain
\begin{equation}
\label{eq: theorem-density-coarse-distance}
 \|\bm u_{t-1}^\coupling-\bm u_0^\coupling\|_\infty
 \leq
 \sqrt{\frac{2n}{\sigma}
 (D_0^\coupling-D_{t-1}^\coupling)}
 \leq\sqrt{\frac{4n\vmax\iota}{\sigma}}
 \leq a.
\end{equation}
Since every coordinate of \(\bm u_0^\coupling\) is at least \(4a\), we conclude
that \(\bm u_{t-1}^\coupling\geq3a\bm1\).  We then apply
\eqref{eq: theorem-density-two-sample} at \(s=t-1\) and use \(2\varepsilon_0\leq
a\), which gives \(\bm u_{t-1}'\geq2a\bm1\).
{The re-solving program \(P_t^\resolving\) and the auxiliary program \(P_t'\) are
obtained from \(P_{t-1}'\) by replacing the item \(\bm v_t^\historic\) with
\(\bm v_t^\online\) and by adding the endowment \(\bm v_t^\online\cdot\bm x_t^\dagger\)
after deleting that item, respectively, and \(P_t^\coupling\) is obtained from
\(P_{t-1}^\coupling\) by deleting \(\bm v_t^\online\) and adding the same
endowment, whose mass is at most \(\vmax\).  Since \(\kappa\geq1\), we have
\(\vmax/T\leq\kappa\vmax/T\leq a\), and
\Cref{lemma: one-sided-bound} applied at \(\bm u_{t-1}'\geq2a\bm1\) and at
\(\bm u_{t-1}^\coupling\geq3a\bm1\) shows, before any conditional concentration at
round \(t\), that \(\bm u_t^\resolving,\bm u_t',\bm u_t^\coupling\in[a,\vmax]^n\).
With these bounds in place, \Cref{lemma: concentration-single} applied to the
replaced item and \Cref{lemma: fixed-allocation} applied to the fixed allocation
give \(\|\bm u_t^\resolving-\bm u_{t-1}'\|_\infty\leq\kappa\vmax/T\) and
\(\|\bm u_t'-\bm u_t^\resolving\|_\infty\leq\kappa\vmax/T\).}
We apply \eqref{eq: theorem-density-two-sample} at \(s=t\), add the bound on
\(\|\bm u_t'-\bm u_t^\resolving\|_\infty\) obtained above. This gives, 
\begin{equation}
\label{eq: theorem-density-dual-primal-comparison}
 \|\bm u_t^\coupling-\bm u_t^\resolving\|_\infty
 \leq2\varepsilon_0+\frac{\kappa\vmax}{T}.
\end{equation}
The bounds \eqref{eq: theorem-density-coarse-gap}--\eqref{eq: theorem-density-dual-primal-comparison}
close the induction.  We apply the Lipschitz bound in
\eqref{eq: theorem-density-local-constants} to the corresponding pairs of primal
optimizers and obtain
\begin{equation}
\label{eq: theorem-density-dual-comparisons}
 \|\bm\beta_{t-1}^\coupling-\bm\beta_{t-1}'\|_\infty\leq\iota,\qquad
 \|\bm\beta_t^\dagger-\bm\beta_{t-1}'\|_\infty\leq\iota,\qquad
 \|\bm\beta_t^\coupling-\bm\beta_t^\dagger\|_\infty\leq\iota.
\end{equation}

We now prove the dual comparison used in
\eqref{eq: theorem-density-coarse-gap}.  Let
\(\bm y_t=\bm v_t^\actual\cdot\bm x_t^\dagger
=\bm W_t^\dagger-\bm W_{t-1}^\dagger\).  We evaluate \(D_{t-1}^\coupling\) at
\(\bm\beta_t^\coupling\), the optimal multiplier of \(D_t^\coupling\), subtract
the two dual objectives, and obtain
\begin{align}
 0\leq D_{t-1}^\coupling-D_t^\coupling
 &\leq\frac1T\left(
 \max_i\beta_{t,i}^\coupling v_{t,i}^\actual
 -\langle\bm\beta_t^\coupling,\bm y_t\rangle\right)\nonumber\\
 &\leq\frac{2\vmax}{T}
 \|\bm\beta_t^\coupling-\bm\beta_t^\dagger\|_\infty
 \leq\frac{2\vmax\iota}{T}.
\label{eq: theorem-density-coupling-increment}
\end{align}
The second inequality uses
\(\langle\bm\beta_t^\dagger,\bm y_t\rangle
=\max_i\beta_{t,i}^\dagger v_{t,i}^\actual\), which holds for every allowed
tie-breaking rule.  We sum \eqref{eq: theorem-density-coupling-increment}
over the verified rounds and reach \eqref{eq: theorem-density-coarse-gap}.

\paragraph{Bounding the number of expected disagreements.}

We consider the filtration with the complete history and sequential online values, $ \mathcal F_{t-1}
 =\sigma(\bm v_{1:T}^\historic,\bm v_{1:t-1}^\actual)$.\footnote{In the case where the algorithm uses randomized tie-breaking, this additionally includes the random bits used by the algorithm} The multiplier \(\bm\beta_{t-1}'\) is
\(\mathcal F_{t-1}\)-measurable, while \(\bm v_t^\actual\) is conditionally
independent of \(\mathcal F_{t-1}\).

 {We set
\(X_t=\indi{\bm v_t^\actual\notin\mathcal S(\bm\beta_{t-1}';\iota)}\) when
\(\bm u_{t-1}'\in[a,\vmax]^n\), and \(X_t=0\) otherwise.} We will show that $X_t = 1$ captures a ``safe'' event that implies agreement between the re-solving and the coupling program on the current allocation. 
By \eqref{eq: theorem-density-local-constants}, each coordinate of
\(\bm\beta_{t-1}'\) is at least \(b/n\) whenever \(\bm u_{t-1}'\in[a,\vmax]^n\).
On this event we combine the density bound \(\rho/\vmax^n\) with
\Cref{lem: safe-volume} and obtain
\(\mathbb E[X_t\mid\mathcal F_{t-1}]\leq2\rho n^2\iota/b\).  When
\(\bm u_{t-1}'\notin[a,\vmax]^n\), we have \(X_t=0\) and this conditional
expectation vanishes.  In both cases,
\begin{equation}
\label{eq: unsafe-counts}
 \mathbb E[X_t\mid\mathcal F_{t-1}]
 \leq\frac{2\rho n^2\iota}{b}.
\end{equation}

On \(\mathcal E_{\mathrm{uc}}\) we have \(\bm u_{t-1}'\in[a,\vmax]^n\) for every
round \(t\), so \(X_t\) is the unsafe indicator itself.  Suppose \(X_t=0\).  We
combine the first two bounds in
\eqref{eq: theorem-density-dual-comparisons} with
\Cref{definition: safe-region} and conclude that \(D_{t-1}^\coupling\) and the
algorithm select the same winner for item \(t\).  Conditional on
\(\mathcal F_{t-1}\), ties under \(\bm\beta_{t-1}'\) have probability zero, so
fixing the algorithm's decision does not change the optimizer of the coupling
program, and \(D_{t-1}^\coupling=D_t^\coupling\).  We combine this fact with
\eqref{eq: theorem-density-coupling-increment} and obtain, almost surely,
\begin{align}
 \mathcal R_{\log f}(\mathcal A,\bm v^\actual)
 &=D_0^\coupling-D_T^\coupling
 =\sum_{t=1}^T(D_{t-1}^\coupling-D_t^\coupling)\nonumber\\
 &\leq\frac{2\vmax\iota}{T}\sum_{t=1}^T X_t
 \qquad\text{on }\mathcal E_{\mathrm{uc}}.
\label{eq: theorem-density-log-regret}
\end{align}

On the complement of \(\mathcal E_{\mathrm{uc}}\), the welfare regret is at most
\(f(\vmax\bm1)\).  On \(\mathcal E_{\mathrm{uc}}\), we apply the regret
conversion of \Cref{proposition: regret-to-ratio}, then the telescoping bound
\eqref{eq: theorem-density-log-regret}, and then the conditional estimate
\eqref{eq: unsafe-counts}, and obtain
\begin{align}
 \mathbb E[\mathcal R_f(\mathcal A,\bm v^\actual)]
 &\leq f(\vmax\bm1)\Pr(\mathcal E_{\mathrm{uc}}^c)
 +\frac{2f(\vmax\bm1)\vmax\iota}{T}\sum_{t=1}^T\mathbb E[X_t]\nonumber\\
 &\leq\frac{f(\vmax\bm1)}{2T}
 +\frac{4f(\vmax\bm1)\vmax\rho n^2}{b}\,\iota^2\nonumber\\
 &=\frac{f(\vmax\bm1)}{2T}
 +\frac{4f(\vmax\bm1)\vmax\rho L^2}{b}
 \left(2\varepsilon_0+\frac{\kappa\vmax}{T}\right)^2.
\label{eq: theorem-density-final-bound}
\end{align}

For \(T\geq2\), we use \(\eta=1/(4T)\) in the definition of
\(\varepsilon_0\) (see \Cref{lem: uc}) and obtain
\(2\varepsilon_0+\kappa\vmax/T\leq c_\varepsilon\sqrt{n\log T/T}\), where
\(c_\varepsilon\) depends only on \(f,\ulower^*\) and \(\vmax\).  If the previous two conditions \(2\varepsilon_0+2\kappa\vmax/T\leq a\) and
\(4n\vmax\iota\leq\sigma a^2\) hold, we substitute this estimate into
\eqref{eq: theorem-density-final-bound} and use \(1/T\leq n\log T/T\), which
gives \(\mathbb E[\mathcal R_f(\mathcal A,\bm v^\actual)]\leq C\rho n\log T/T\).

It remains to treat the case in which one of the two conditions fails.  The
constant \(c_0=\min\{a/2,\;a^2\sigma/(4\vmax L)\}\) is positive, and
\(2\varepsilon_0+\kappa\vmax/T\leq c_0\) implies both conditions.  If one of them
fails, then \(2\varepsilon_0+\kappa\vmax/T>c_0\), and the preceding estimate
gives \(n\log T/T>c_0^2/c_\varepsilon^2\).  Since \(\rho\geq1\), the quantity
\(\rho n\log T/T\) is bounded below by the positive constant
\(c_0^2/c_\varepsilon^2\), which is independent of \(\rho,n\) and \(T\), so we
absorb the bound \(\mathcal R_f\leq f(\vmax\bm1)\) by increasing \(C\).  This
proves the theorem.

\(\hfill\square\)
\section{Missing Proofs in \Cref{sec: robust}}
\label{proof: robust}

\subsection{The Wasserstein Coupling and the Drift Term}

\paragraph{Proof of \Cref{proposition: wasserstein}.}
For each \(t\in[T]\), since the space \([0,\vmax]^n\) is compact and the $\ell_1$ transportation cost is continuous, there exists an optimal coupling
\(\Gamma_t^*\) of \(\mathcal P_t^\actual\) and
\(\mathcal P_t^\historic\) that minimizes the \(\ell_1\) transportation. Take the product coupling
\(\Gamma^*=\bigotimes_{t=1}^T\Gamma_t^*\).  The two marginal sequence distributions are 
\(\mathcal P^\actual\) and \(\mathcal P^\historic\), respectively. By the definition of the Wasserstein discrepancy, 
\begin{align*}
 \mathbb E_{\Gamma^*}[\delta_{1:T}]
 &=\frac1T\sum_{t=1}^T
   \mathbb E_{\Gamma_t^*}
   [\|\bm v_t^\coupling-\bm v_t^\actual\|_1]\\
 &=\frac1T\sum_{t=1}^T
   \mathcal W_1(\mathcal P_t^\actual,\mathcal P_t^\historic)
 =\mathcal W.
\end{align*}
This proves \Cref{proposition: wasserstein}.

\(\hfill\square\)

\begin{restatable}{lemma}{sensitivity}
\label{lemma: r2-and-r3}
If
\(\delta_{1:T}=T^{-1}\sum_{t=1}^T
\|\bm v_t^\coupling-\bm v_t^\actual\|_1
\leq\bar\delta<\ulower^*\), then under
\Cref{assumption: lower-bound} there is a constant
\(B_{\bar\delta}>0\), independent of \(n,T\), such that
\begin{equation}
\label{eq: sensitivity-statement}
 P^*(\mathbf v^\actual)-P^*(\mathbf v^\coupling)
 \leq\frac{B_{\bar\delta}}{n}\,\delta_{1:T}.
\end{equation}
The constant depends only on the fixed welfare parameter and on
\(\ulower^*,\vmax,\bar\delta\).
\end{restatable}

\paragraph{Proof.}
Consider applying a hindsight optimal allocation for \(P^*(\mathbf v^\actual)\) to the coupling sequence
\(\mathbf v^\coupling\), and denote the resulting time-averaged utility by
\(\bm y\).  Applying the same allocation to the two sequences gives
\begin{align}
 \|\bm u^*(\mathbf v^\actual)-\bm y\|_1
 &\leq\frac1T\sum_{t=1}^T\sum_{i=1}^n
 |v_{t,i}^\actual-v_{t,i}^\coupling|x_{t,i}^*
 \leq\delta_{1:T}.
\label{eq: sensitivity-utility}
\end{align}
In particular, the segment joining these two utility vectors lies in
\([\ulower^*-\bar\delta,\vmax]^n\). 

The subgradient of the logarithmic objective can be computed as follows, 
\begin{equation*}
     \partial_i\log f(\bm u)
 = \begin{cases}\frac{u_i^{p-1}}{\sum_{j=1}^n u_j^p}, & (p\neq0)\\
    1/(nu_i), & (p=0)
 \end{cases}
\end{equation*}

Since every coordinate is in a
fixed compact subinterval of \((0,\infty)\), there is a constant
\(B_{\bar\delta}\), independent of \(n,T\), for which
\[
 \max_{i\in[n]}\partial_i\log f(\bm u)
 \leq\frac{B_{\bar\delta}}n
 \qquad
 \text{on }[\ulower^*-\bar\delta,\vmax]^n.
\]
Consider the segment joining the utility vectors $\bm u^*(\bf v^\online)$ and $y$. By the fundamental theorem of calculus along the segment and
\eqref{eq: sensitivity-utility}, 
\begin{equation}
\label{eq: sensitivity-objective}
 \left|P^*(\mathbf v^\actual)-\log f(\bm y)\right|
 \leq\frac{B_{\bar\delta}}n
 \|\bm u^*(\mathbf v^\actual)-\bm y\|_1
 \leq\frac{B_{\bar\delta}}n\,\delta_{1:T}.
\end{equation}
The vector \(\bm y\) is feasible for \(P^*(\mathbf v^\coupling)\), so
\(\log f(\bm y)\leq P^*(\mathbf v^\coupling)\).  Combining this fact with
\eqref{eq: sensitivity-objective} proves
\eqref{lemma: r2-and-r3}.

\(\hfill\square\)

\subsection{Proof of \Cref{thm: adaptive}}
\adaptiveA*

\paragraph{Proof.}
We use the coupling sequence specified in \Cref{proposition: wasserstein}, and use \(\Gamma^*\) to generate the pair
\((\mathbf v^\actual,\mathbf v^\coupling)\) jointly.  The observed
historical sequence is generated independently from \(\mathcal P^\historic\).  Moreover,
\(\mathbf v^\coupling\) and \(\mathbf v^\historic\) are independent sequences
with the same historical product law, and by
\Cref{proposition: wasserstein} we have \(\mathbb E[\delta_{1:T}]=\mathcal W\).

We apply
\Cref{lem: uc} on \([a,\vmax]^n\) with failure probability \(\eta=1/(4T)\) and lower bound $a = \ulower^*/4$, and let \(\varepsilon_0\) and \(\kappa\) be given as in \Cref{lem: uc}. For \(s\in\{0,\ldots,T\}\), define the auxiliary historical-tail program by
\begin{equation}
\label{eq: adaptive-auxiliary}
 P_s'=P_s^*(\bm W_s^\dagger;\bm v_{s+1:T}^\historic),
 \qquad
 (\bm u_s',\bm\beta_s')
 =(\bm u_s^*(\bm W_s^\dagger;\bm v_{s+1:T}^\historic),
   \bm\beta_s^*(\bm W_s^\dagger;\bm v_{s+1:T}^\historic)).
\end{equation}

Similar to \eqref{eq: theorem-density-local-constants}, there are constants
\(B,L,\sigma>0\), independent of
\(n,T\), such that, for all \(\bm u,\bm z\) in the box $[a/2, \vmax]^n$,
\begin{align}
 \max_{i\in[n]}\partial_i\log f(\bm u)&\leq\frac Bn,\nonumber\\
 \|\nabla\log f(\bm u)-\nabla\log f(\bm z)\|_\infty
 &\leq\frac Ln\|\bm u-\bm z\|_\infty,\nonumber\\
 \log f(\bm z)&\leq\log f(\bm u)
 +\langle\nabla\log f(\bm u),\bm z-\bm u\rangle
 -\frac{\sigma}{2n}\|\bm z-\bm u\|_\infty^2.
\label{eq: adaptive-local-constants}
\end{align}

We set \(\bar\delta=\min\{a/2,\;\sigma a^2/(48B)\}\) and
\(c_0=\min\{a,\;\sigma a^2/(32\vmax L)\}\), and we first prove the result under
the condition \(2\varepsilon_0+\kappa\vmax/T\leq c_0\); we later show that the condition will hold automatically when $T \geq n c' \log T$.  Since \(\kappa\geq1\)
and \(c_0\leq a\), this condition implies \(2\varepsilon_0\leq a\) and
\(\vmax/T\leq\kappa\vmax/T\leq a\).

We apply \Cref{lem: uc} separately to
\(\mathbf v^\historic\) and to \(\mathbf v^\coupling\).  Their deterministic
centers coincide because their product laws coincide.  Let
\(\mathcal E_{\rm uc}\) be the intersection of the two events on which the
conclusion of \Cref{lem: uc} holds.  Each of these two events fails with
probability at most \(\eta\), so a union bound gives
\(\Pr(\mathcal E_{\rm uc})\geq1-1/(2T)\).  On \(\mathcal E_{\rm uc}\), each of
the two optimizers stays within \(\varepsilon_0\) of the common deterministic center, so we apply the triangle inequality and obtain, for every
\(s\in\{0,\ldots,T\}\),
\begin{equation}
\label{eq: adaptive-two-sample}
 \bm u_s',\bm u_s^\coupling\in[a,\vmax]^n
 \implies
 \|\bm u_s'-\bm u_s^\coupling\|_\infty\leq2\varepsilon_0.
\end{equation}

Applying Markov's inequality to the average $\ell_1$ discrepancy $\delta_{1:T}$, whose expectation is $\mathcal{W}$ by \Cref{proposition: wasserstein},  
\begin{equation*}
    \Pr(\mathcal E^c)\leq1/(2T)+\mathcal W/\bar\delta.
\end{equation*}
Let
\(\mathcal E=\mathcal E_{\rm uc}\cap
\{\delta_{1:T}\leq\bar\delta\}\) be the ``good'' event in this proof. 
We next show inductively that on \(\mathcal E\), every re-solving, auxiliary, and coupling
program used in the argument has its optimizer in \([a,\vmax]^n\).

At \(s=0\), we apply \Cref{assumption: lower-bound} to obtain
\(\bm u_0',\bm u_0^\coupling\in[4a,\vmax]^n\).  Suppose inductively that
\(\bm u_s',\bm u_s^\coupling\in[a,\vmax]^n\) for \(s\leq t-1\), and that
\(\bm u_\tau^\resolving\in[a,\vmax]^n\) together with
\(\|\bm u_\tau^\coupling-\bm u_\tau^\resolving\|_\infty
\leq2\varepsilon_0+\kappa\vmax/T\)
for \(\tau\leq t-1\).
We combine the KKT conditions with \eqref{eq: adaptive-local-constants} and
obtain for $\tau \leq t-1$,
\begin{equation}
\label{eq: adaptive-dual-comparison}
 \|\bm\beta_\tau^\coupling-\bm\beta_\tau^\dagger\|_\infty
 \leq\frac{L}{n}\left(2\varepsilon_0+\frac{\kappa\vmax}{T}\right),
 \qquad
 \|\bm\beta_\tau^\coupling\|_\infty\leq\frac Bn.
\end{equation}

We next apply \Cref{lem: coupling-main} to $\tau \leq t-1$.  Since
\(\|\bm v_\tau^\coupling-\bm v_\tau^\actual\|_\infty
\leq\|\bm v_\tau^\coupling-\bm v_\tau^\actual\|_1\), we sum
\eqref{eq: adaptive-dual-comparison} over \(\tau\leq t-1\) and obtain
\begin{align}
 P^*(\mathbf v^\coupling)-\log f(\bm u_{t-1}^\coupling)
 &=D_0^\coupling-D_{t-1}^\coupling\nonumber\\
 &\leq\frac{B}{nT}\sum_{\tau=1}^{t-1}
 \|\bm v_\tau^\coupling-\bm v_\tau^\actual\|_\infty
 +\frac{2\vmax L(t-1)}{nT}\left(2\varepsilon_0+\frac{\kappa\vmax}{T}\right)
 \nonumber\\
 &\leq\frac{B\delta_{1:T}}{n}
 +\frac{2\vmax L}{n}\left(2\varepsilon_0+\frac{\kappa\vmax}{T}\right).
\label{eq: adaptive-coupling-prefix}
\end{align}
On \(\mathcal E\) we repeat the proof of \Cref{lemma: r2-and-r3} with the first
bound in \eqref{eq: adaptive-local-constants} and obtain
\begin{equation}
\label{eq: adaptive-sensitivity-use}
 P^*(\mathbf v^\actual)-P^*(\mathbf v^\coupling)
 \leq\frac Bn\,\delta_{1:T}.
\end{equation}

We fix an optimal allocation of the \((t-1)\)-st coupling program, keep its
allocations on the first \(t-1\) actual items, and apply its remaining
allocations to \(\bm v_{t:T}^\actual\) instead of
\(\bm v_{t:T}^\coupling\).  The resulting utility \(\bm y_{t-1}\) is feasible
for the full actual hindsight program, and
\begin{align}
 \|\bm u_{t-1}^\coupling-\bm y_{t-1}\|_1
 &\leq\frac1T\sum_{\tau=t}^T\sum_{i=1}^n
 |v_{\tau,i}^\coupling-v_{\tau,i}^\actual|x_{\tau,i}^\coupling
 \leq\delta_{1:T}.
\label{eq: adaptive-transferred-utility}
\end{align}
By the induction hypothesis and
\(\delta_{1:T}\leq\bar\delta\leq a/2\), the entire segment between these
two vectors lies in \([a/2,\vmax]^n\), so we apply the first bound in
\eqref{eq: adaptive-local-constants} and obtain
\(|\log f(\bm u_{t-1}^\coupling)-\log f(\bm y_{t-1})|\leq B\delta_{1:T}/n\).
We combine this inequality with
\eqref{eq: adaptive-coupling-prefix} and
\eqref{eq: adaptive-sensitivity-use}, use the feasibility of
\(\bm y_{t-1}\), and obtain
\begin{equation}
\label{eq: adaptive-actual-gap}
 0\leq P^*(\mathbf v^\actual)-\log f(\bm y_{t-1})
 \leq\frac{3B\delta_{1:T}}{n}
 +\frac{2\vmax L}{n}\left(2\varepsilon_0+\frac{\kappa\vmax}{T}\right).
\end{equation}

Both the actual hindsight optimizer and \(\bm y_{t-1}\) lie in
\([a/2,\vmax]^n\), so we combine the first-order optimality condition at
\(\bm u^*(\mathbf v^\actual)\) with the strong-concavity inequality in
\eqref{eq: adaptive-local-constants} and obtain
\(P^*(\mathbf v^\actual)-\log f(\bm y_{t-1})
\geq\frac{\sigma}{2n}\|\bm y_{t-1}-\bm u^*(\mathbf v^\actual)\|_\infty^2\).
By the choices of \(\bar\delta\) and \(c_0\), we have
\(3B\delta_{1:T}\leq\sigma a^2/16\) and
\(2\vmax L(2\varepsilon_0+\kappa\vmax/T)\leq\sigma a^2/16\).  Therefore,
\begin{align}
 \|\bm u_{t-1}^\coupling-\bm u^*(\mathbf v^\actual)\|_\infty
 &\leq\delta_{1:T}
 +\sqrt{\frac{2}{\sigma}\left(3B\delta_{1:T}
 +2\vmax L\left(2\varepsilon_0+\frac{\kappa\vmax}{T}\right)\right)}
 \leq a.
\label{eq: adaptive-interior}
\end{align}
Since every coordinate of the actual hindsight optimizer is at least \(4a\), we
conclude that \(\bm u_{t-1}^\coupling\geq3a\bm1\).  Both
\(\bm u_{t-1}'\) and \(\bm u_{t-1}^\coupling\) lie in \([a,\vmax]^n\), so we
apply \eqref{eq: adaptive-two-sample} at \(s=t-1\) and use
\(2\varepsilon_0\leq a\) to obtain \(\bm u_{t-1}'\geq2a\bm1\).

{The re-solving program \(P_t^\resolving\) and the
auxiliary program \(P_t'\) are obtained from \(P_{t-1}'\) by replacing the item
\(\bm v_t^\historic\) with \(\bm v_t^\online\) and by adding the endowment
\(\bm v_t^\online\cdot\bm x_t^\dagger\) after deleting that item, respectively, and
\(P_t^\coupling\) is obtained from \(P_{t-1}^\coupling\) by deleting
\(\bm v_t^\coupling\) and adding the same endowment, whose mass is at most
\(\vmax\).  Since \(\vmax/T\leq a\),
\Cref{lemma: one-sided-bound} applied at \(\bm u_{t-1}'\geq2a\bm1\) and at
\(\bm u_{t-1}^\coupling\geq3a\bm1\) proves, without using concentration at \(t\), that}
\begin{equation}
\label{eq: adaptive-current-bounded}
 \bm u_t^\resolving,\ \bm u_t',\ \bm u_t^\coupling
 \in[a,\vmax]^n.
\end{equation}
{\Cref{lemma: fixed-allocation} then gives
\(\|\bm u_t'-\bm u_t^\resolving\|_\infty\leq\kappa\vmax/T\).}
Now that \eqref{eq: adaptive-current-bounded} is available, we apply
\eqref{eq: adaptive-two-sample} at \(s=t\) and obtain
\begin{equation}
\label{eq: adaptive-current-comparison}
 \|\bm u_t^\coupling-\bm u_t^\resolving\|_\infty
 \leq
 \|\bm u_t^\coupling-\bm u_t'\|_\infty
 +\|\bm u_t'-\bm u_t^\resolving\|_\infty
 \leq2\varepsilon_0+\frac{\kappa\vmax}{T}.
\end{equation}
The bounds \eqref{eq: adaptive-current-bounded} and
\eqref{eq: adaptive-current-comparison} close the induction.  In particular,
on \(\mathcal E\) every round-\(t\) re-solving program has an optimal
multiplier, so the exceptional outcomes on which the algorithm has no dual
solution are contained in \(\mathcal E^c\), whose probability is at most
\(1/(2T)+\mathcal W/\bar\delta\).

At \(t=T\), the last coupling program contains exactly the algorithm's realized
utility.  We sum \Cref{lem: coupling-main} over all \(T\) rounds, use the dual
comparison \eqref{eq: adaptive-dual-comparison} and the sensitivity bound
\eqref{eq: adaptive-sensitivity-use}, and obtain the logarithmic regret bound
\begin{align}
 \mathcal R_{\log f}(\mathcal A,\mathbf v^\actual)
 &=P^*(\mathbf v^\actual)-D_T^\coupling\nonumber\\
 &=
 P^*(\mathbf v^\actual)-P^*(\mathbf v^\coupling)
 +\sum_{t=1}^T(D_{t-1}^\coupling-D_t^\coupling)\nonumber\\
 &\leq\frac{2B\delta_{1:T}}{n}
 +\frac{2\vmax L}{n}\left(2\varepsilon_0+\frac{\kappa\vmax}{T}\right)
 \qquad\text{on }\mathcal E.
\label{eq: adaptive-log-regret}
\end{align}
On \(\mathcal E^c\), the regret on generalized-mean welfare objective is at most \(f(\vmax\bm1)\) by boundedness of values.  On
\(\mathcal E\), we apply the regret conversion of
\Cref{proposition: regret-to-ratio} and the logarithmic regret bound
\eqref{eq: adaptive-log-regret}, and we then use the probability bound
\(\Pr(\mathcal E^c)\leq1/(2T)+\mathcal W/\bar\delta\) together with
\(\mathbb E[\delta_{1:T}]=\mathcal W\).  We obtain
\begin{align}
 \mathbb E[\mathcal R_f(\mathcal A,\mathbf v^\actual)]
 &\leq f(\vmax\bm1)\left(\frac1{2T}+\frac{\mathcal W}{\bar\delta}\right)
 +\frac{2f(\vmax\bm1)B}{n}\mathcal W
 +\frac{2f(\vmax\bm1)\vmax L}{n}
 \left(2\varepsilon_0+\frac{\kappa\vmax}{T}\right)\nonumber\\
 &\leq C_4\left(\frac1T+\mathcal W
 +\frac1n\left(2\varepsilon_0+\frac{\kappa\vmax}{T}\right)\right),
\label{eq: adaptive-expected-regret}
\end{align}
where \(C_4\) is independent of \(n,T\) and \(\mathcal W\).

Recall the definition of $\varepsilon_0$. Since $n\leq T$, we obtain 
\begin{align*}
    2\varepsilon_0 + \kappa \vmax / T & \leq \kappa\vmax\left(2\sqrt{\frac{n\log(T+1)+\log(8nT^2)}{2T}}+\frac{3n}{T}\right)\\
    &\leq \kappa\vmax\left(2\sqrt{\frac{8n\log T}{2T}}+\frac{3}{\sqrt{\log2}}\sqrt{\frac{n\log T}{T}}\right)\\
    &= \left(4+\frac{3}{\sqrt{\log2}}\right)\kappa\vmax\sqrt{\frac{n\log T}{T}}
    \leq 8\kappa\vmax\sqrt{\frac{n\log T}{T}},
\end{align*}
We then let $c' = (c_0 / (8\kappa \vmax ))^2$. Since $n\log T / T \leq c'$, we have $2\varepsilon_0+\kappa\vmax/T\leq 8\kappa \vmax \sqrt{c'}=c_0$, so the bound \eqref{eq: adaptive-expected-regret} holds. We substitute the estimate above into \eqref{eq: adaptive-expected-regret} and use $1/T\leq2\sqrt{\log T/(nT)}$, a consequence of $n\leq4T\log T$, to obtain $\mathbb E[\mathcal R_f(\mathcal A,\mathbf v^\actual)]\leq(2+8\kappa \vmax)C_4\sqrt{\log T/(nT)}+C_4\mathcal W$. This proves the theorem with $C_3=(2+8\kappa \vmax)C_4$.

\(\hfill\square\)

\section{Missing Experiment Results}

\label{proof-appendix-experiments}
\subsection{Datasets}
\paragraph{Instagram Notification.} The Instagram notifications dataset~\citep{kroer2023fair} contains $m=41072$ timestamped notification events from user arrivals and $n=4$ notification types (``Comment Subscribed'', ``Feed Suite Organic Campaign'', ``Like'', ``Story Daily Digest''). Each event specifies the utility that a type receives when its notification is pushed to the corresponding user. Following~\citet{kroer2023fair}, we model each user arrival as a unit-capacity impression (at most one notification is sent), and study online fair allocation of these impressions to notification types under the observed arrival order.

\paragraph{Recommendation with MovieLens Ratings.} We use the MovieLens ratings data with $m=610$ users, and their ratings on $9742$ movies events, which takes place over $T= 100,836$ time steps~\citep{harper2015movielens}. We construct a fair online allocation instance where each user visit is an impression opportunity and the agents are the $n=10$ most frequent genres. We map each user's average rating for a genre to an acceptance probability, complete the resulting user--genre matrix via matrix completion, and define each genre's utility as its expected number of acceptances. 

\subsection{Results for $p = -1$ and $p = 0.5$}

Here we report the simulation results for different generalized-mean welfare objectives with $p = -1$ (the harmonic mean) and $p = 0.5$ (a choice between the utilitarian and Nash metrics). In these simulations, the overall pattern we describe in the $p=0$ case still holds. 

\begin{figure*}[!htbp] 
    \centering
    \begin{subfigure}
	\centering
\includegraphics[width=0.49\linewidth]{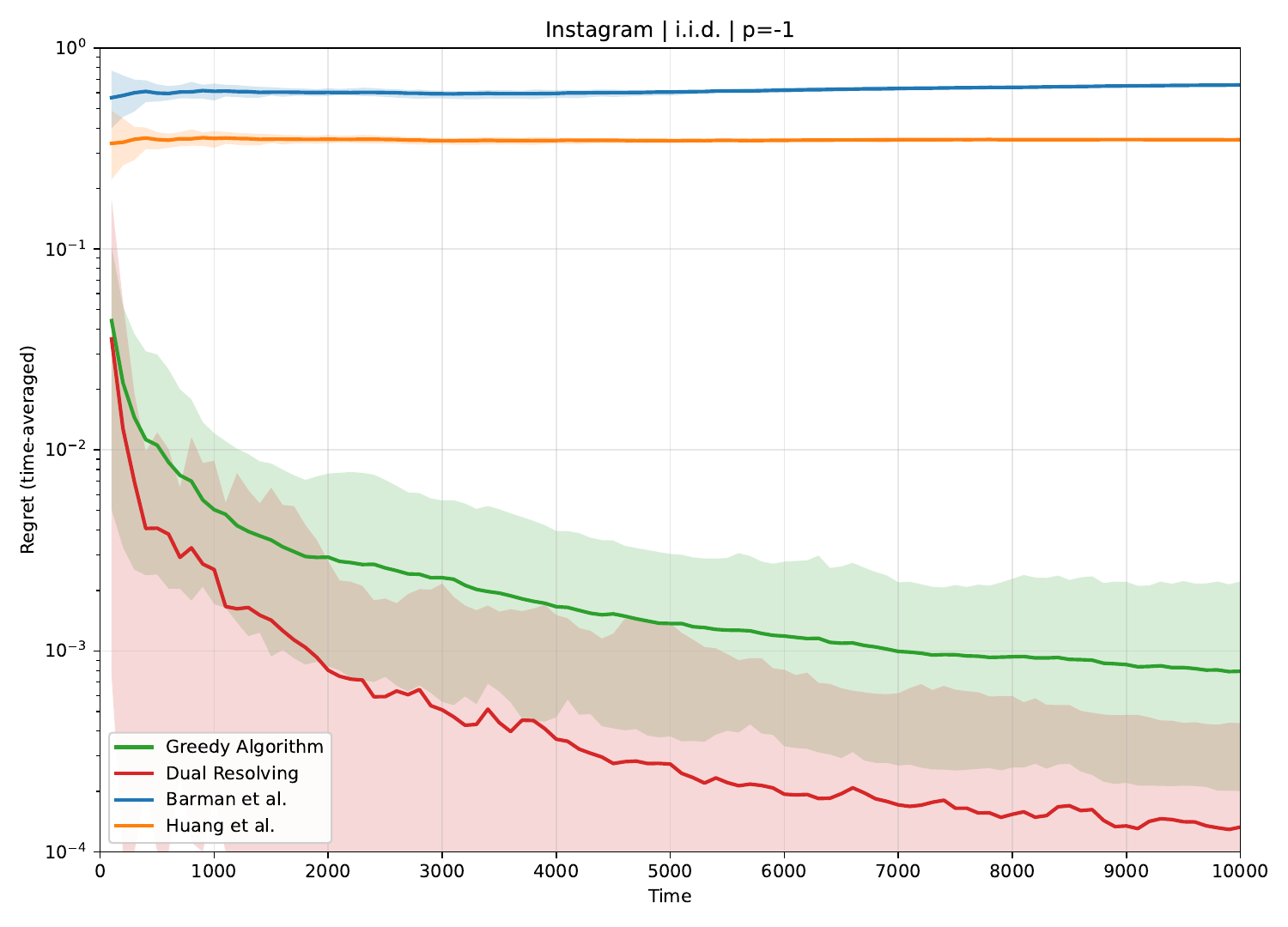}
	\end{subfigure}
     \begin{subfigure}
	\centering
\includegraphics[width=0.49\linewidth]{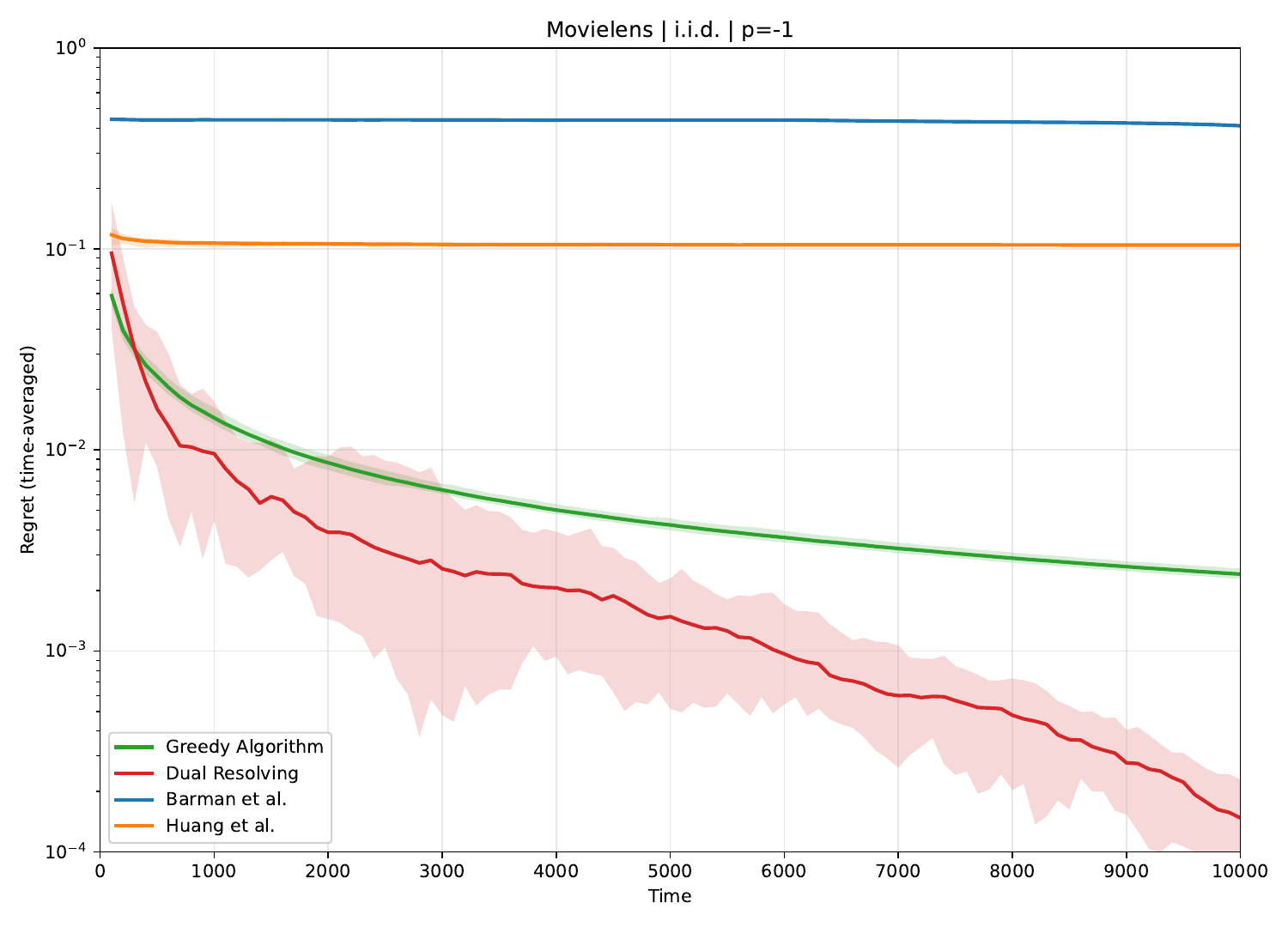}
	\end{subfigure}
     \begin{subfigure}
	\centering
\includegraphics[width=0.49\linewidth]{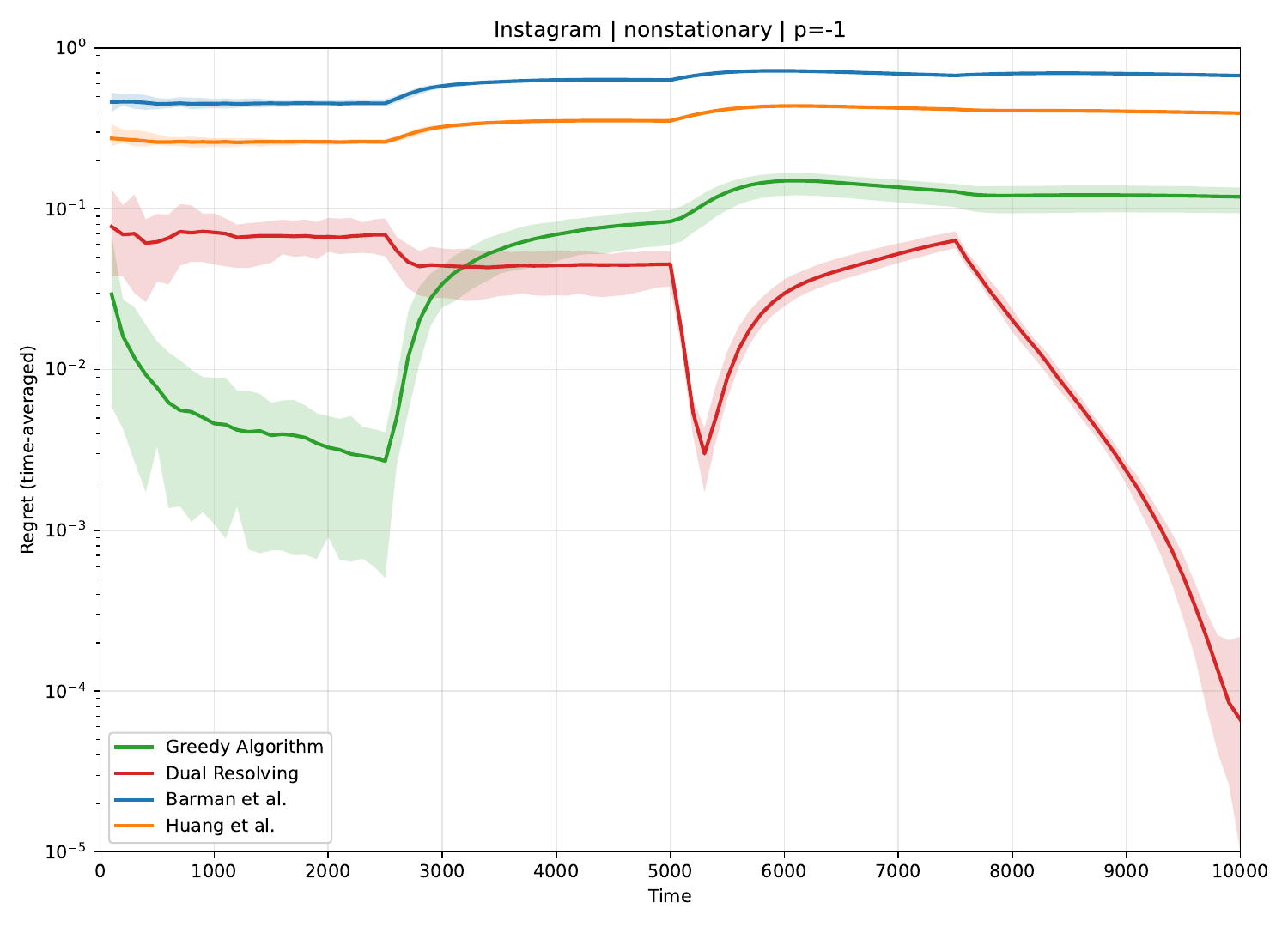}
	\end{subfigure}
     \begin{subfigure}
	\centering
\includegraphics[width=0.49\linewidth]{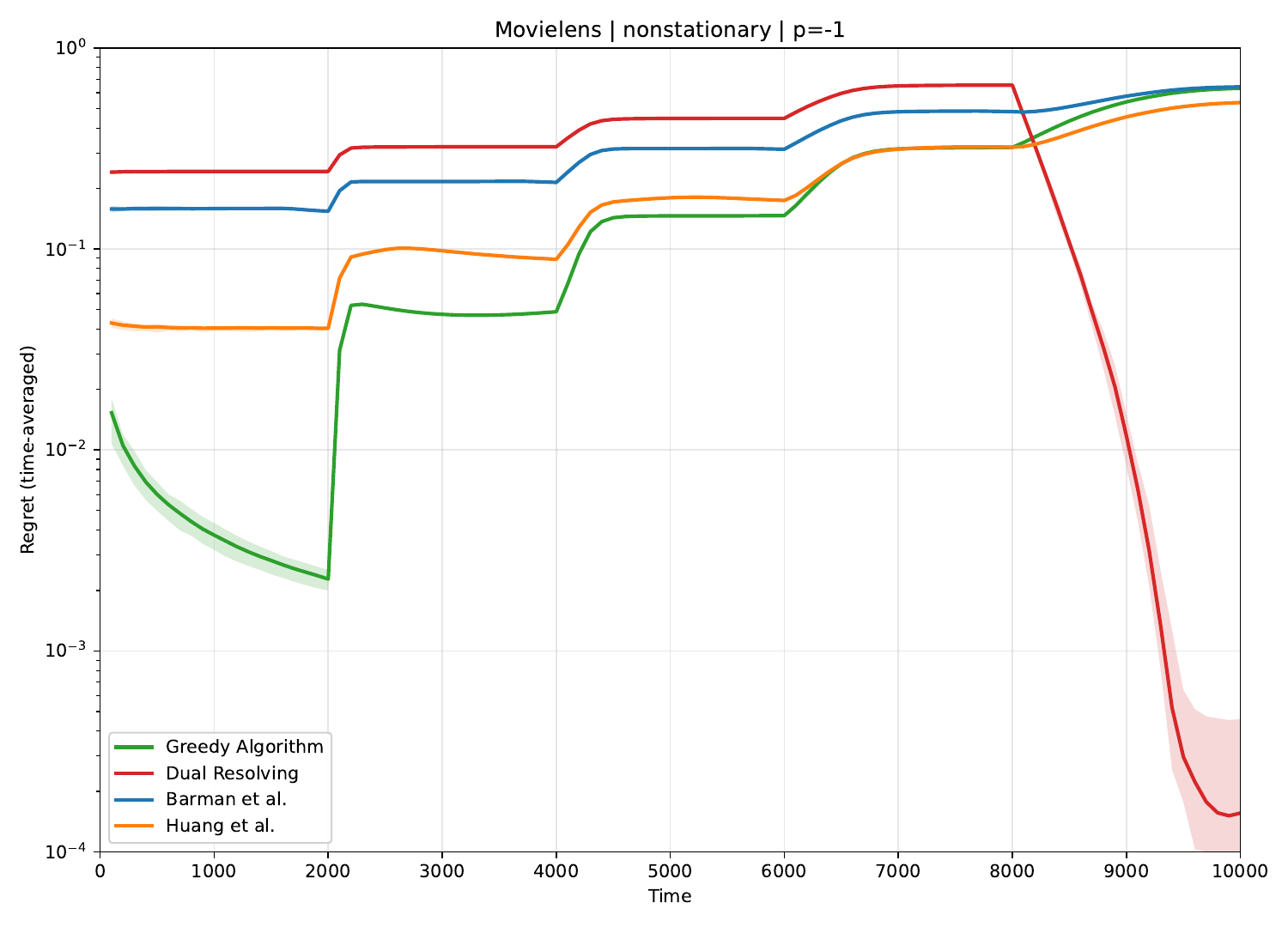}
	\end{subfigure}
     \begin{subfigure}
	\centering
\includegraphics[width=0.49\linewidth]{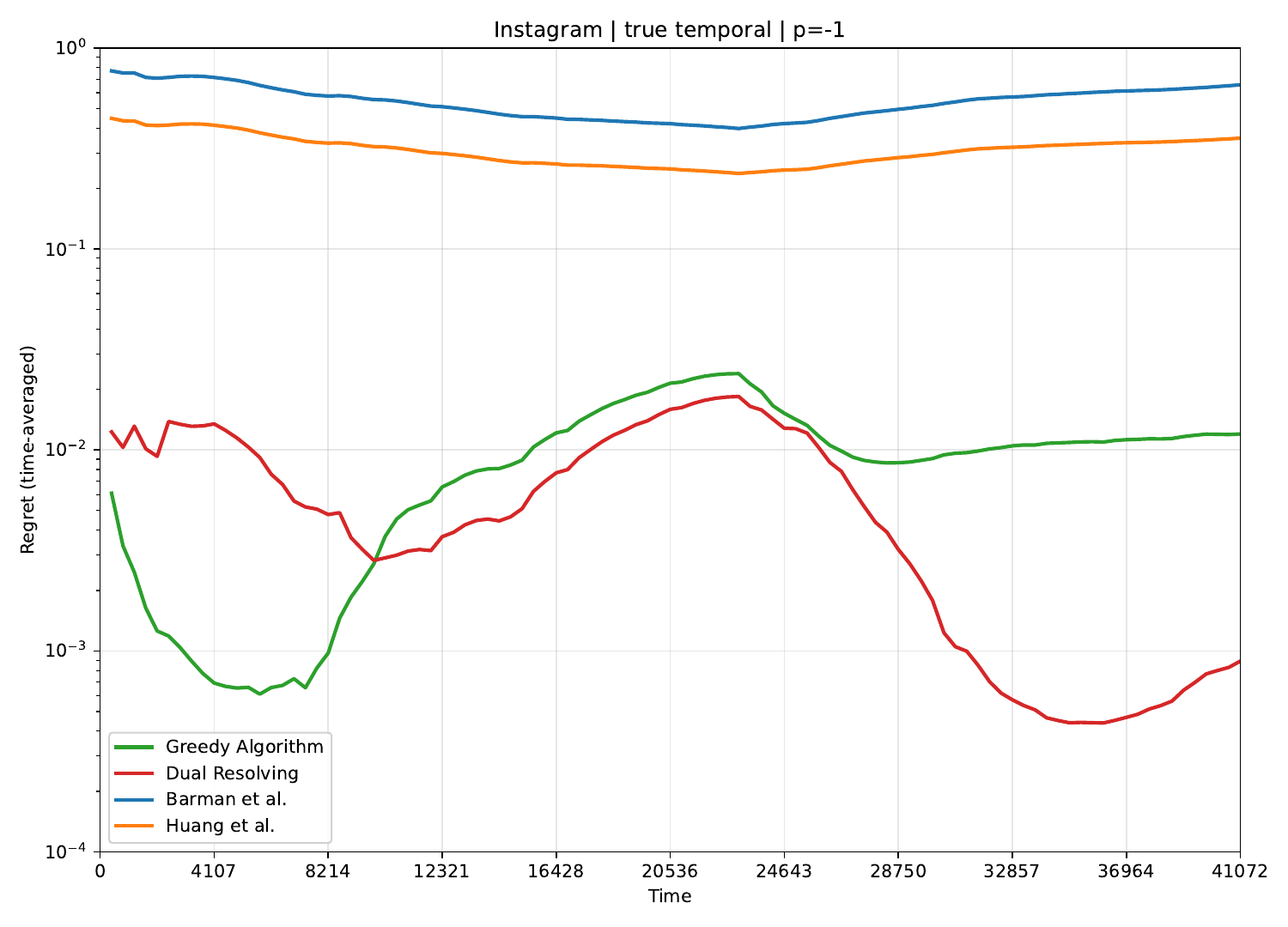}
	\end{subfigure}
     \begin{subfigure}
	\centering
\includegraphics[width=0.49\linewidth]{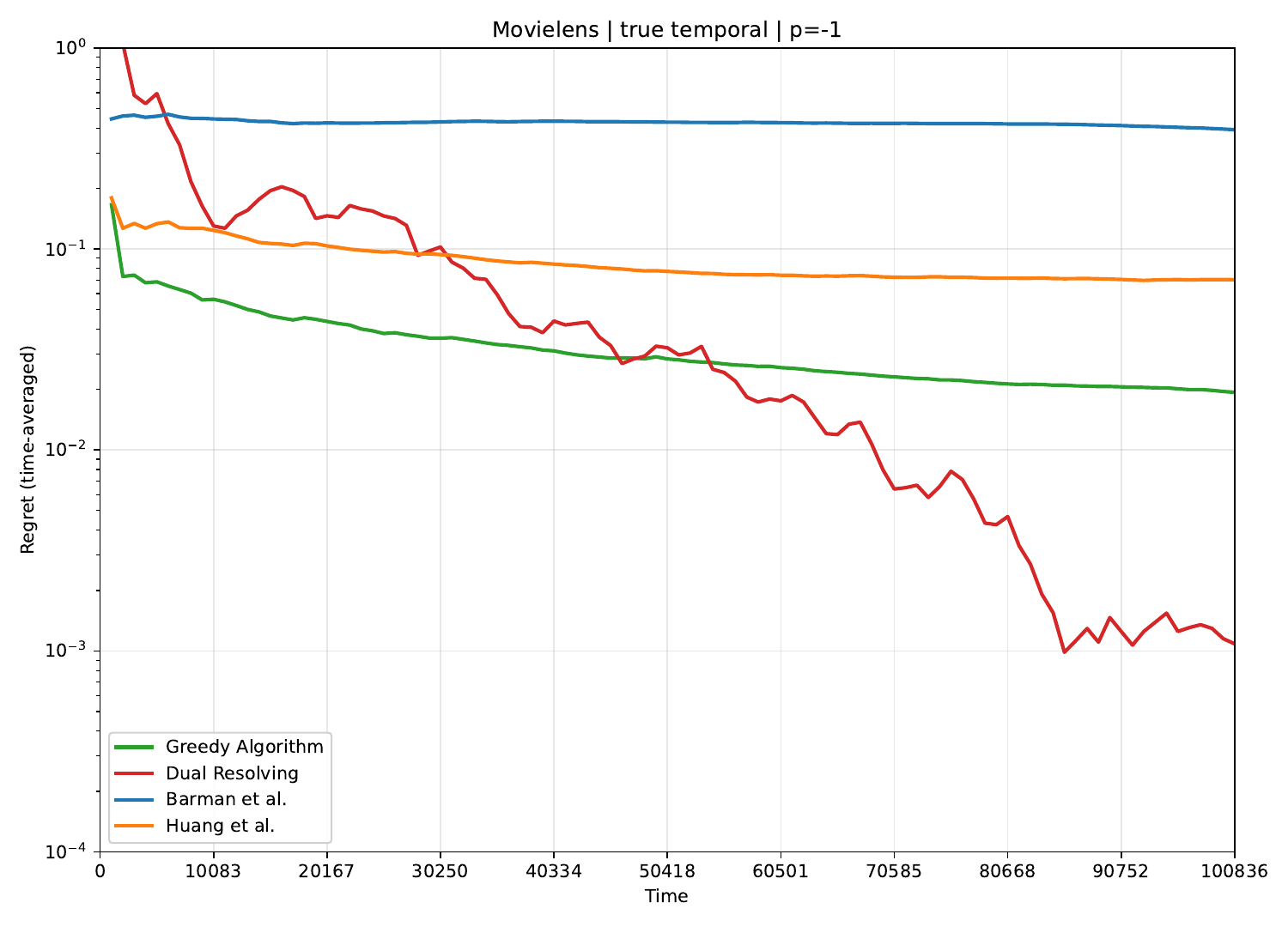}
	\end{subfigure}

    \caption{\small Simulations of the greedy algorithm and the re-solving algorithm on the Instagram notification dataset \textit{(left)} and the MovieLens dataset \textit{(right)} with the $p=-1$ in the objective, under three inputs models: \textit{i.i.d.}\ \textit{(top)}, periodic \textit{(middle)}, and true temporal \textit{(bottom)}.}
    \label{fig: p=-1}
\end{figure*}

\begin{figure*}[!htbp] 
    \centering
    \begin{subfigure}
	\centering
\includegraphics[width=0.49\linewidth]{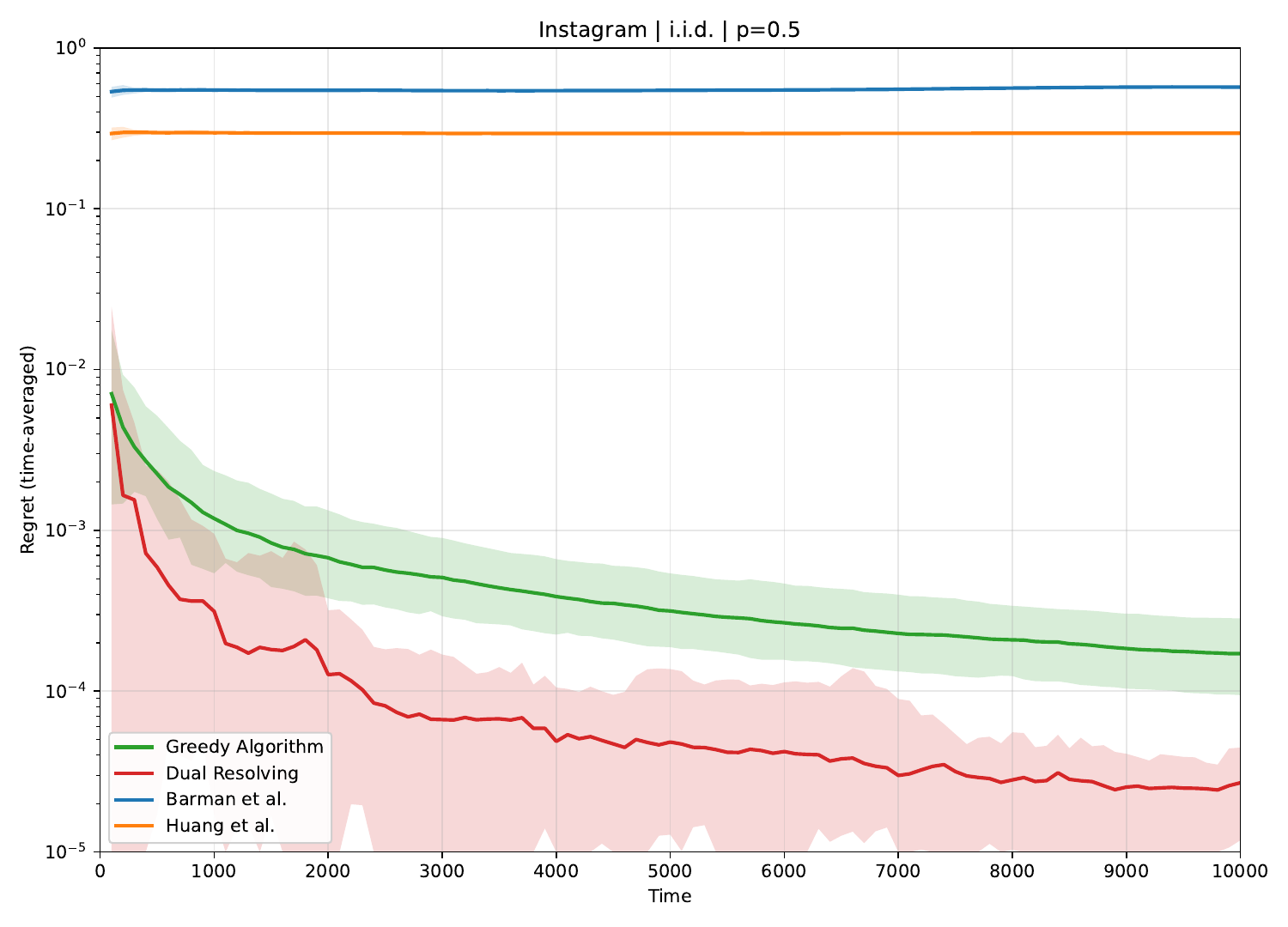}
	\end{subfigure}
     \begin{subfigure}
	\centering
\includegraphics[width=0.49\linewidth]{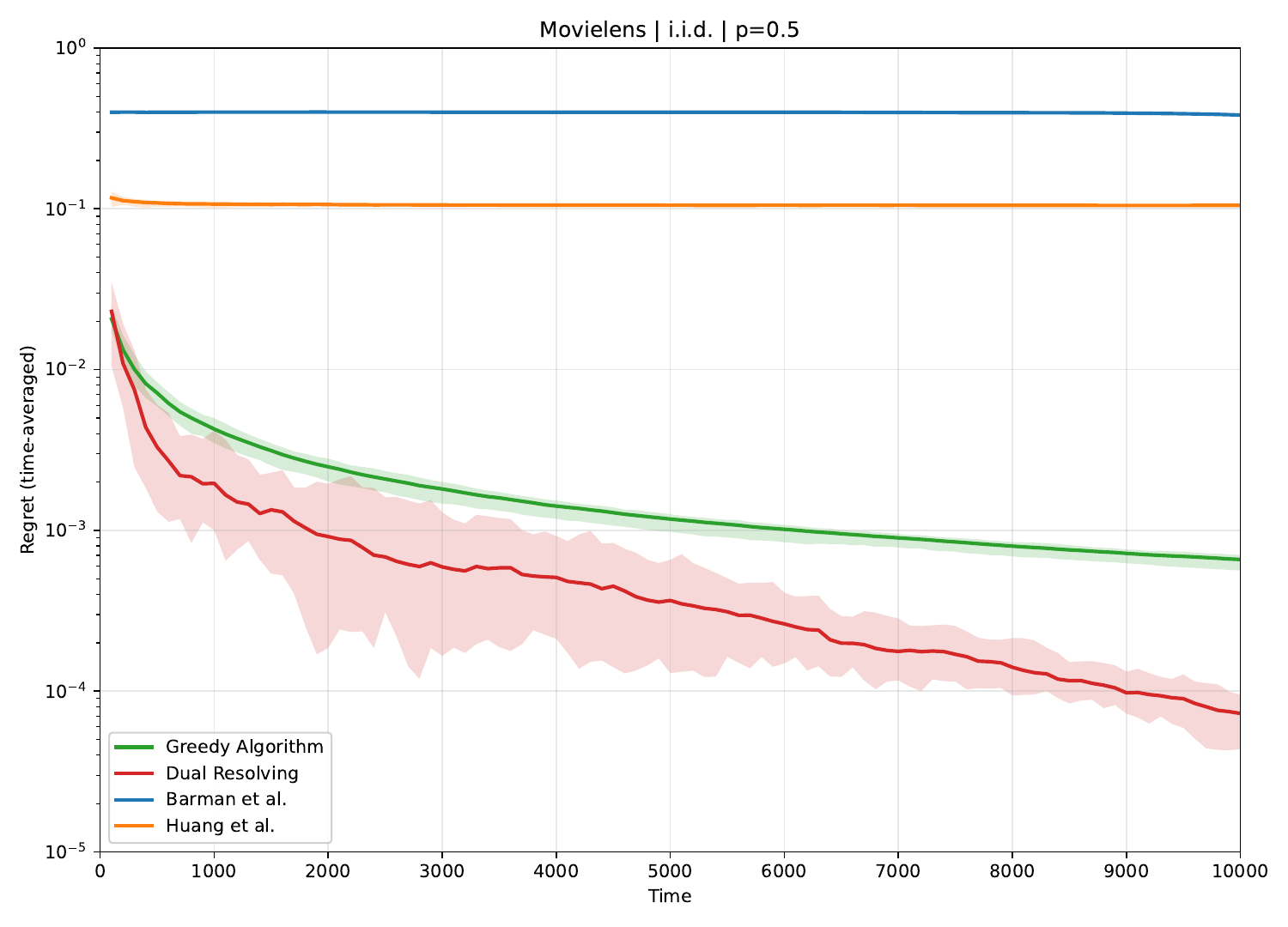}
	\end{subfigure}
     \begin{subfigure}
	\centering
\includegraphics[width=0.49\linewidth]{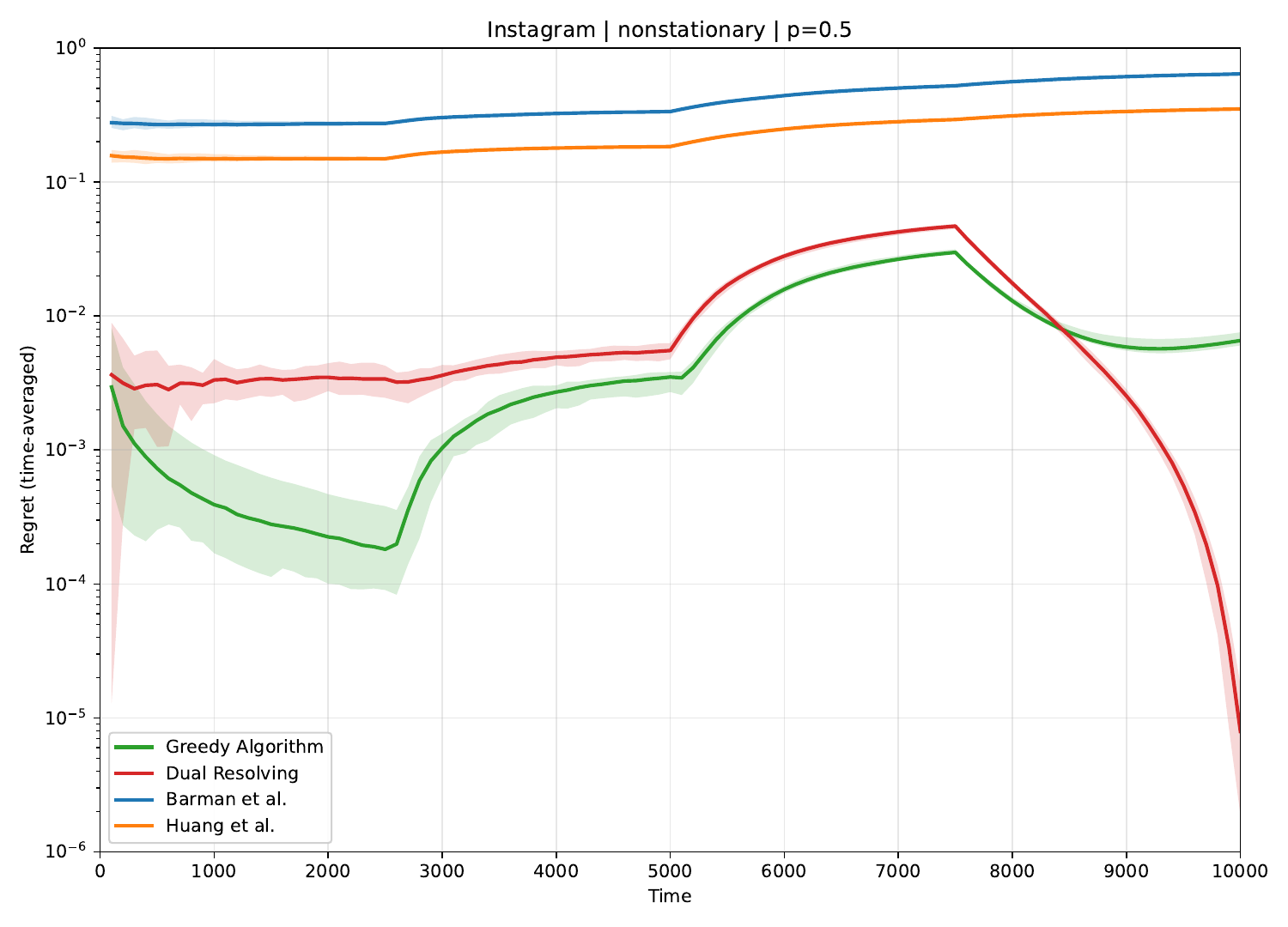}
	\end{subfigure}
     \begin{subfigure}
	\centering
\includegraphics[width=0.49\linewidth]{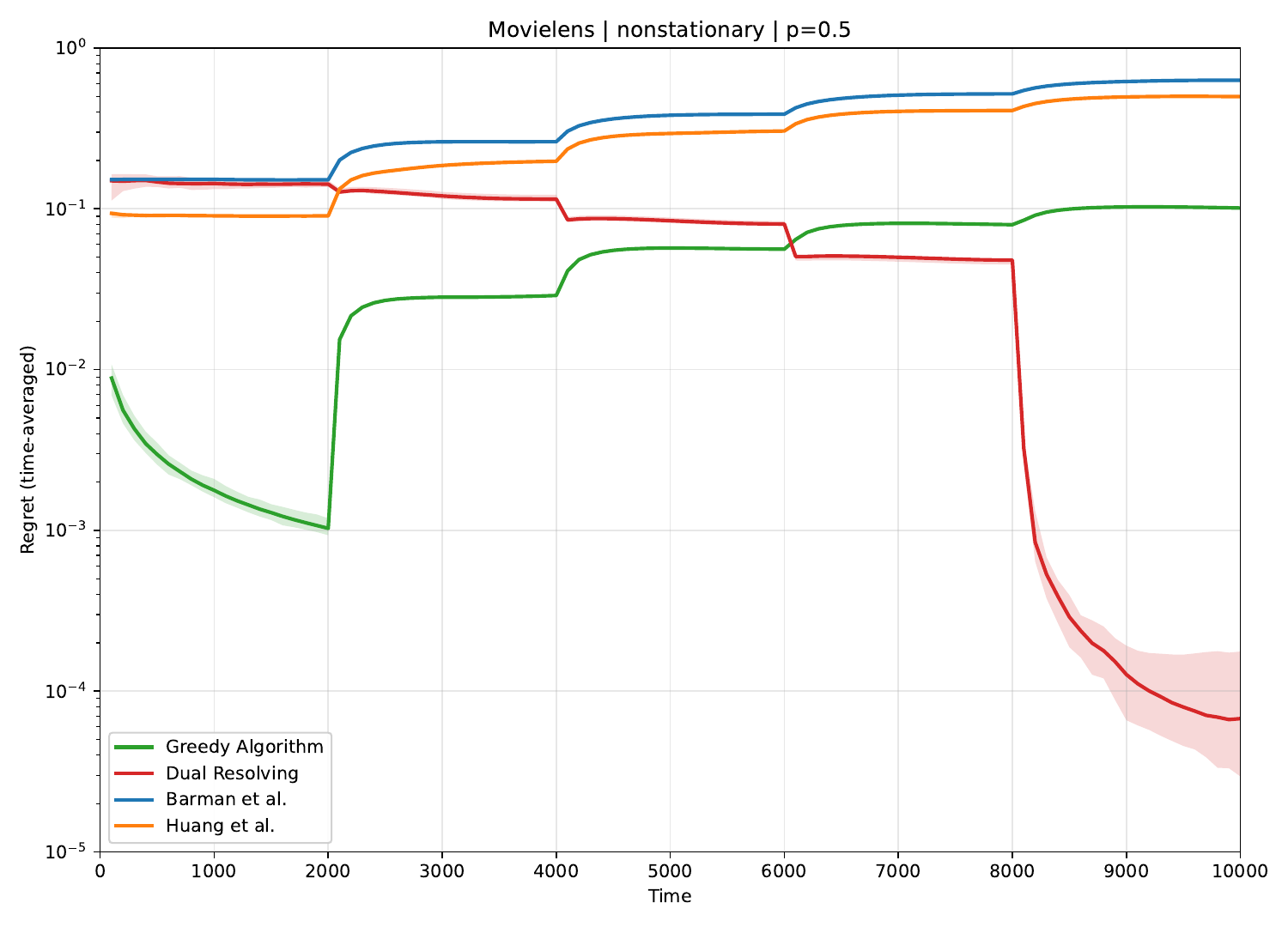}
	\end{subfigure}
     \begin{subfigure}
	\centering
\includegraphics[width=0.49\linewidth]{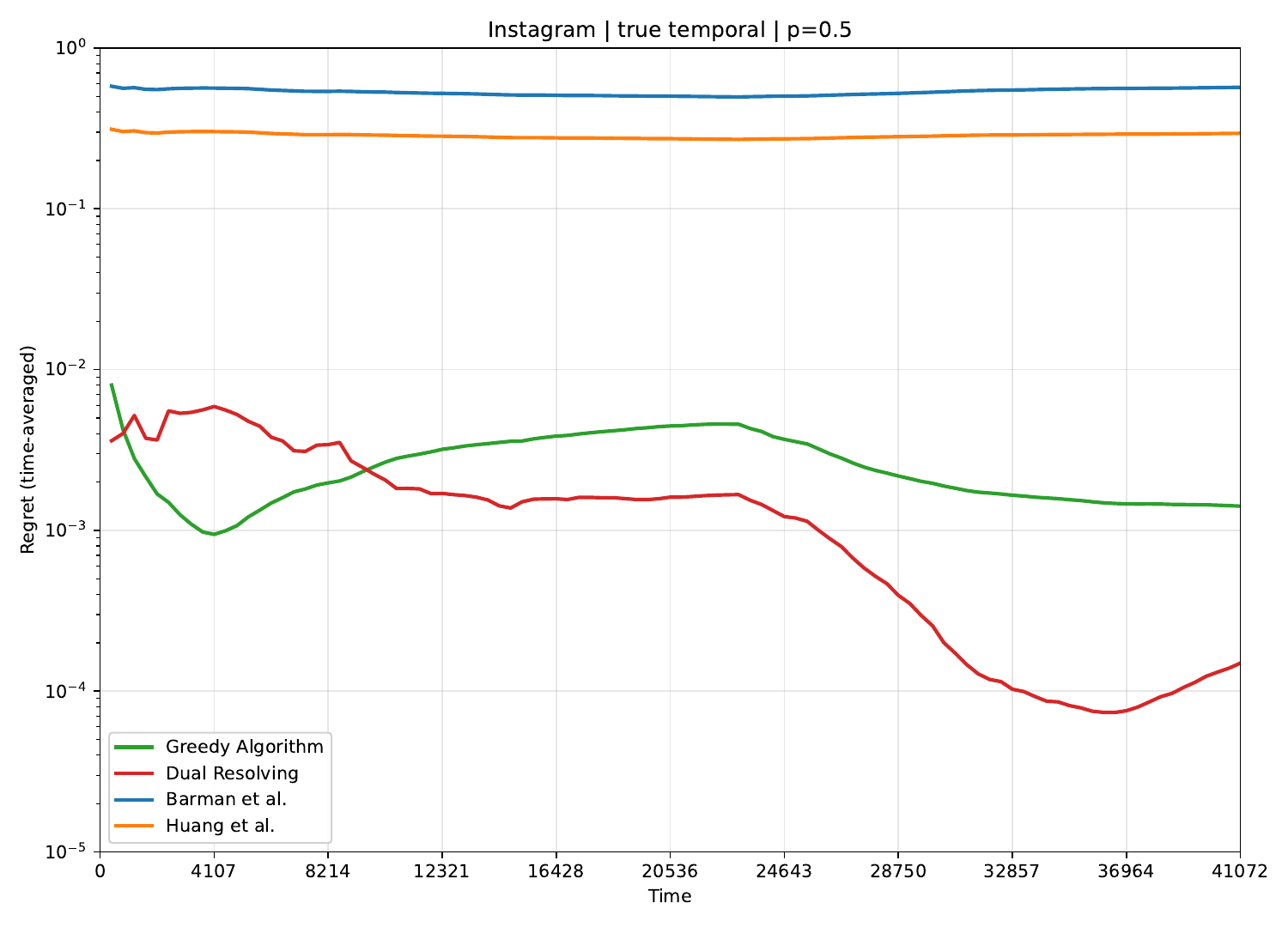}
	\end{subfigure}
     \begin{subfigure}
	\centering
\includegraphics[width=0.49\linewidth]{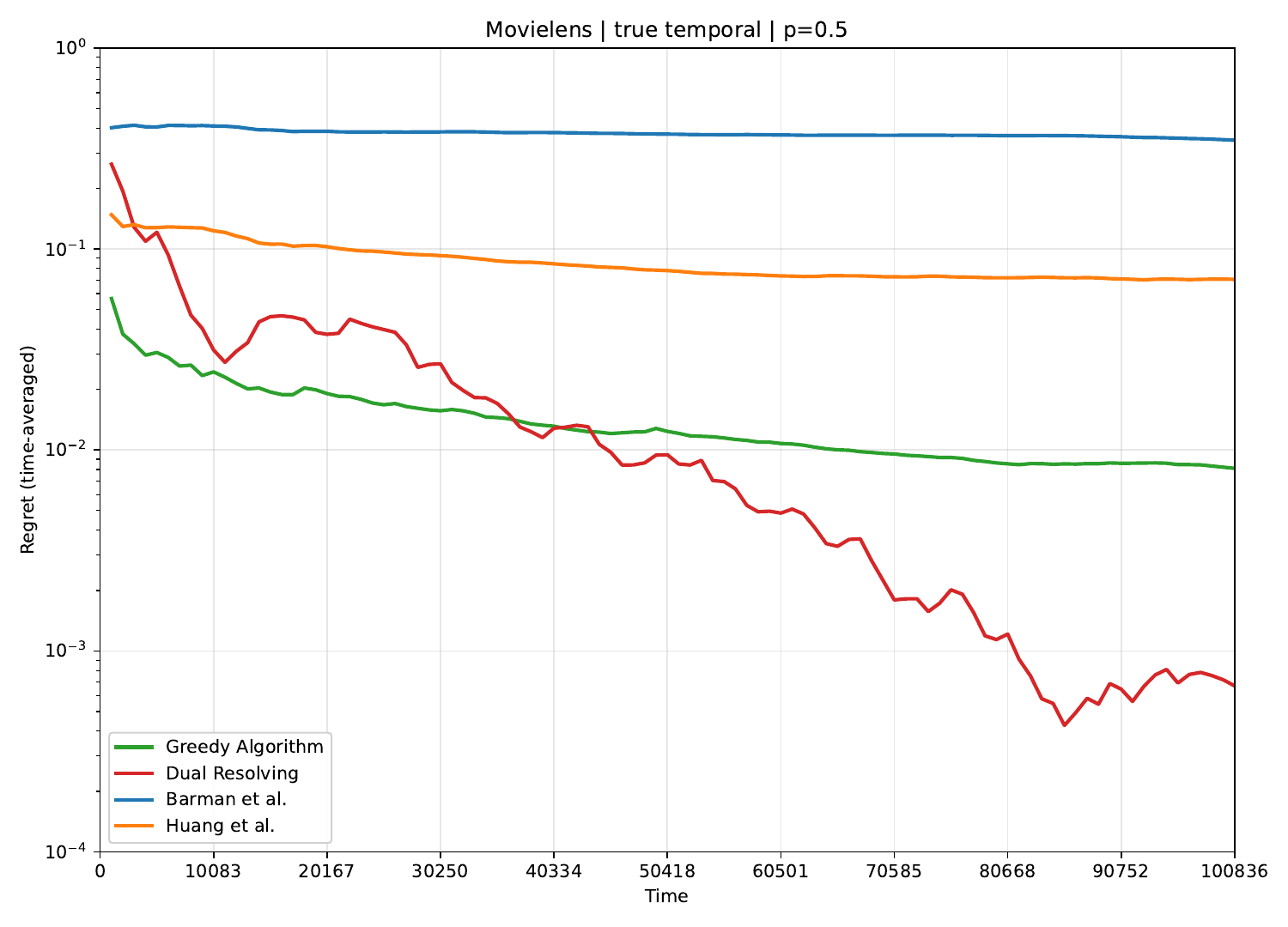}
	\end{subfigure}

    \caption{\small Simulations of the greedy algorithm and the re-solving algorithm on the Instagram notification dataset \textit{(left)} and the MovieLens dataset \textit{(right)} with the $p=0.5$ in the objective, under three inputs models: \textit{i.i.d.}\ \textit{(top)}, periodic \textit{(middle)}, and true temporal \textit{(bottom)}.}
    \label{fig: p=0.5}
\end{figure*}

\end{document}